\documentclass[acmlarge,screen,nonacm,10pt]{acmart}

\usepackage{multicol}
\usepackage{amsmath}
\usepackage{ulem}
\usepackage{subfigure}
\usepackage{algorithm2e}
\usepackage{algorithmic}
\usepackage{multirow}
\usepackage{makecell}
\usepackage{tabularx}
\usepackage{multicol}
\usepackage{nicematrix}
\usepackage{array}
\usepackage{tabularray}
\usepackage{enumitem}
\setlist[itemize]{leftmargin=*}
\usepackage{adjustbox}
\usepackage{marginnote}
\usepackage{titlesec}
\usepackage{amsthm}
\usepackage{bm}
\usepackage{diagbox}
\usepackage{longtable}
\usepackage{mathtools,leftindex,leftidx,tensor,mhchem}
\setcopyright{none}

\newcommand{\segset}{\mathcal{S}}
\newcommand{\seg}{\mathbf{s}}
\definecolor{ao(english)}{rgb}{0.0, 0.5, 0.0}
\titleformat{\paragraph}
{\normalfont\normalsize\bfseries}{\theparagraph}{1em}{}
\titlespacing*{\paragraph}
{0pt}{0.25ex plus 1ex minus .2ex}{0.5ex plus .2ex}
\usepackage{relsize}

\newcolumntype{Y}{>{\hsize=\hsize\centering\arraybackslash}X}
\newcolumntype{Z}{>{\hsize=1.5\hsize\centering\arraybackslash}X}

\usepackage{geometry}
\usepackage{titlesec}
\titleformat{\section}
  {\LARGE\bfseries}{\thesection}{1em}{}
  
\titleformat{\subsection}
  {\Large\bfseries}{\thesubsection}{.5em}{}

\titleformat{\subsubsection}{\large\bfseries}{\thesubsubsection}{.5em}{}

\titleformat{\paragraph}{\normalsize\bfseries}{}{}{}

\usepackage{xcolor,colortbl}
\newcommand{\ps}[1]{\cellcolor{green!25}#1}
\newcommand{\pu}[1]{\cellcolor{olive!35}#1}
\newcommand{\ac}[1]{\cellcolor{red!25}#1}
\newcommand{\cs}[1]{\cellcolor{cyan!25}#1}
\newcommand{\cu}[1]{\cellcolor{violet!25}#1}
\newcommand{\fsw}[1]{\cellcolor{orange!25}#1}
\newcommand{\blk}[1]{\cellcolor{gray!25}#1}

\newcommand{\ct}[1]{{\bf\color{red}#1}}

\begin{document}

\title{Rigorous Simulation-based Testing for Autonomous Driving Systems – Targeting the Achilles' Heel of Four Open Autopilots 
}


\author{
\tabcolsep 0.3cm
\MakeTextLowercase{
\begin{tabular}{cccc}
    \MakeTextUppercase{C}hangwen \MakeTextUppercase{L}i$^\dag$ & \MakeTextUppercase{J}oseph \MakeTextUppercase{S}ifakis$^*$ & \MakeTextUppercase{R}ongjie \MakeTextUppercase{Y}an$^\dag$ &
    \MakeTextUppercase{J}ian \MakeTextUppercase{Z}hang$^\dag$\\
    {licw@ios.ac.cn} & {Joseph.Sifakis@univ-grenoble-alpes.fr} & {yrj@ios.ac.cn} & {zj@ios.ac.cn}
\end{tabular}}
}

\renewcommand{\shortauthors}{Changwen Li, Joseph Sifakis, Rongjie Yan, Jian Zhang}

\authorsaddresses{$\dag$ Key Laboratory of System Software (Chinese Academy of Sciences) and State Key Laboratory of Computer Science, Institute of Software, Chinese Academy of Sciences. \\
$*$ Univ. Grenoble Alpes, CNRS, Grenoble INP, VERIMAG and SUSTECH/RITAS}

\begin{abstract}
{\bf Abstract. }
Simulation-based testing remains the main approach for validating Autonomous Driving Systems. We propose a rigorous test method based on breaking down scenarios into simple ones, taking into account the fact that autopilots make decisions according to traffic rules whose application depends on local knowledge and context. This leads us to consider the autopilot as a dynamic system receiving three different types of vistas as input, each characterizing a specific driving operation and a corresponding control policy.

~
 
The test method for the considered vista types  generates test cases for critical configurations that place the vehicle under test in critical situations characterized by the transition from cautious behavior to progression in order to clear an obstacle. The test cases thus generated are realistic, i.e., they determine the initial conditions from which safe control policies are possible, based on knowledge of the vehicle's dynamic characteristics.
Constraint analysis identifies the most critical test cases, whose success implies the validity of less critical ones. Test coverage can therefore be greatly simplified. Critical test cases reveal major defects in Apollo, Autoware, and the Carla and LGSVL autopilots. Defects include accidents, software failures, and traffic rule violations that would be difficult to detect by random simulation, as the test cases lead to situations characterized by finely-tuned parameters of the vehicles involved, such as their relative position and speed.

Our results corroborate real-life observations and confirm that autonomous driving systems still have a long way to go before offering acceptable safety guarantees.

\end{abstract}

\renewcommand{\keywordsname}{\textbf{Keywords}}


\maketitle

\section{Introduction}

Validating the safety of Autonomous Driving Systems (ADSs) is of paramount importance for their acceptance, as they are critical systems for which strong trustworthiness guarantees are required~\cite{koopman2016challenges}. Validating ADSs using model-based techniques such as verification and static analysis is virtually impossible, as their modeling is hampered by two obstacles. The first is the overwhelming complexity of these systems and of their environment. The second reason is the inevitable integration of artificial intelligence components, e.g., for perception, which are regarded as black boxes. Therefore, testing remains the only possible validation method.

Testing is an experimental process that consists in validating a property $P(x,y)$ linking the input $x$ to the response $y$ of a system. Unlike verification, which involves reasoning about a system model to confirm the validity of a property, it can only be used to detect property violations. Consequently, test methods can only estimate the probability of a property being valid after a large number of experiments. 

\begin{figure}[h]
\centering
\includegraphics[width=0.85\textwidth]{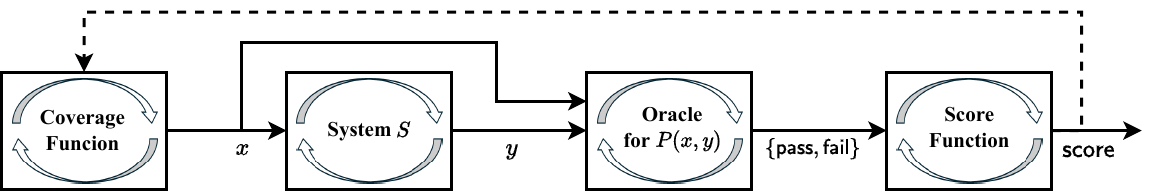}
\caption{Test method }\label{fig:testmethod}
\end{figure}

Given the property $P(x,y)$ of a system $S$, $y=S(x)$ producing output $y$ for input $x$, a test environment integrates the system under test $S$ and an Oracle which is an agent that evaluates the property $P(x,y)$ and produces verdicts (Fig.~\ref{fig:testmethod}). 

A test method~\cite{sifakis2023testing} provides guidelines for choosing among possible inputs $x$, called test cases, and deciding on the degree of validity of the property. It usually relies on a coverage function, $coverage(X)\in [0,1] $ that measures the extent to which the set of test cases $X$ explores the characteristics of the system's behavior in relation to the property $P(x,y)$. It also uses a score function, $score(X,Y)$, that provides for a test set $(X, Y)$, where $Y$ is the set of responses corresponding to $X$, the likelihood that $S$ meets $P$.

The coverage function is usually defined by reasoning on an abstract system model classifying the test cases so as to reduce the complexity of the exploration of the system’s behavior space. For example, in structural software testing, e.g., \cite{wegener2001evolutionary}, coverage functions give, for a set $X$ of test cases, the percentage of source code lines executed or the percentage of edges visited in the software's control flow graph. For functional testing e.g., ~\cite{jard2005tgv}, coverage functions indicate the success rate in satisfying a set of test purposes characterizing essential system properties. The score functions for software systems can be as simple as the success rate, or based on more sophisticated statistical criteria~\cite{denise2004generic,gouraud2001new}.

Note that for a test method an important requirement is reproducibility of results: If $(X_1,Y_1)$, $(X_2,Y_2)$ are two sets of tests then: $coverage(X_1)=coverage(X_2)$ implies $score(X_1,Y_1) \sim score(X_2,Y_2)$. This means that the scores of two different test sets with the same coverage are the same within some margin that should be estimated by the theory.  

ADS testing is mainly carried out by simulation, in particular to validate behavior in critical situations that are both dangerous and difficult to reproduce in real life. Developing rigorous test methods for ADSs poses extremely difficult problems, not least because it is hard to define adequate coverage criteria.
ADS testing has been the subject of numerous studies, ranging from the types of test cases to the methodologies used to generate tests and analyze their results, whether data-driven, knowledge-based or adversarial~\cite{ding2023survey}, and their implementation in simulation or real-world environments. There are a number of surveys on ADS test methods covering many facets of the subject~\cite{huang2016autonomous,lou2022testing,tang2023survey}. 

To test safety at system level, simulation techniques have been applied to complement and accelerate real-life testing. Simulation is essential for validating the safety properties of ADSs, in particular by applying scenarios that correspond to critical situations and are virtually impossible to test on highly regulated public roads. However, the realism of modeling and applied scenarios is an essential requirement for simulation approaches, to reflect reproducible real-life conditions as far as possible. In particular, scenarios must take into account both the context in which they take place and the ability of vehicles to perform the specified operations. To achieve this, scenario generators need to be coupled to the simulator, so that scenario choreography is compatible with the constraints of the physical environment, induced by its topology and traffic rules. In addition, it is important that the tests enable in-depth analysis and diagnosis, not only of accidents, but also of regulatory violations. To this end, test methods must enable us to establish that a failure could have been avoided under real-life conditions, and thus to establish the responsibility of vehicle autopilots. 

There is an extensive literature on simulation-based system testing, for example~\cite{kaur2021survey,zhong2021survey}. The present paper proposes a test method that reduces the complexity of the test cases by decomposing them into sequences of minimal realistic scenario types. The decomposition is based on the locality of context and knowledge, and the implied limited responsibility of the autopilot. The method calculates realistic and critical scenarios based on the analysis of safety constraints. These characterize the positions and relative speeds of the vehicles involved in a scenario according to their dynamic characteristics, which determine their ability to perform the corresponding operation. Critical scenarios are calculated as parameter configurations satisfying boundary conditions under rationality assumptions that any well-designed autopilot must satisfy. 

We provide test results around critical configurations revealing a large number of defects in four widely used open-source autopilots: Apollo~\cite{team2017apollo}, LGSVL~\cite{rong2020lgsvl}, Carla~\cite{dosovitskiy2017carla} and Autoware~\cite{autoware2024}. Identified defects lead to accidents or traffic violations, as well as to undesirable performance deterioration.       Consequently, the proposed test method aims to satisfy both route coverage and weight coverage criteria~\cite{tang2023survey}.

Intensive use of open-source autopilots reveals a large number of bugs of various causes reported in GitHub repositories and analyzed in articles such as~\cite{garcia2020comprehensive,tang2021issue}. From this point of view, our results come as no surprise. However, our method enables us to systematically pinpoint malfunctions and identify the sources of design errors that manifest themselves in configuration intervals that are difficult to detect by random testing.

The proposed test method considers that ADSs are dynamic systems with a number of vehicles and objects moving dynamically in a physical environment. The latter is a traffic system made up of roads with signaling equipment that defines the traffic rules to be respected by mobile agents. So, there is an explosive number of test cases, since positions and speeds are reals in given intervals. In addition, vehicle routes can intersect in different ways and be arbitrarily long. 

To reduce the complexity of the problem, it is factorized in three dimensions. 

{\bf Locality of context}: ADSs operate in complex environments taking on a wide variety of configurations, each of which can affect the behavior of the system. Therefore, ADS safety is strongly dependent on the context in which vehicles operate. The traffic infrastructure can be seen as the composition of a finite number of patterns comprising different types of roads and junctions with their signaling equipment. Each vehicle’s safety policy highly depends on the type of road or junction and their signaling equipment. We can therefore consider a road network as a composition of basic models for which different traffic rules are applicable, such as a section of freeway, a traffic circle, an intersection or a merger of roads.

{\bf Locality of knowledge}: A vehicle’s driving policy is based only on local knowledge of the system state due to limited visibility. 
It must therefore drive safely, taking into account any obstacles in its zone of visibility. 
In this way, the collective behavior of vehicles in an ADS can be understood and analyzed as the composition of smaller sets of vehicles grouped according to proximity and visibility criteria.

{\bf Rights-based responsibility}: 
ADSs are a special kind of distributed systems where each agent is responsible for managing a free space defined by traffic rules in its planned route. Hence, there are no explicit interaction rules between vehicles.
Traffic rules guarantee that if each vehicle drives safely in the space determined dynamically by its rights, then the whole system is safe. This principle of rights-based responsibility greatly simplifies the validation problem as the interaction between vehicles is unidirectional (flow-oriented)~\cite{shalev2017formal,hasuo2022responsibility}. It is enough to show that a vehicle drives safely in different contexts and configurations involving a limited number of other vehicles and objects. 

~

This factorization reduces the general problem to the testing of a single vehicle in different contexts characterized both by the type of road on which it travels and by the configuration of obstacles in its zone of visibility.

We call \textit{vista} the input of an autopilot generated by its perception function after analysis and interpretation of the information provided by the sensors. 
In this way, a vista defines the state of a vehicle's environment, including the positions and kinetic attributes of objects, as well as information on signaling equipment in the vehicle's visibility zone.
A vista also defines the limits of the responsibility of the vehicle. Based on the information provided by a vista, the autopilot should drive safely, acting reactively and assuming that the objects in the vista behave in accordance with traffic regulations.	

A vehicle's vista takes into account visibility parameters that depend on various factors in its environment, including topology and physical obstacles, as well as weather and light conditions. 

A simple analysis shows that driving policies depend on three different road patterns corresponding to three different types of vistas:	
\begin{enumerate}
\item 
\textit{road vistas} where there are no crossroads in the vehicle's area of visibility, and the autopilot is tasked with taking into account the obstacles in its route ahead; 	
\item  \textit{merging vistas} when the vehicle's route joins a road or lane where arriving  vehicles have a higher priority and it must therefore give way to these vehicles;	
\item  \textit{crossing vistas} where the vehicle's route crosses a junction accessible to other vehicles, and therefore, the vehicle must comply with the traffic rules applicable in this context.
\end{enumerate}

A key idea of this work is that to validate an ADS, it is sufficient to validate the behavior of its different vehicle types for relevant configurations of its environment in the different vista types. To fully support this idea, we need a compositional result showing that safe and legal travel along a route is guaranteed by safe and legal travel in scenarios corresponding to the successive vistas encountered. This result is demonstrated in a recently published paper on safe by design autonomous driving systems~\cite{bozga2024safe}.

Another fundamental idea is that each vista requires a specific operation involving the responsibility of the vehicle under test, called ego vehicle, in order to overcome a potential conflict. Our analysis shows that for each vista, critical situations can be characterized by configurations involving, in addition to the ego vehicle, an oncoming vehicle whose route may intersect that of the ego vehicle and a forward vehicle located after the intersection on the ego vehicle's route. 	

The main result of the paper is the identification of critical configurations between these vehicles for each type of vista, following an analysis that takes into account the dynamic characteristics of the vehicles, in particular their braking and acceleration  capacity, as well as the relationships between their speeds and their distances from locations where collisions can occur. 

The identified deficiencies include accidents mainly due to an overestimation of the vehicle's capabilities during the transition from carefully approaching a critical zone to attempting to cross it. Other minor deficiencies are due to over-cautious behavior that preserves safety but has an impact on performance, such as poor road occupancy and potential congestion. The analysis also reveals unrealistic assumptions about acceleration and deceleration rates, as well as control policies that depart from common-sense engineering principles.

Validating high-risk scenarios is a common idea, and one recommended by~\cite{najm2007pre}, which proposes a typology of pre-crash scenarios. However, our method is based on a technical classification that does not take into account the type of accidents possible, but rather the identification of configurations that are difficult to manage, characterized by constraints on the vehicle's environment and its kinetic state.

These configurations are formalized as relationships between two types of parameters. 	

The first type of parameters comprises the actual positions, speeds and routes of the vehicles concerned, defined in a mathematical model of the physical environment in the form of a map with a suitable scenario concept~\cite{li2023simulation,bozga2022specification}. 	

The second type of parameters consists of functions which characterize the acceleration and deceleration capabilities of the vehicle concerned. Assessing the dynamic characteristics of vehicles can require considerable experimental work, not least to integrate autopilots into a simulation and test environment. 

The calculated configurations identify realistic critical scenarios: for any failure detected by testing, it is possible to show that a correct and feasible control policy exists. From this point of view, it has certain similarities with ~\cite{djoudi2020simulation,mezali2022design}where the oracle applies measurable comparison criteria between the behavior of the vehicle under test and the behavior of a precise reference controller. However, these works address the problem for simple scenarios involving an ego vehicle and a front  vehicle. In addition, they characterize by constraints only those initial states from which an ideal reference controller would exhibit safe behavior without explicitly modeling its behavior. 
Finally, let's highlight the existence of testing approaches where the oracle uses a reference human behavior model~\cite{favaro2023building,kusano2022collision}  to define acceptance criteria. Tests of autopilots against reference human behavior focus on a ``positive risk balance''~\cite{favaro2021exploring}, demonstrating that autopilots can do better than human drivers, but may ultimately overlook cases where autopilots may fail while human drivers may succeed.

The paper is structured as follows.	
Section~\ref{se:approach} presents the overall approach including definition of basic concepts, the three types of vistas and corresponding control policy principles.	
Section~\ref{se:policy} presents the method for generating critical test cases.  Section~\ref{se:setting} presents the test environment, its implementation and the context of test case generation and tested properties.  Section~\ref{se:results} presents the test results for the four autopilots considered, and an analysis of the different types of defects revealed.  Section~\ref{se:discussion} discusses conclusions and future work directions.

\section{The Approach}\label{se:approach}
\subsection{Simulation-based test environment }

\begin{figure}[h]
\centering
\includegraphics[width=0.9\textwidth]{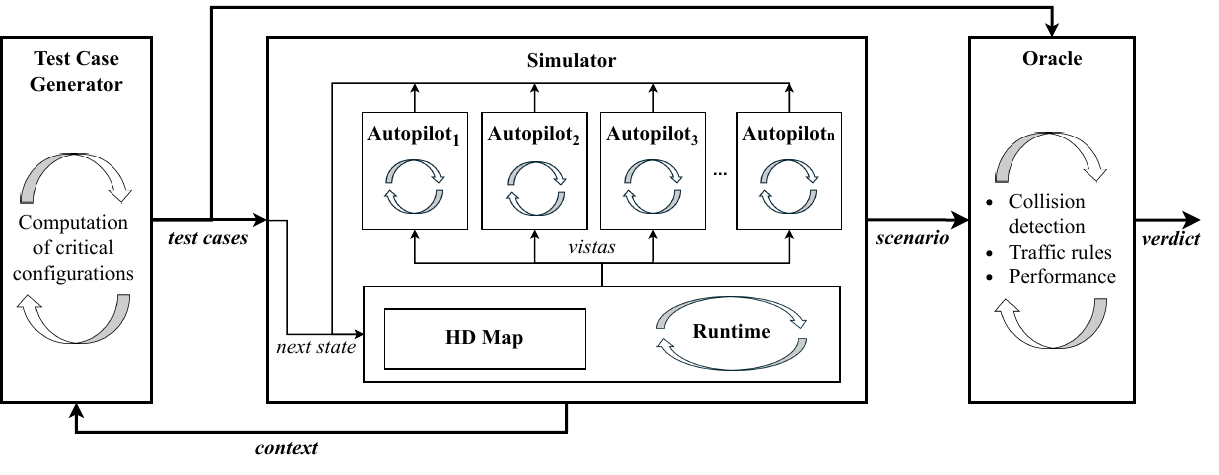}
\caption{Test environment}\label{fig:environment}
\end{figure}

The test environment for ADSs integrates three collaborating tools: 1) a Simulator; 2) a Test Case Generator; and 3) an Oracle (Fig.~\ref{fig:environment}). 

The Simulator executes an ADS model obtained as the composition of two entities. 	

The first entity is a model of the environment, which is usually represented by an HD map containing all the information relevant to determining the state of the system. The information includes vehicle positions and their kinetic attributes, obstacles around each vehicle, and signalling information used to enforce traffic rules. 

The second entity consists of behavioral models of the vehicle autopilots and their possible interactions. The Simulator's runtime exhibits cyclic behavior, alternating between concurrent executions of the vehicle's autopilots for a given period of time and calculation of the resulting system state on the environment model. At the beginning of a cycle, it provides the autopilots with their corresponding vistas that model the perceived state of the external environment. Then the autopilots compute for a predefined lapse of time, their new speeds and positions. The cycle ends by calculating the state of the resulting system and updating the kinetic state of the vehicles on the map.

The Simulator generates scenarios that are sequences of system states analyzed by the Oracle.

The Test Case Generator provides test cases that define initial system states designed to explore system behavior by selecting relevant configurations for a given simulation context. The configurations specify relative positions and speeds as well as the routes of the vehicles. Test cases must be realistic, i.e., they must define states from which at least one safe control policy exists for the vehicle involved.

For a given test case and the corresponding scenario, the Oracle checks that the system under test satisfies the given properties and provides verdicts. Verdicts indicate success or failure, including the possibility of accidents, traffic violations, performance degradation or other system defects. 

We define below basic concepts for the simulation-based test environment. 

\subsection{Environment modeling with maps}

We consider that a map is represented as a \textit{metric graph} $G= (V, E, \segset)$ \cite{bozga2022specification} where 
\begin{itemize}

\item $V$ is a finite set of vertices, 
\item $\segset$ is a set of segments defined below,
\item $E\subseteq V\times \segset\times V$ is a finite set of edges labeled by non-zero length segments in $\segset$.

\end{itemize}

Segments model abstraction of roads and characterize their geometric characteristics at different levels of abstraction. The highest level may ignore the form for the segment and give only its length. The lowest level can be a two-dimensional area. An intermediate level can be a curve showing the form of the road and making abstraction of its width. 
The set of segments S is equipped with a partial concatenation operator $\cdot: \segset\times \segset\rightarrow S\cup \{\bot\}$ and a length norm $||\cdot ||: S\rightarrow R_{\geq 0}$ satisfying the following properties:
\begin{enumerate}
\item \textit{associativity}: for any segments $\seg_1, \seg_2, \seg_3$ either both $(\seg_1\cdot \seg_2)\cdot \seg_3$ and $s_1\cdot (\seg_2\cdot \seg_3 )$ are defined and equal, or both undefined;
\item \textit{length additivity wrt concatenation}: for any segments $\seg_1, \seg_2$ whenever $\seg_1, \seg_2$ are defined it holds $||\seg_1\cdot \seg_2|| = ||\seg_1|| + ||\seg_2||$;
\item \textit{segment split}: for any segment $\seg$ and non-negative $a_1, a_2$, such that $||\seg|| = a_1 + a_2$ there exist unique $\seg_1, \seg_2$ such that $\seg = \seg_1\cdot \seg_2$, $||\seg_1|| =a_1,  ||\seg_2|| =a_2$.
\end{enumerate}

Note that for curves the concatenation $\seg_1\cdot \seg_2$ is defined only  if the derivative at the end point of $\seg_1$ is equal to the derivative at the beginning point of $\seg_2$.


A position $p$ of $G$ is uniquely defined as a point of a segment $\seg$ associated with the edge $e$ of $G$. We adopt the following notations.  
\begin{itemize}
\item If $p$ is a position of a segment $\seg$ at distance $d$ from the origin of $\seg$, we write $d=\seg(p)$.
\item If $p$ and $p'$ are two positions of the same segment $\seg$ such that $d=\seg(p), d'=\seg(p')$ and $d\#d'$   where $\#$ is a relational symbol in set $\{=, <, \leq\}$, then we write $p \#_\seg p'$. Additionally, we write $p'= p +_\seg  d''$ or equivalently $p'-_\seg p=d''$ to express the fact that $p'$ is ahead of $p$ at distance $d''=d'-d$.  
\item If $p<_{\seg} p'$, we write $\seg[p,p']$ to denote the sub-segment of $\seg$ delimited by positions $p$ and $p'$.

\item We write $p \leftindex_\seg=_{\seg'} p'$ if positions $p$ and $p'$, respectively on $\seg$ and $\seg'$, are coincide.

\end{itemize}


We consider a map to be a metric graph, where the edges represent the building blocks of a traffic infrastructure with their attributes. Depending on the abstraction level, an edge can represent a road, which may consist of several lanes. 

Driving through junctions is subject to specific rules and corresponding operations depending on their type: roundabout, crossroad, highway entry, etc.  Additionally, each road element can be equipped with signaling
features such as stop, yield, road works, and traffic lights.


A weakly connected metric graph $G = (V, \segset, E)$ can be interpreted as a map with a set of roads $R$ and a set of junctions $J$, defined in the following manner:
\begin{itemize}
\item a road $r$ of $G$ is a maximal directed path $r = (v_0, \seg_1,v_1) (v_1,\seg_2,v_2)\cdots$ $(v_{n-1},\seg_n,v_n)$, where all the vertices $v_1,\ldots ,v_{n-1}$ have indegree and outdegree equal to one. We say that $v_0$ is the entrance and $v_n$ is the exit of $r$. 	
Let $R =\{r_i\}_{i\in I}$ be the set of roads of $G$.
\item	a junction $j$ of $G$ is any maximal subgraph $G'$ of $G$, obtained from $G$ by removing from its roads all the vertices (and connecting edges) except their entrances and exits. Additionally, we require that from any vertex of indegree zero there is a path to a vertex of outdegree zero. 
Let $J = \{j_m\}_{m\in M}$ be the set of junctions of $G$.
\end{itemize}

Note that any metric graph $G$ specifying a map, is the union of the subgraphs representing its roads and junctions. 
We assume that maps include information about features of roads and junctions that are relevant to traffic regulations. In particular, roads and junctions can be typed. Road types can be highway, built-up area roads, carriage roads, etc. Junction types can be roundabouts, crossroads, highway exit, highway entrance, etc.

\subsection{ADSs as dynamic systems}

An ADS is a \textit{dynamic system} involving a set of vehicles $C$, a set of objects $O$, and a map $G$ that is the abstraction of the environment where the objects are located and the vehicles can move. 
The state $q$ of an ADS is the union of the states of its vehicles and objects: $q= qC\cup qO$, where  $qC=\{q_c\}_{c\in C}$ and $qO= \{q_o\}_{o\in O}$ where
\begin{itemize}
\item The state of a vehicle $c$, is a tuple $q_c=\langle \seg_c, p_c, v_c, \ldots \rangle$, where $\seg_c$ is a segment representing the route of $c$, $p_c$ is its current position on $s_c$ and $v_c$ is its speed. 
\item	The state of an object $o$, is a tuple $q_o= \langle p_o, a_o, \ldots \rangle$, where $p_o$ is its position and $a_o$ denotes its attributes. We consider the objects to be signaling equipment of three different types: 
\begin{itemize}
\item	Speed limits, such that $a_o= vl$ is the enforced speed; 
\item Stop or yield signs guarding junctions and characterized by their position and type attribute;
\item	Lights with an attribute \textit{color} taking values ``red'' or ``green”.
\end{itemize}
\end{itemize}
An ADS evolves from an initial state $q_0=qC_0\cup qO_0$ and through states $q_i=qC_i\cup qO_i$, with $q_i \xrightarrow{\Delta t} q_{(i+1)}$ where $\Delta t$ is an adequately chosen time step. The latter can be the period of the Simulator.

\begin{figure}[h]
\centering
\includegraphics[width=0.45\textwidth]{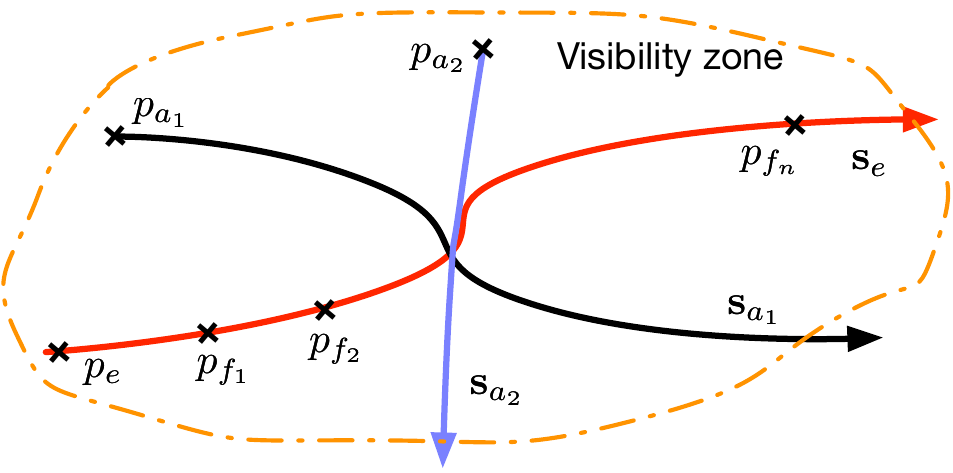}
\caption{Vista of an ego vehicle with position $p_e$ and route $\seg_e$ and two arriving vehicles}\label{fig:vistaview}
\end{figure}

\vspace{10pt}
\noindent{\bf Visibility zone}:
For a reference vehicle called \textit{ego vehicle}, we define the concept of \textit{vista} characterized by its state and the states of the obstacles in its visibility zone. The latter is defined by an area of the map around the ego vehicle using two types of parameters (Fig.~\ref{fig:vistaview}):
\begin{itemize}
\item	The frontal visibility of the ego vehicle on its route up to a front distance $fd(q_e)$ delimiting an interval $\seg_e[p_e,p_e+_{\seg_e}fd(q_e)]$ in which the ego vehicle’s autopilot can perceive the objects on its route. The frontal visibility limit $p_e+_{\seg_e}fd(q_e)$ depends on factors such as road curvature, obstacles in view, and weather conditions at position $p_e$.
\item	The lateral visibility of an ego vehicle whose route joins a road or lane of the road on which it is travelling, is the distance from the junction point where the ego vehicle can perceive the vehicles travelling on this road or lane. If $r$ is the name of the road or the lane then we denote by $ld_r(q_e)$ this distance. 
\end{itemize}

These parameters delimit the visibility zone by points on the route of the ego vehicle and also on the possibly intersecting routes from which vehicles can arrive.

\vspace{10pt}
\noindent {\bf Vistas}: Given an ADS, a vista for an ego vehicle with state $q_e$ is a triple $vs=\langle q_e, qF, qA\rangle$ where	
\begin{itemize}
\item	$q_e$ is the state of the ego vehicle.
\item	$qF$ is the ordered set of the states of \textit{front obstacles} $ F=\{f_1,\ldots, f_n\}\subseteq C\cup O$ located in the front visibility zone of the ego vehicle such that $p_e< p_{f_1}<_{\seg_e} …. <_{\seg_e} p_{f_n}=p_e+_{\seg_e}fd(q_e)$. We consider that the last visible obstacle is a fictitious obstacle at the front visibility limit.
\item	$qA$ is the ordered set of the states of the \textit{arriving vehicles} $A \subseteq C$, at most one per entrance of a junction within the forward visibility zone of the vehicle. If there is no real vehicle $a_i$ in the visible segment to the $i$th entrance, then we consider a fictitious vehicle $a_i$ with state $\langle \seg_{a_i}, v_{a_i}\rangle $ defined by: 
\begin{itemize}
\item	$\seg_{a_i}$ is the segment of a road of length $|\seg_{a_i}|= ld_i(q_e)$ that ends at the $i$th entrance of the junction;
\item	$v_{a_i}$ is the speed limit enforced on this road.	
\end{itemize}
Thus, for arriving vehicles $a_i$ fictitious or not, we have: $\exists p, p_i>0, p_e<_{\seg_e}p, ~p_i-_{\seg_{a_i}}ld_i(q_e)\leq_{\seg_{a_i}} p_{a_i}<_{\seg_{a_i}} p_i$ and $p \leftindex_{\seg_e}{=}_{\seg_{a_i}}p_i$; that is the itinerary of each $a_i$ intersects $\seg_e$ at the same point $p$.
\end{itemize}
	
\vspace{5pt}
Note that while the vista $vs$ includes the status of all visible obstacles in the ego vehicle's route, it only includes the state of a single arriving vehicle crossing its route within the forward visibility limit.
The precautionary principle requires us to consider fictitious objects at the limits of the visibility zone. These fictitious objects can be a frontal obstacle at distance $fd(q_e)$, or a vehicle arriving from a road or track joining its route at distance $ld(q_e)$. In this way, visibility constraints are implicitly taken into account in a vista.

An ADS with $m$ vehicles is a dynamic system using a map $G$ and having states $q(t)$ that can change after time $\Delta t$ to $q(t+\Delta t)$. 	

\begin{figure}[h]
\centering
\includegraphics[width=0.8\textwidth]{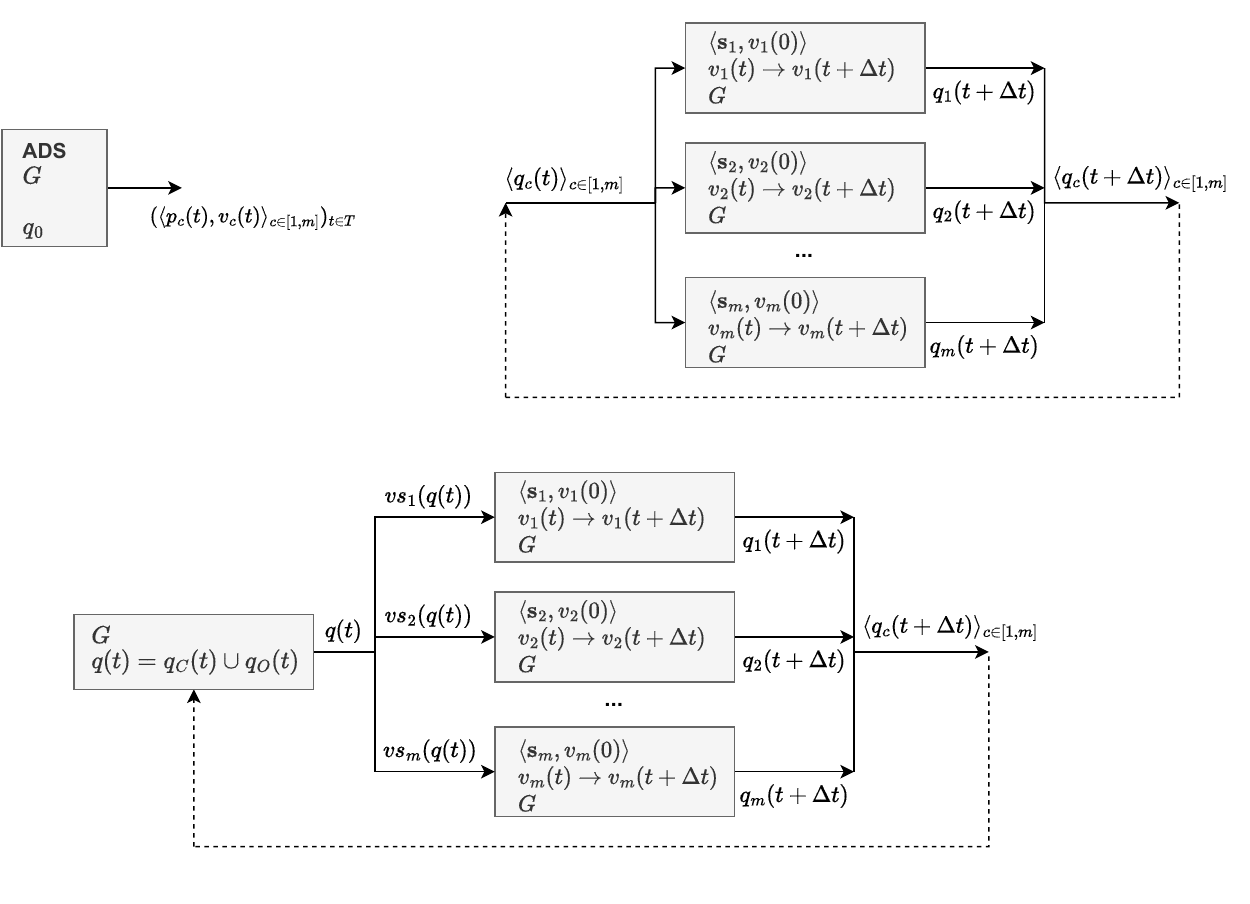} 
\caption{An ADS as a dynamic system}\label{fig:dynsys}
\end{figure}

Fig.~\ref{fig:dynsys}, top left, shows an ADS represented by a single component which, for a given map $G$ and starting from the initial state {$q_0$}, produces a timed sequence of vehicle positions and speeds. 

At the top right of this figure is a decomposition of the ADS as a dynamic system with autopilots, one for each vehicle $c_i$, which for given global system state $q(t)$ at time $t$ and map $G$, 
compute their new state $q_i(t+ \Delta t)$ at time $t+ \Delta t$.

In the lower part of the same figure, this architecture is refined by the addition of a component representing the vehicle's environment, which alone knows the overall state of the system and communicates to each vehicle only its vista, from which it calculates its next state.
The environment component receives the states of the vehicles at the end of each cycle and computes the global state of the ADS using the map and the knowledge of the state of the objects. Furthermore, it produces for the $i$-th autopilot the corresponding vista $vs_i(q(t))$, taking into account visibility parameters computed from the overall state of the ADS. 
Note that this architecture is adopted by the Simulator of Fig.~\ref{fig:environment}.

\begin{figure}[h]
\centering
\includegraphics[width=\textwidth]{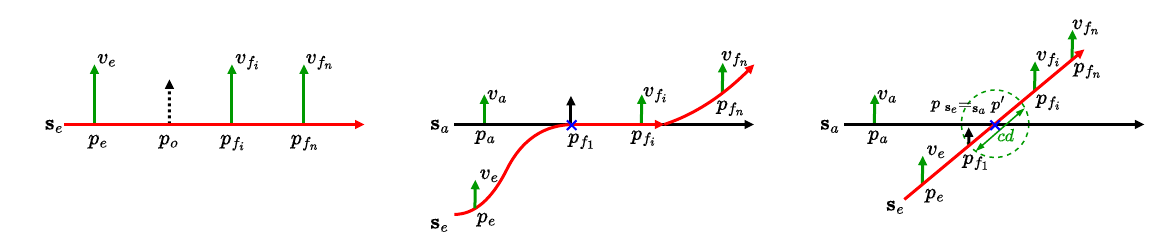}
\caption{Road, merging and crossing vistas}\label{fig:variousvista}
\end{figure}

\subsection{The three basic vista types}

As explained in Introduction, we consider that an autopilot receives basic types of vistas (Fig.~\ref{fig:variousvista}) as input, each requiring specific operations implemented by the corresponding control policies:
\begin{enumerate}

\item	\textit{Road vistas} of the form $rvs =\langle q_e, qF, \emptyset \rangle$ where there is no junction in the forward visibility zone of the ego vehicle. 	
Note that road vistas can be simplified in the following manner:	

\begin{itemize}
\item	if the first obstacle is a vehicle $f$ then what matters for the autopilot of the ego vehicle is the simplified road vista $srvs=\langle q_e, q_f, \emptyset \rangle$ as its responsibility is limited to the space up to vehicle $f$. 
\item	if the first obstacle is an object $o$, e.g. a speed limit signal with state $q_o=\langle p_o, v_l\rangle $ then the simplified vista is $srvs=\langle\{q_e, q_o,q_f\}, \emptyset\rangle$ where $f$ is the first vehicle after $p_o$ if there is such a visible vehicle. Otherwise, $q_f$ is the state of a fictitious vehicle at the limit of the front visibility zone $p_e+_{\seg_e} fd(q_e)$. Note that if there is more than one speed limit signal between the ego vehicle and $f$, we assume that the distance between two successive signals is large enough to allow speed adjustment. For example, for two successive signals at positions $p_{o_1}$ and $p_{o_2}$ with decreasing imposed speed limits $vl_1$ and $vl_2$, we assume that the ego vehicle has sufficient space to reduce its speed from $vl_1$ to $vl_2$. 
\end{itemize}

\item	\textit{Merging vistas} describe situations where {the route of the ego vehicle merges into a main road}. They are of the form $mvs =\langle q_e, qF, \langle \seg_a, v_a \rangle \rangle$ such that $f_1$ is a yield sign at position $p_{f_1}$ and $\exists d,p'.~ \seg_e[p_{f_1} +_{\seg_e} d] = \seg_a[p', p' +_{\seg_a} d]$ with $p_a <_{\seg_e} p'$.
A merging vista can be simplified by replacing $qF$ by $qF’$ where $\langle q_e, qF’, \emptyset\rangle$ is the simplified  road vista for $\langle q_e, qF, \emptyset\rangle$. 
Note that lane change is a special case of merging vista where the ego vehicle should be cautious towards the vehicle on the outer lane. Furthermore, overtaking involves two successive merging operations: one consists of moving from the initial lane to an adjacent lane, and the other of returning to the initial lane after a phase of driving in a straight line to ensure that it is far enough away from the overtaken vehicle.

\item	\textit{Crossing vistas} describe situations where the route of the ego vehicle crosses a main road. They are of the form $cvs =\langle q_e, qF, qA\rangle$, such that $\exists p$ on $\seg_e$, $p_i$ on $\seg_{a_i}$  with $p \leftindex_{\seg_e}=_{\seg_{a_i}}p_i$ for all $a_i$. Furthermore, we assume that $f_1$ is an object such as a traffic light or a yield sign. Additionally, for a crossing vista, we consider that from the map, we can associate a critical distance parameter $cd$ that is the length of $\seg_e$ in the intersection.
A simplified crossing vista $cvs =\langle q_e, qF', qA\rangle$ is such that $ \langle q_e, qF',\emptyset\rangle$ is a simplified road vista. 
We can define subtypes of crossing vistas. One is when $f_1$ is a traffic light, another when $f_1$ is a yield or a stop sign. 

\end{enumerate}

\section{Test Policy Principles for Vistas}\label{se:policy}

In order to test an autopilot, we need some knowledge about its
deceleration and acceleration  capabilities. This is necessary to assess
the feasibility of the control: in the event of a problem, the
vehicle is only liable if there is a feasible safe policy. 
For example, if a test case sets a vehicle's initial speed at a sufficiently high level, while a fixed obstacle is nearby, an accident will occur for which the vehicle is not responsible. It is thus important to define test cases for which there is evidence that there exists a feasible control policy based on the knowledge of the capabilities of the vehicle to change its kinetic state by accelerating or decelerating.

\subsection{Realistic test cases and feasibility of control policies}\label{se:realisitic}

\begin{figure}[h]
\centering
\includegraphics[width=0.55\textwidth]{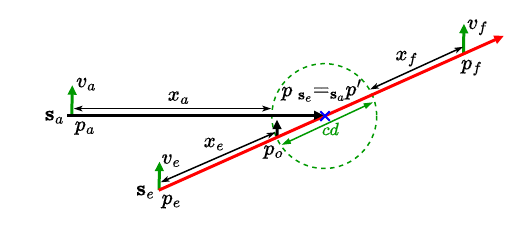}
\caption{A vista and its characteristic parameters}\label{fig:vistapara}
\end{figure}

As shown in Fig. \ref{fig:vistapara}, a vista for the ego vehicle describes a situation involving a potential obstacle on its route and calling for operation implemented by a specific control policy. In its simplest and general form, a vista involves:
\begin{enumerate}
\item	the ego vehicle at position $p_e$, with its route $\seg_e$ and speed $v_e$, that encounters an obstacle located at position $p_o$ ahead, i.e. at a distance $x_e = p_o -_{\seg_e} p_e$.

\item	an arriving vehicle at speed $v_a$, located at position $p_a$ on its route $\seg_a$, which encounters $\seg_e$ at position $p'$ such that $p' \leftindex_{\seg_a}{=}_{\seg_{e}}p$. The intersection of $\seg_e$ and $\seg_a$ determines a critical zone. The zone covers a critical distance $cd$ on $\seg_e$. We denote by $x_a$ the distance from $p_a$ to the entrance of the critical zone.

\item	a front vehicle at position $p_f$ on the route $\seg_e$ located after the obstacle at a distance $p_f-_{\seg_e}p_o$. We denote by $x_f = (p_f -_{\seg_e} p) - cd$ the distance from the exit of the critical zone to the front vehicle.
\end{enumerate}

It is easy to check that the three different types of vistas match this model. 	

For road vistas, the obstacle is just a speed limit signal with no cross-roads. Henceforth, we do not consider road  vistas, as the driving operations they involve are less complex and, to some extent, covered by the operations of the other types of vistas.  	

Merging vistas have a critical distance $cd=0$ where the critical zone degenerates to a merging point. They cover two cases.
The first is when the obstacle can be a yield sign, and the road of the ego vehicle merges into a road of higher priority where the arriving vehicle is traveling. The second case is where the ego vehicle moves from one lane to another in which the arriving vehicle is traveling. In both cases, we assume that there is a front vehicle after the merging point.	

Crossing vistas cover also two cases. The first is when the obstacle is the traffic lights protecting an intersection, in which case arriving vehicles are irrelevant. The second is an intersection where the crossing ego vehicle faces a yield sign.

Each vista requires a specific operation when the ego vehicle approaches the obstacle. The aim of the operation is to clear the obstacle safely, respecting traffic regulations and, of course, avoiding accidents with the arriving vehicle and the front vehicle.

The operation corresponding to a vista is logically characterized by scenarios comprising two successive phases:
\begin{enumerate}
\item A \textit{caution phase} during which the ego vehicle approaches the obstacle, reducing its speed if necessary, and waiting for conditions to be favorable to clear the obstacle, e.g., approaching a merge, approaching a crossing, or remaining in the same lane before overtaking.
\item A \textit{progress phase} during which the ego vehicle clears the obstacle after checking that there is no risk of collision with the arriving vehicle or the vehicle in front, e.g., to overtake a vehicle, enter a main road, or cross an intersection. The progress phase therefore, consists of moving as quickly as possible to avoid collision with the arriving vehicle, while retaining the possibility of avoiding a collision with the front vehicle.
\end{enumerate}

We derive conditions characterizing feasible policies for the different types of vistas by distinguishing the caution and progress phases in the corresponding scenario. 

Since the initial state of the ADS under test is determined by the test case generator, we need to ensure that the test cases are realistic, i.e., that the autopilots can generate safe control policies. Unrealistic test cases can lead to violations of safety properties for which the autopilot is not responsible.
For this, we need to know how a vehicle, as an electromechanical system, reacts to braking or acceleration commands from its autopilot.
To avoid detailed modeling of a vehicle as a dynamic system,
see for example~\cite{li2021estimation}, we assume that we know for each vehicle the three following functions that are sufficient to decide feasibility. 
\begin{enumerate}
\item	The \textit{braking function} $B(v)$ that gives the distance needed to brake from speed $v$ to speed 0. 
\item	The \textit{acceleration time function} $AT(v,x)$ that gives the time needed to cover distance $x$ by accelerating from speed $v$.
\item	The \textit{acceleration speed function} $AV(v,x)$ that gives the speed reached from speed $v$ after covering distance $x$.  
\end{enumerate}

The following properties of these functions that for simplicity we call A/D functions (acceleration/deceleration functions), are useful for our analysis: 
\begin{enumerate}
\item \textit{Strictness}: $B(0) =0$; $AT(v,0) =0$; $AV(v,0) =v$; i.e., there are “neutral values” of the arguments, implying no change in the kinetic state.
\item	\textit{Monotonicity}: 
\begin{itemize}
\item	$B(v_1)\leq B(v_2)$, if $v_1\leq v_2$
\item	$AT(v, x_1)\leq AT(v, x_2)$ if $x_1\leq x_2$
\item	$AV(v, x_1)\leq  AV(v, x_2)$ if $x_1\leq x_2$ and $AV(v_1, x)\leq AV(v_2, x)$ if $v_1\leq v_2$
\end{itemize}
\end{enumerate}

Note that the function $B(v)$ can be generated from a function $\mathbf{B}(v,v')$ that gives the distance needed to brake from speed $v$ to speed $v'$ $(v'\leq v)$.  In that case, we have a simple additivity relationship between the successive speeds and distances reached when braking: 
$B(v)= \mathbf{B}(v,v')+ B(v')$. This means that if the system brakes from the initial speed $v$ and its speed decreases to $v'$, its behavior from $v'$ will be the same as when braking from the initial speed $v'$. Additivity implies determinism and greatly simplifies testing.

The aim is to determine realistic test cases, which correspond to situations that a vehicle can safely cope with on the basis of its braking and acceleration capabilities. This problem appears in hardware or software testing in a simpler form: the test cases are chosen from the domain of input variable values. 

We show that using A/D functions, we can characterize realistic test cases by constraints on the parameters $x_e, v_e, x_a, v_a, x_f$ and $v_f$ for the different types of vistas and corresponding phases, caution or progress.  The constraints are derived based on the assumption that the autopilots of the ADS adequately use the available features of the vehicles for acceleration and deceleration. In particular, they conform to regulations, and they timely brake to avoid collision with obstacles on their route, constantly evaluating the free space ahead. Therefore, for test cases meeting the constraints, there is good reason to believe that safe policies exist, and any violation implies faulty autopilot behavior in the vehicle concerned.

In addition, to simplify our analysis, we  assume that autopilots follow a \textit{principle of rationality}~\cite{verschure2003real,newell1980physical}: ``If the system wants to attain goal $G$ and knows that to do act $A$ will lead to attaining $G$, then it will do $A$.''  
We apply this principle, which dictates that a system works in the best possible way, to judge the rationality of control policies by observing  that for a given distance $x_e$ and speed $v_e$ of the ego vehicle, the constraints induced by $x_a$ and $x_f$ relax as these variables increase. 
In other words, if $\langle x_{a_1}, x_{f_1}\rangle \le \langle x_{a_2}, x_{f_2}\rangle$ and safe progress is possible for $\langle x_{a_1}, x_{f_1}\rangle$, then safe  progress is possible for $\langle x_{a_2},x_{f_2} \rangle$.

Note that the test results reveal that, this property may not be satisfied in two possible ways, as we can observe: 1) safety issues for $\langle x_{a_2},x_{f_2} \rangle$ such as accidents or traffic violations; 2) performance issues when the autopilot is cautious for $\langle x_{a_2},x_{f_2} \rangle$ while progress is perfectly possible from $\langle x_{a_1},x_{f_1} \rangle$.

\subsection{Constraints characterizing realistic critical test cases}\label{se:constraints}

\subsubsection{Merging vista with a yield sign}

\begin{figure}[h]
\centering
\includegraphics[width=0.65\textwidth]{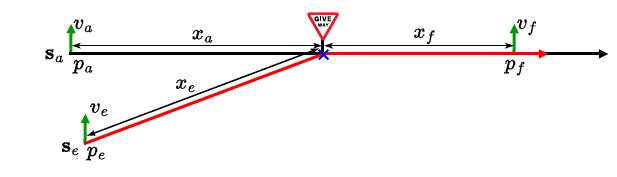}
\caption{Merging vista with a yield sign}\label{fig:merging}
\end{figure}

In the caution phase the ego vehicle approaches the merging point with a speed $v_e$. As long as the conditions are not right for progress, the speed of the vehicle should be such that $B(v_e)\leq x_e$. If the braking function is monotonic, this constraint determines the maximal speed $v_{{max}_e}$ for realistic test cases:  $B(v_{{max}_e})=x_e$. 

In the progress phase, two constraints are implied respectively from the arriving vehicle $a$ and the front vehicle $f$.

For the arriving vehicle we assume that it can drive at the maximal speed limit $vl$ and it is at distance $x_a$ from the merging point. So if the ego vehicle decides to progress and reach the merging point by accelerating, it will need time $AT(v_e, x_e)$. Within this time the arriving vehicle will have covered a distance $vl * AT(v_e, x_e)$. So, the remaining distance from the merging point is $x_a- vl * AT(v_e, x_e)$. This distance should be large enough for a safe brake. Therefore, $B(vl) + vl* AT(v_e, x_e) \leq  x_a$.

For the front vehicle $f$, the ego vehicle should avoid collision in the worst case where $v_f=0$. So, when reaching the merger point, its speed will be $AV(v_e,x_e)$ and the distance needed to brake safely is $B (AV(v_e,x_e))$. Thus the second progress constraint is $B(AV(v_e,x_e)) \leq  x_f$.

Therefore, realistic test cases for progress are characterized by the constraints 
$B(vl) + vl * AT(v_e, x_e) \leq  x_a$ and $B(AV(v_e,x_e)) \leq  x_f$.

The rationality principle 
suggests that there exists an order of criticality on the tuples $\langle x_e, v_e, x_a, x_f \rangle $. Feasible cautious policies for a given $x_e$ are possible for speeds $v_e \le \hat v_e$ such that $B(\hat v_e)= x_e$. Feasible progress policies for given $x_e$ and $v_e$ are possible for tuples $\langle x_a, x_f\rangle \le \langle \hat x_a, \hat x_f \rangle$ such that $B(vl) + vl * AT ( v_e, x_e ) = \hat x_a$ and $B(AV (  v_e, x_e )) = \hat x_f$.

\subsubsection{Lane change vista}
For this vista, we assume that the ego vehicle is traveling cautiously in a lane with a vehicle in front of it at distance $x_{f’}$. The decision to perform a lane change operation is subject to two constraints: one from the possibly arriving vehicles and the other from a front vehicle on the external lane situated respectively at distance $x_a$ and $x_f$ behind and ahead of the estimated merging point. We assume that we have an estimate of the distance $x_e$ to travel to reach the external lane. Furthermore, the ego vehicle does not accelerate and keeps its distance constant.

\begin{figure}[h]
\centering
\includegraphics[width=0.65\textwidth]{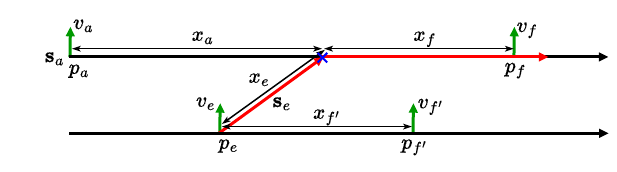}
\caption{Lane change vista}\label{fig:changing}
\end{figure}

In the caution phase, the ego vehicle approaches the leading vehicle on the inside lane at speed $v_e$. Before the ego vehicle makes the lane change, its speed must satisfy the safety constraint $B(v_e) \le x_{f'}$

The time needed by the ego vehicle to reach the merging point is $x_e/v_e$, and in this time, the arriving vehicle will have traveled a distance $vl*(x_e/v_e)$ such that $vl* (x_e/v_e) \leq x_a - B(vl)$. In other words, the distance $x_a$ must be large enough for the arriving vehicle to travel at the maximum permitted speed and, if necessary, brake to avoid collision with the ego vehicle in the outside lane: $vl* (x_e/v_e) + B(vl) \leq x_a$. 

{The constraint for the vehicle ahead is simply $B(v_e)\leq x_f$.
The rationality principle gives for progress from $x_e$, $v_e$ the most critical test cases as configurations $\langle x_e,v_e, \hat x_a, \hat x_f\rangle$}
such that $vl*({x}_{e}/{v}_{e}) + B(vl) =\hat{x}_{a}$ and $B({v}_{e}) \leq \hat{x}_{f}$. Passing these test cases would logically imply that the ego vehicle can safely change lanes in all the less critical situations. 

\subsubsection{Crossing with yield sign vista}

The vista involves the ego vehicle traveling at speed $v_e$ at a distance $x_e$ from the point of intersection with a main road protected by a yield sign. In this operation, the ego vehicle should approach cautiously, moderating its speed until it decides to progress if there is no risk of collision with some arriving vehicle at a distance $x_a$ and with estimated maximal speed $vl$. Additionally, it should avoid collision with a front vehicle at distance $x_f$ after the intersection. We suppose that around the intersection point, there is a critical zone delimited by a critical distance $cd$ such that the presence of two vehicles in this zone is considered a potential accident.

\begin{figure}[h]
\centering
\includegraphics[width=0.65\textwidth]{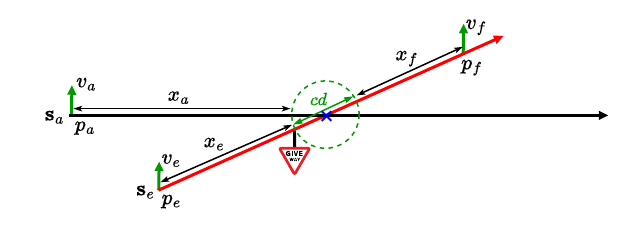}
\caption{Crossing with yield sign vista}\label{fig:crossing}
\end{figure}

Thus, realistic test cases are characterized by the following constraints:

For the caution phase, we have the constraint $B(v_e) \leq  x_e$.
 
 For the progress phase, we assume that the ego vehicle starts by accelerating from speed $v_e$ when it estimates that an arriving vehicle at speed $vl$ is far enough to avoid collision.
 
The time needed by the ego vehicle to cross and get out of the critical zone is $AT(v_e, x_e+cd)$. At this time, the arriving vehicle will have traveled a maximal distance $vl* AT(v_e, x_e+cd)$. The arriving vehicle's distance to the critical zone should be larger than the maximal traveled distance at speed $vl$ before the ego vehicle exits the zone which gives $vl * AT(v_e, x_e + cd) \le x_a$.

When the ego vehicle leaves the critical zone, it will have speed $AV(v_e, x_e+cd)$ from which it should be able to brake safely to avoid collision with the obstacle ahead. Thus $B(AV(v_e, x_e+cd)) \leq  x_f$. 

{
Also for this case, the rationality principle  results in configurations $\langle x_e, v_e, \hat x_a, \hat x_f \rangle$, satisfying the constraints: $vl * AT ( v_e , x_e + cd) = \hat x_a$  and $B(AV (v_e,  x_e + cd)) = \hat x_f$.
}

\subsubsection{Crossing with traffic light vista}

\begin{figure}[h]
\centering
\includegraphics[width=0.65\textwidth]{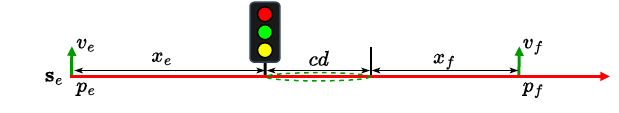}
\caption{Crossing with traffic light vista}\label{fig:trafficlight}
\end{figure}

The vista involves the ego vehicle approaching at speed $v_e$ and at distance $x_e$ from a traffic-light protected intersection of width $cd$. The traffic light has a state variable taking values ``red”, ``yellow,” and ``green”. We assume that we know the duration $ty$ of the yellow light. Furthermore, the lights of the intersection have an ``all red” phase of duration $tar$ where all the lights are red before some light passes from red to green. These constants are very important for respecting safety regulations requiring that when the ego vehicle enters the critical zone the lights should be either green or yellow. In addition, a vehicle entering the intersection must exit before a light turns green on a transverse road.

The constraint for the caution phase is $B(v_e) \leq  x_e$.

The constraint for progress is that the ego vehicle 
must not see red light which means that even if the lights switch to yellow right after the decision to cross is taken, it will reach the entrance to the intersection before the lights turn to red. That is $AT(v_e, x_e) \leq  ty$.

Additionally, the time needed to cross the critical zone, that is to cover distance $x_e+cd$ will be less than $ty+tar: AT(v_e, x_e+cd) \leq  ty+tar$. 

To these constraints, we add the one induced by the presence of a front obstacle at $x_f$ from the intersection: $B(AV(v_e, x_e+cd))\leq  x_f$.

{For given $x_e,~v_e$ and contextual parameters $ty, tar, cd$, if safe progress is possible, i.e., $AT(v_e, x_e)\le ty$ and $AT(v_e, x_e + cd) \le ty + tar$, the rationality principle indicates the most critical test cases as configurations $\langle x_e, v_e, \hat x_f \rangle $ such that $B(AV(v_e, x_e + cd)) = \hat x_f$.}

\section{The Testing Environment}\label{se:setting}

\subsection{Test case parameters and tested properties}

For the considered four types of vistas, we define the following context parameters for the experiments:
\begin{itemize}
\item 
\textbf{Merging vista.} The ego vehicle is on a road that merges into a main road protected by a yield sign. The speed limit on the main road is 80 km/h.
\item 
\textbf{Lane Change vista.} We consider a two-lane straight road with a speed limit of 80 km/h. The ego vehicle will change from the inside lane to the outside lane with traveling distance $x_e =13.5$ m.
\item 
\textbf{Crossing with yield sign vista.} We consider an intersection with a speed limit of 80 km/h. The critical zone covers 24 meters of the ego vehicle's route. The entrance for the ego vehicle is protected by a yield sign, while the entrance for the arriving vehicle is unprotected.
\item 
\textbf{Crossing with traffic light vista.} We consider an intersection protected by traffic lights. The critical zone covers 24 meters of the ego vehicle's route. The light for the ego vehicle is initially green, then immediately changes to yellow at the next time step. Thus, the initial state of the ego vehicle is the state where the light changes from green to yellow. The duration of the yellow light is 3 seconds, and the duration of the all-red phase is 2 seconds. 
\end{itemize}

\vspace{10pt}

The autopilot compares the scenarios generated from the test cases with the properties listed in Tab. \ref{tab:safety-properties} and delivers the following types of verdicts.

\begin{itemize}
\item \texttt{PS} and \texttt{PU}$p$, respectively for safe and unsafe progress violating property $p$. 

\item \texttt{CS}, \texttt{CU}$p$, respectively for safe caution and unsafe caution violating property $p$.

\item \texttt{Ae} and \texttt{Aa} respectively, for accidents in which the ego vehicle collides with the arriving vehicle and the arriving vehicle collides with the ego vehicle. 
\item \texttt{Fsw} for a failure of the autopilot’s software.
\item \texttt{Blk} for the ego vehicle, which runs aground midway and blocks the path of the arriving vehicle.
\end{itemize}

\begin{table}[H]
    \centering
    \caption{Safety Properties}
    \label{tab:safety-properties}
    \vspace{-10pt}
    \begin{tabular}{p{0.95\linewidth}}
    \hline
         \textbf{For crossing vistas (with yield signs or traffic lights)} \\
         $p_1$:  Two vehicles must not be in the critical zone at the same time. \\
         $p_2$: The ego vehicle must not stop inside the critical zone.\\
         \textbf{For crossing with traffic light vistas}: \\
         $p_3$: The ego vehicle must not enter the critical zone when the light is red.\\
         $p_4$: The ego vehicle must not be in the critical zone when a side light is green.\\
    \hline
    \end{tabular}
\end{table}

In the experiments, we choose different initial speeds for the ego vehicle
to achieve uniform coverage. Furthermore, we take the distance of
the ego vehicle such that $x_e = B(v_e)$ and apply test cases for configurations of values $x_f$ and $x_a$ around the critical values $\hat{x}_{a}$ and $\hat{x}_{f}$.

For merging and yield-sign crossing vistas, the ego vehicle is expected to accelerate to pass the critical zone. We consider that the initial speed $v_e$ takes values between 0 m/s and 15 m/s, with a step of 5 m/s.

In the case of a lane change, the ego vehicle is supposed to progress by
maintaining its speed. We consider speeds $v_e$ from 5 m/s to 20 m/s with a step of 5m/s.

For traffic light vistas, we consider a wider range of $v_e$ from 0.0 m/s to 20.0 m/s where the ego vehicle should properly decide to progress or be cautious.

For $x_f$ and $x_a$, we consider values in the interval from 0.0 to 320.0 meters.
We consider distances greater than 320.0 meters to be large enough not to give rise to critical situations. 
The exploration strategy initially considers a step of 40 meters and
refines the intervals at which the verdict changes from caution to progress.

In addition, we eliminate cases where $x_a + x_f < B(vl )$ for
merging and lane change vistas, to ensure that the arriving vehicle
can brake safely before the front vehicle.

\subsection{Estimating a vehicle's dynamic parameters}

We show how we can obtain specifications of the A/D functions for a given autopilot.

Some autopilots, such as Apollo, Autoware, and LGSVL, provide explicit definitions of acceleration and deceleration rates as continuous piecewise functions $a(v, T, t)$ and $b(v, T, t)$, where $v$ is the initial speed, $T$ indicates the total time of acceleration or deceleration, $t$ measures the time elapsed since the start of the process. It is then possible to obtain analytic specifications of the A/D functions as follows. 

For the deceleration function $B(v)$ with given initial speed $v$, we formulate the speed reached at time $\tau$, after deceleration, as $v - \smallint_0^\tau b(v, T, t)dt$. Then, we can compute the time $T$ such that the resulting speed is equal to 0, i.e., {$v - \smallint_0^{T} b(v, T, \tau)d\tau = 0$}. Finally, for a given time $T$, 
the function $B(v)$ is estimated as $\smallint_0^{T}(v-\smallint_0^\tau b(v, T, t)dt)d\tau$.

For the acceleration functions $AT(v, x)$ and $AV(v, x)$ {with given initial speed $v$ and distance $x$}, we formulate the accelerated speed at time $\tau$ as $v + \smallint_0^\tau a(v, T, t)dt$ and the traveled distance at time $\tau'$ as $\smallint_0^{\tau'}(v + \smallint_0^\tau a(v, T, t) dt)d\tau$. Then, we can compute the time $T$ such that the accelerated distance is equal to $x$, i.e., $\smallint_0^{T}(v + \smallint_0^\tau a(v, T, t) dt)d\tau = x$. 
Finally, 
we have $AT (v, x) = T$ and $AV (v, x) = v + \smallint_0^{T} a(v, t )dt$. 

In absence of explicit definition of the acceleration and deceleration rates applied by Carla’s autopilot, we make an empirical estimation of the A/D functions as explained below.

To estimate the function $B(v)$, we consider scenarios with an initial speed $v$ and a front obstacle in distance $x$. The simulation runs until the vehicle either stops safely or collides with the obstacle. We determine the value of $B(v)$ as the {minimal} distance $x$ where the vehicle stops without collision. This minimal distance is obtained by progressively increasing $x$ from 0 until the value at which a safe stop occurs.

To estimate the functions $AT(v, x)$ and $AV(v, x)$, we consider scenarios where a vehicle starts at speed $v$ on a clear road with a speed limit $v’ (v’ > v)$. The simulation produces  time $t$ and distance $d$ taken by the vehicle to accelerate from $v$ to $v’$. We adjust the value of $v’ $ to ensure that the accelerated distance $d$ equals to the given distance $x ~(d=x)$. Consequently, $AT(v, x)$ can be determined as the time $t$, and $AV(v, x)$ as the speed $v’$. The required $v’$ can be obtained by progressively increasing it from $v$ until the simulation yields the distance $d$ such that $d=x$.

Note that the estimation method assumes that the A/D functions are strict and monotonic.

Of the four autopilots, only LGSVL has an additive braking function that ensures that if the system brakes safely
from a speed $v$ to avoid an obstacle at distance $x$, and in this process when it reaches a speed
$v' \le v$, it is at a distance $x'$ from the obstacle, then it can brake safely from speed $v'$ to avoid an obstacle at a distance $x'$. This property, which greatly simplifies test coverage, is not valid for the other autopilots. 
For example, for the Apollo autopilot, we have $B(20.0) = 50.0$ m and in this braking process from a speed of 15.0 m/s, it can stop at a distance of 20.4 m. However, when braking from 15.0 m/s, the vehicle needs at least $B(15.0) = 31.7 > B(20.0) - \mathbf{B}(20.0, 15.0) = 20.4$ m to stop safely. 
Similarly, for the Autoware autopilot, we have $B(15.0) = 29.8 > B(20.0)-\mathbf{B}(20.0, 15.0) = 23.6$ m,  and for the Carla autopilot, we have $B(15.0) = 15.8 > B(20.0) - \mathbf{B}(20.0, 15.0) = 13.8$ m.

Note that the observed lack of additivity is due to the fact that the braking rate $b$ depends not only on speed but also on the time elapsed since the start of the braking process.

\subsection{Implementation of the testing environment}

\begin{figure}[htbp]
    \centering
    \includegraphics[width=0.65\linewidth]{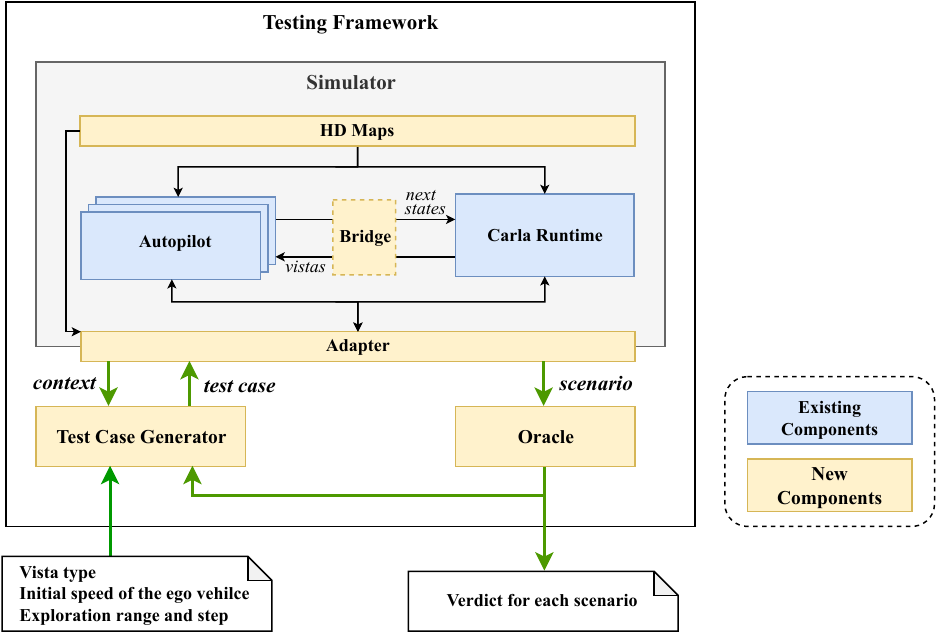}
    \caption{The testing framework for Apollo, Autoware, and Carla autopilots}
    \label{fig:framework-impl}
\end{figure}

We have implemented the proposed test method for four open autopilots, including Apollo 8.0, Autoware.universe 1.0, Carla 0.9.15, and LGSVL 2021.3. The test framework integrates a Test Case Generator, an Oracle, and a Simulator for ADS. 

The Test Case Generator generates test cases corresponding to critical configurations for which safe control policies exist. The Simulator takes the generated test cases as input and configures the autopilots and the Runtime through an Adapter. It then produces a scenario in the form of a simulated sequence of ADS states. The Oracle checks the generated scenarios against safety properties and provides verdicts.

To test LGSVL, we took the RvADS validation framework proposed by \cite{li2021estimation} and integrated the new Test Case Generator and the Oracle into its framework. 

For Apollo, Autoware, and Carla autopilots, we built a new testing framework as shown in Fig. \ref{fig:framework-impl}.
The Simulator in this framework integrates the autopilots with the Carla Runtime. It includes an Adapter for the interaction between the internal components, such as autopilots and the Runtime, and the external components, such as the Test Case Generator and the Oracle. During simulation, the autopilots and the Runtime 
access contextual information from HD maps, and interact cyclically. At each cycle, the Runtime provides the vistas for each autopilot, and the autopilots compute the next states for the corresponding vehicles. If an accident happens, the Runtime reports it, and the simulation will be interrupted.

The link between the Carla autopilot and the Carla Runtime is
straightforward. 
However, as the integration of Apollo and Autoware autopilots is not supported by the Carla Runtime, we developed a Bridge.
{Apollo and Autoware employ middleware (Cyber RT for Apollo and ROS2 for Autoware) to receive the state from the external runtime, manage the execution of the internal components, and return commands for state updates.} The Carla Runtime manages vehicle and traffic light states. It offers an API for state queries and updates. The Bridge fetches the world state from the Carla API, passes it to the autopilots, and translates the autopilots' output back to the runtime for state updates. This process only involves calling interfaces of the autopilots and the runtime, and does not affect their internal behavior. Note that the Bridge allows the simulation of several vehicles, each controlled by an autopilot. {This feature for creating} realistic critical situations is not offered by most existing simulators, which only allow a single vehicle equipped with an autopilot.

Furthermore, different autopilots require HD maps in various formats: Carla uses OpenDrive, Apollo uses Apollo HD Map, Autoware uses Lanelet2, and LGSVL uses the LGSVL Map Model. Existing maps provided by Carla lack comprehensive support for all these formats. To address this issue, we created a collection of HD Maps that defines the contexts for the four vistas, i.e., merging, lane changing, crossing with yield signs, and crossing with traffic lights, in all four formats.

The Test Case Generator receives inputs specifying the vista type, the initial speed of the ego vehicle, and the range and step for exploring the parameters around the critical values. It first generates the critical parameters based on the analysis of the vehicle dynamics and context parameters accessed from the Simulator, then generates the parameters exploring the given range and step, and finally refines the intervals where the verdict given by the oracle changes from caution to progress.

The Oracle checks that a scenario satisfies the safety properties and, in the event of an accident, considers the vehicle whose front end collides with another vehicle to be at fault. 

In addition, the Oracle analyzes the behavior of the ego vehicle to detect caution or progress for a given vista. In the case of merging or crossing, progress means reaching the critical zone before the arriving vehicle. For intersections with traffic lights, progress means accelerating to cross the intersection.

The safety properties listed in Tab. \ref{tab:safety-properties} are satisfied if at each simulation tick, the state of the system satisfies the described situation. 
The properties depend on three parameters that can be consulted from the {generated} scenario: 1) the positions of the ego vehicle and the arriving vehicle, and in particular, whether or not they are in the critical zone; 2) the speed of the ego vehicle; 3) the state of the signal light, if applicable.
In this way, the Oracle checks properties by simply browsing the system states and checking their satisfaction.

Note that the four autopilots can have non-deterministic behavior. Two of them, Autoware and Apollo, use scheduling policies to coordinate the execution of their modules which, for a given vista, can produce different scenarios. Autoware does not use a strictly synchronous execution policy. It has a speed-dependent behavior that does not strictly follow the order of data flow between Panning and Control in the propagation of input values.  Apollo can be equipped with a multithreading option that introduces non-deterministic behavior as a function of module execution speed.  
To obtain reproducible results, we modified Autoware's scheduling policy, ensuring that it adheres to a strictly synchronous execution policy. However, this restriction produces a possible behavior of the Autoware autopilot. For the same reason, we have disabled the multithreading option for Apollo. 

For the other two autopilots, Carla and LGSVL, non-determinism is due to the use of randomly assigned parameters taking values in given intervals. Deterministic behavior is achieved by setting the random seed value to 1 for the Carla autopilot and by setting the aggression parameter to 1 for LGSVL.

\section{Test results}\label{se:results}
\subsection{Test methodology and interpretation of the results}\label{se:overview}

\begin{figure}[t]
\centering
\subfigure[For merging, lane change, and crossing with yield sign vistas]{
    \includegraphics[width=.28\linewidth]{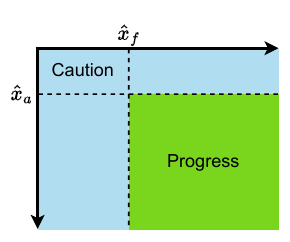}
}
\hspace{1cm}
\subfigure[For crossing with traffic light vistas]{
    \centering
    \includegraphics[width=0.26\linewidth]{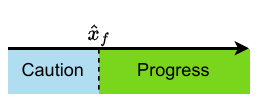}
}
\vspace{-10pt}
\caption{Caution/progress partition for worst-case safe policies, 
for given distance $x_e$ and speed $v_e$}\label{fig:ideal}
\end{figure}

\begin{figure}[t]
    \centering
    \begin{minipage}{0.305\textwidth}
    \includegraphics[width=0.98\linewidth]{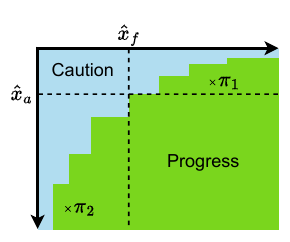}   
    \caption{Caution/progress theoretical partition of feasible safe policies, for given distance $x_e$ and speed $v_e$}
    \label{fig:feasible}
    \end{minipage}
    \hspace{0.5cm}
    \begin{minipage}{0.305\textwidth}
    \includegraphics[width=0.98\linewidth]{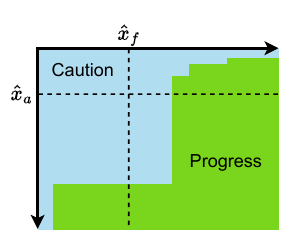}
    \caption{Caution/progress partition of feasible safe and possibly overcautious policies, for given distance $x_e$ and speed $v_e$}
    \label{fig:rational}
    \end{minipage}
    \hspace{0.5cm}
    \begin{minipage}{0.305\textwidth}
        \includegraphics[width=0.98\linewidth]{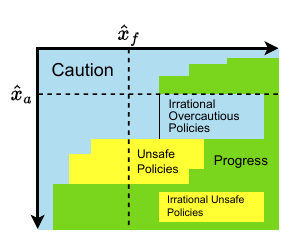}
        \caption{Caution/progress partition with unsafe and irrational policies, for given distance $x_e$ and speed $v_e$}
        \label{fig:undesirable}
    \end{minipage}
\end{figure}

According to the results presented in Section~\ref{se:constraints}, for mergings and crossings with yield signs, the transition from cautious behavior to progress is safe if both conditions $x_a \ge \hat{x}_a$ and $x_f\ge \hat{x}_f$ are true, where $\hat{x}_a$ and $\hat{x}_f$ are the critical values predicted by the theory. This defines a partition between cautious and progress configurations shown in Fig. \ref{fig:ideal} (a). For crossings with traffic lights, the transition from cautious to progress depends only on the condition $x_f \ge \hat{x}_f$ for a given speed of the ego vehicle as shown in Fig. \ref{fig:ideal} (b).

Clearly, the critical values $\hat{x}_a$ and $\hat{x}_f$ estimated are theoretical and they determine a range of values around which
the transition between caution and progress can take place. Furthermore, these values are computed under the following assumptions that determine worst-case situations that may not be followed by the tested autopilots: 	
\begin{enumerate}
    \item the ego vehicle in the caution phase travels at the maximal safe speed relating its distance from the critical zone $x_e$ to its speed $v_e$, $x_e=B(v_e)$;
    \item the ego vehicle will accelerate at the maximal speed until it crosses the critical zone and then it decelerates to avoid collision with the front vehicle;
    \item the arriving vehicle drives at the maximal speed allowed by the speed limit $vl$ and the space available until the vehicle in front, and may brake only when the ego vehicle reaches the critical zone. 
\end{enumerate}

If the above assumptions do not hold, as shown in Fig. \ref{fig:feasible}, there may be feasible safe policies for $x_a \le \hat{x}_a$ or $x_f \le \hat{x}_f$. For example, a safe policy $\pi_1$ for $x_a \le \hat{x}_a$, if the arriving vehicle decelerates cautiously as it approaches a critical zone, allowing the ego vehicle to progress safely. Symmetrically, there may be a safe policy $\pi_2$ for $x_f \le\hat{x}_f$, if the ego vehicle does not accelerate to maximum speed or cautiously approaches the critical zone at less than the maximum speed for safe braking. 
Note that if there is a safe progress policy for $\langle x_a, x_f\rangle$ then there is a safe policy for any $ \langle x_a', x_{f}'\rangle \geq \langle x_a, x_f \rangle $. Therefore, the safe progress area is theoretically the union of rectangular areas as shown in Fig. \ref{fig:feasible}. 

Safe autopilot controllers can adopt cautious policies when $x_a$ and $x_f$ are above critical values, as shown in Fig.~\ref{fig:rational}. However, the rationality of their policies is an essential criterion: if progress is observed for a pair of values $\langle x_a, x_f\rangle$, it is possible to progress by increasing these values.

Our tests reveal that the transition from caution to progress may involve undesirable situations of mainly two types, as shown in Fig. \ref{fig:undesirable}: either over-cautious policies that are safe but make poor use of the available space, or over-optimistic policies that lead to accidents or traffic violations.	

Overcautious policies neglect the possibility of progress, which can degrade performance and lead to poor road occupancy and bottlenecks. An area with obviously overcautious policies is shown in Fig. \ref{fig:undesirable} where progress is possible as it lies below a progress area whose policies are subject to stricter constraints. Here are a few examples of overcautious control policies observed in our experiments: not overtaking a stationary vehicle in front when the ego vehicle’s performance allows it and the outside lane is clear; not advancing when the front vehicle or arriving vehicle is far enough to do it.

Over-optimistic policies decide to go ahead on the basis of a poor estimate of the available free space in relation to the vehicle's acceleration and braking capacities, or by ignoring the applicable traffic rules. They may lead to accidents and property violations. Such policies could be avoided by following cautious policies.

\subsection{Testing the Apollo Autopilot}
\subsubsection{Apollo autopilot: Estimating the A/D functions}

The Apollo autopilot accommodates continuous changes in acceleration and deceleration rates, adhering to interval constraints of [-4.0 m/s$^3$, 2.0 m/s$^3$] for \textit{jerk}. It also limits the maximum acceleration to 2 m/s$^2$ and the maximum deceleration to 6 m/s$^2$. The acceleration and deceleration rate functions, $a(v, T, t)$ and $b(v, T, t)$, are formulated as piecewise linear functions that begin at zero, increase linearly to their respective maximum values, and then decrease linearly back to zero. 

Applying the analytical method, we obtain the values of the A/D functions for Apollo autopilot listed in Tab.~\ref{tab:apollo-braking} and Tab.~\ref{tab:apollo-av}. The values indicate the monotonicity of these functions.

\begin{table}[H]
    \centering
    \caption{$B(v)$ for Apollo autopilot}
    \vspace{-10pt}
    \label{tab:apollo-braking}
    \begin{tabular}{|c|c|c|c|c|c|}
    \hline
        $\bm v$ & 0.0 & 5.0 & 10.0 & 15.0 & 20.0 \\
    \hline
        $\bm{B(v)}$ & 0.0 & 6.1 & 17.3 & 31.7 & 50.0 \\
    \hline
    \end{tabular}
\end{table}

\begin{table}[H]
\caption{$AV(v, x)$, $AT(v, x)$ for Apollo autopilot}
\label{tab:apollo-av}
\vspace{-10pt}
        \begin{tabular}{|c|c|c|c|c|c|c|c|}
        \hline
        \diagbox[width=3em,height=2em]{~$\bm v$}{$\bm x$~} 
             & 0.0 & 10.0 & 20.0 & 30.0 & 40.0 & 50.0 & 60.0 \\
        \hline
          0.0& 0.0, 0.0& 5.8, 3.7& 8.4, 5.0& 10.5, 6.0& 12.1, 6.8& 13.6, 7.6& 15.0, 8.2 \\
          \hline
         5.0& 5.0, 0.0& 6.9, 1.7& 9.2, 2.9& 11.1, 3.8& 12.7, 4.6& 14.2, 5.3& 15.5, 6.0 \\
         \hline
         10.0& 10.0, 0.0& 10.6, 1.0& 12.1, 1.8& 13.6, 2.6& 15.0, 3.2& 16.2, 3.9& 17.4, 4.4 \\
         \hline
         15.0& 15.0, 0.0& 15.3, 0.7& 16.1, 1.3& 17.2, 1.9& 18.3, 2.4& 19.4, 2.9& 20.4, 3.4 \\
        \hline
        \end{tabular}
\end{table}

\vspace{2em}
\subsubsection{Apollo autopilot: Results and analysis}

The test results for the Apollo autopilot are shown in Tab. \ref{tab:apollo-merging} to Tab. \ref{tab:apollo-traffic_light} where the critical values $\hat{x}_a$ and $\hat{x}_f$ are indicated in red.

\begin{minipage}{\linewidth}
\begin{table}[H]
    \centering
    \caption{Apollo autopilot: Verdicts for merging scenarios}
    \label{tab:apollo-merging}
\end{table}
\vspace{-20pt}
\begin{center}
    Table \ref{tab:apollo-merging} (a) Apollo autopilot: Verdicts for merging scenarios with $v_e=0.0$
    \footnotesize 
    \adjustbox{max width=\textwidth}{\begin{tabular}{|c|c|c|c|c|c|c|c|c|c|c|c|c|c|c|c|c|c|}
\hline
\diagbox{$\bm{xa}$}{$\bm{xf}$}
& \ct{0.0} & 5.0 & 10.0 & 20.0 & 30.0 & 35.0 & 40.0 & 80.0 & 85.0 & 90.0 & 100.0 & 120.0 & 160.0 & 200.0 & 240.0 & 280.0 & 320.0 \\
\hline
0.0 & - & - & - & - & - & - & - & \cs{\textsf{CS}} & \cs{\textsf{CS}} & \cs{\textsf{CS}} & \cs{\textsf{CS}} & \cs{\textsf{CS}} & \cs{\textsf{CS}} & \cs{\textsf{CS}} & \cs{\textsf{CS}} & \cs{\textsf{CS}} & \cs{\textsf{CS}} \\
\hline
40.0 & - & - & - & \cs{\textsf{CS}} & \cs{\textsf{CS}} & \cs{\textsf{CS}} & \cs{\textsf{CS}} & \cs{\textsf{CS}} & \cs{\textsf{CS}} & \cs{\textsf{CS}} & \cs{\textsf{CS}} & \cs{\textsf{CS}} & \cs{\textsf{CS}} & \cs{\textsf{CS}} & \cs{\textsf{CS}} & \cs{\textsf{CS}} & \cs{\textsf{CS}} \\
\hline
\ct{59.5} & \cs{\textsf{CS}} & \cs{\textsf{CS}} & \cs{\textsf{CS}} & \cs{\textsf{CS}} & \cs{\textsf{CS}} & \cs{\textsf{CS}} & \cs{\textsf{CS}} & \cs{\textsf{CS}} & \cs{\textsf{CS}} & \cs{\textsf{CS}} & \cs{\textsf{CS}} & \cs{\textsf{CS}} & \cs{\textsf{CS}} & \cs{\textsf{CS}} & \cs{\textsf{CS}} & \cs{\textsf{CS}} & \cs{\textsf{CS}} \\
\hline
80.0 & \cs{\textsf{CS}} & \cs{\textsf{CS}} & \cs{\textsf{CS}} & \cs{\textsf{CS}} & \cs{\textsf{CS}} & \cs{\textsf{CS}} & \cs{\textsf{CS}} & \cs{\textsf{CS}} & \cs{\textsf{CS}} & \cs{\textsf{CS}} & \cs{\textsf{CS}} & \cs{\textsf{CS}} & \cs{\textsf{CS}} & \cs{\textsf{CS}} & \cs{\textsf{CS}} & \cs{\textsf{CS}} & \cs{\textsf{CS}} \\
\hline
85.0 & \cs{\textsf{CS}} & \cs{\textsf{CS}} & \cs{\textsf{CS}} & \cs{\textsf{CS}} & \cs{\textsf{CS}} & \cs{\textsf{CS}} & \cs{\textsf{CS}} & \cs{\textsf{CS}} & \cs{\textsf{CS}} & \cs{\textsf{CS}} & \cs{\textsf{CS}} & \cs{\textsf{CS}} & \cs{\textsf{CS}} & \cs{\textsf{CS}} & \cs{\textsf{CS}} & \cs{\textsf{CS}} & \cs{\textsf{CS}} \\
\hline
\end{tabular}}
\end{center}
\end{minipage}
\vspace{10pt}
\begin{center}
    Table \ref{tab:apollo-merging} (b) Apollo autopilot: Verdicts for merging scenarios with $v_e=5.0$
    \footnotesize 
    \adjustbox{max width=\textwidth}{\begin{tabular}{|c|c|c|c|c|c|c|c|c|c|c|c|c|c|c|c|c|}
\hline
\diagbox{$\bm{xa}$}{$\bm{xf}$}
& 0.0 & 5.0 & \ct{7.6 }& 10.0 & 20.0 & 40.0 & 50.0 & 55.0 & 60.0 & 80.0 & 120.0 & 160.0 & 200.0 & 240.0 & 280.0 & 320.0 \\
\hline
0.0 & - & - & - & - & - & - & - & - & \cs{\textsf{CS}} & \cs{\textsf{CS}} & \cs{\textsf{CS}} & \cs{\textsf{CS}} & \cs{\textsf{CS}} & \cs{\textsf{CS}} & \cs{\textsf{CS}} & \cs{\textsf{CS}} \\
\hline
40.0 & - & - & - & - & \cs{\textsf{CS}} & \cs{\textsf{CS}} & \cs{\textsf{CS}} & \cs{\textsf{CS}} & \cs{\textsf{CS}} & \cs{\textsf{CS}} & \cs{\textsf{CS}} & \cs{\textsf{CS}} & \cs{\textsf{CS}} & \cs{\textsf{CS}} & \cs{\textsf{CS}} & \cs{\textsf{CS}} \\
\hline
80.0 & \cs{\textsf{CS}} & \cs{\textsf{CS}} & \cs{\textsf{CS}} & \cs{\textsf{CS}} & \cs{\textsf{CS}} & \cs{\textsf{CS}} & \cs{\textsf{CS}} & \cs{\textsf{CS}} & \cs{\textsf{CS}} & \cs{\textsf{CS}} & \cs{\textsf{CS}} & \cs{\textsf{CS}} & \cs{\textsf{CS}} & \cs{\textsf{CS}} & \cs{\textsf{CS}} & \cs{\textsf{CS}} \\
\hline
\ct{84.1} & \cs{\textsf{CS}} & \cs{\textsf{CS}} & \cs{\textsf{CS}} & \cs{\textsf{CS}} & \cs{\textsf{CS}} & \cs{\textsf{CS}} & \cs{\textsf{CS}} & \cs{\textsf{CS}} & \cs{\textsf{CS}} & \cs{\textsf{CS}} & \cs{\textsf{CS}} & \cs{\textsf{CS}} & \cs{\textsf{CS}} & \cs{\textsf{CS}} & \cs{\textsf{CS}} & \cs{\textsf{CS}} \\
\hline
85.0 & \cs{\textsf{CS}} & \ps{\textsf{PS}} & \ps{\textsf{PS}} & \ps{\textsf{PS}} & \ps{\textsf{PS}} & \ps{\textsf{PS}} & \ps{\textsf{PS}} & \ps{\textsf{PS}} & \cs{\textsf{CS}} & \cs{\textsf{CS}} & \cs{\textsf{CS}} & \cs{\textsf{CS}} & \cs{\textsf{CS}} & \cs{\textsf{CS}} & \cs{\textsf{CS}} & \cs{\textsf{CS}} \\
\hline
90.0 & \cs{\textsf{CS}} & \ps{\textsf{PS}} & \ps{\textsf{PS}} & \ps{\textsf{PS}} & \ps{\textsf{PS}} & \ps{\textsf{PS}} & \ps{\textsf{PS}} & \ps{\textsf{PS}} & \ps{\textsf{PS}} & \ps{\textsf{PS}} & \ps{\textsf{PS}} & \ps{\textsf{PS}} & \ps{\textsf{PS}} & \ps{\textsf{PS}} & \ps{\textsf{PS}} & \ps{\textsf{PS}} \\
\hline
\end{tabular}}
\end{center}
\vspace{10pt}
\begin{center}
    Table \ref{tab:apollo-merging} (c) Apollo autopilot: Verdicts for merging scenarios with $v_e=10.0$
    \footnotesize 
    \adjustbox{max width=\textwidth}{\begin{tabular}{|c|c|c|c|c|c|c|c|c|c|c|c|c|c|}
\hline
\diagbox{$\bm{xa}$}{$\bm{xf}$}
& 0.0 & 5.0 & 10.0 & 20.0 &\ct{ 21.8 }& 40.0 & 80.0 & 120.0 & 160.0 & 200.0 & 240.0 & 280.0 & 320.0 \\
\hline
0.0 & - & - & - & - & - & - & \cs{\textsf{CS}} & \cs{\textsf{CS}} & \cs{\textsf{CS}} & \cs{\textsf{CS}} & \cs{\textsf{CS}} & \cs{\textsf{CS}} & \cs{\textsf{CS}} \\
\hline
40.0 & - & - & - & - & - & \cs{\textsf{CS}} & \cs{\textsf{CS}} & \cs{\textsf{CS}} & \cs{\textsf{CS}} & \cs{\textsf{CS}} & \cs{\textsf{CS}} & \cs{\textsf{CS}} & \cs{\textsf{CS}} \\
\hline
80.0 & \cs{\textsf{CS}} & \cs{\textsf{CS}} & \cs{\textsf{CS}} & \cs{\textsf{CS}} & \cs{\textsf{CS}} & \cs{\textsf{CS}} & \cs{\textsf{CS}} & \cs{\textsf{CS}} & \cs{\textsf{CS}} & \cs{\textsf{CS}} & \cs{\textsf{CS}} & \cs{\textsf{CS}} & \cs{\textsf{CS}} \\
\hline
\ct{95.1} & \cs{\textsf{CS}} & \cs{\textsf{CS}} & \cs{\textsf{CS}} & \cs{\textsf{CS}} & \cs{\textsf{CS}} & \cs{\textsf{CS}} & \cs{\textsf{CS}} & \cs{\textsf{CS}} & \cs{\textsf{CS}} & \cs{\textsf{CS}} & \cs{\textsf{CS}} & \cs{\textsf{CS}} & \cs{\textsf{CS}} \\
\hline
100.0 & \cs{\textsf{CS}} & \cs{\textsf{CS}} & \cs{\textsf{CS}} & \cs{\textsf{CS}} & \cs{\textsf{CS}} & \cs{\textsf{CS}} & \cs{\textsf{CS}} & \cs{\textsf{CS}} & \cs{\textsf{CS}} & \cs{\textsf{CS}} & \cs{\textsf{CS}} & \cs{\textsf{CS}} & \cs{\textsf{CS}} \\
\hline
105.0 & \cs{\textsf{CS}} & \ps{\textsf{PS}} & \ps{\textsf{PS}} & \ps{\textsf{PS}} & \ps{\textsf{PS}} & \ps{\textsf{PS}} & \ps{\textsf{PS}} & \ps{\textsf{PS}} & \ps{\textsf{PS}} & \ps{\textsf{PS}} & \ps{\textsf{PS}} & \ps{\textsf{PS}} & \ps{\textsf{PS}} \\
\hline
110.0 & \cs{\textsf{CS}} & \ps{\textsf{PS}} & \ps{\textsf{PS}} & \ps{\textsf{PS}} & \ps{\textsf{PS}} & \ps{\textsf{PS}} & \ps{\textsf{PS}} & \ps{\textsf{PS}} & \ps{\textsf{PS}} & \ps{\textsf{PS}} & \ps{\textsf{PS}} & \ps{\textsf{PS}} & \ps{\textsf{PS}} \\
\hline
\end{tabular}}
\end{center}
\vspace{10pt}
\begin{center}
    Table \ref{tab:apollo-merging} (d) Apollo autopilot: Verdicts for merging scenarios with $v_e=15.0$
    \footnotesize 
    \adjustbox{max width=\textwidth}{
    \begin{tabular}{|c|c|c|c|c|c|c|c|c|c|c|c|c|c|c|c|c|c|c|c|}
\hline
\diagbox{$\bm{xa}$}{$\bm{xf}$}
& 0.0 & 5.0 & 10.0 & 20.0 & 40.0 & \ct{40.1} & 60.0 & 65.0 & 70.0 & 80.0 & 85.0 & 90.0 & 100.0 & 120.0 & 160.0 & 200.0 & 240.0 & 280.0 & 320.0 \\
\hline
0.0 & - & - & - & - & - & - & \cs{\textsf{CS}} & \cs{\textsf{CS}} & \cs{\textsf{CS}} & \cs{\textsf{CS}} & \cs{\textsf{CS}} & \cs{\textsf{CS}} & \cs{\textsf{CS}} & \cs{\textsf{CS}} & \cs{\textsf{CS}} & \cs{\textsf{CS}} & \cs{\textsf{CS}} & \cs{\textsf{CS}} & \cs{\textsf{CS}} \\
\hline
40.0 & - & - & - & \cs{\textsf{CS}} & \cs{\textsf{CS}} & \cs{\textsf{CS}} & \cs{\textsf{CS}} & \cs{\textsf{CS}} & \cs{\textsf{CS}} & \cs{\textsf{CS}} & \cs{\textsf{CS}} & \cs{\textsf{CS}} & \cs{\textsf{CS}} & \cs{\textsf{CS}} & \cs{\textsf{CS}} & \cs{\textsf{CS}} & \cs{\textsf{CS}} & \cs{\textsf{CS}} & \cs{\textsf{CS}} \\
\hline
60.0 & \cs{\textsf{CS}} & \cs{\textsf{CS}} & \cs{\textsf{CS}} & \cs{\textsf{CS}} & \cs{\textsf{CS}} & \cs{\textsf{CS}} & \cs{\textsf{CS}} & \cs{\textsf{CS}} & \cs{\textsf{CS}} & \cs{\textsf{CS}} & \cs{\textsf{CS}} & \cs{\textsf{CS}} & \cs{\textsf{CS}} & \cs{\textsf{CS}} & \cs{\textsf{CS}} & \cs{\textsf{CS}} & \cs{\textsf{CS}} & \cs{\textsf{CS}} & \cs{\textsf{CS}} \\
\hline
65.0 & \cs{\textsf{CS}} & \ps{\textsf{PS}} & \ps{\textsf{PS}} & \ps{\textsf{PS}} & \ps{\textsf{PS}} & \ps{\textsf{PS}} & \ps{\textsf{PS}} & \ps{\textsf{PS}} & \ps{\textsf{PS}} & \ps{\textsf{PS}} & \ps{\textsf{PS}} & \cs{\textsf{CS}} & \cs{\textsf{CS}} & \cs{\textsf{CS}} & \cs{\textsf{CS}} & \cs{\textsf{CS}} & \cs{\textsf{CS}} & \cs{\textsf{CS}} & \cs{\textsf{CS}} \\
\hline
70.0 & \cs{\textsf{CS}} & \ps{\textsf{PS}} & \ps{\textsf{PS}} & \ps{\textsf{PS}} & \ps{\textsf{PS}} & \ps{\textsf{PS}} & \ps{\textsf{PS}} & \ps{\textsf{PS}} & \ps{\textsf{PS}} & \ps{\textsf{PS}} & \cs{\textsf{CS}} & \ps{\textsf{PS}} & \ps{\textsf{PS}} & \ps{\textsf{PS}} & \ps{\textsf{PS}} & \ps{\textsf{PS}} & \ps{\textsf{PS}} & \ps{\textsf{PS}} & \ps{\textsf{PS}} \\
\hline
\end{tabular}}
\end{center}

\vspace{2em}
\paragraph{1) Apollo autopilot: Analysis for merging scenarios}

Verdicts for merging scenarios and different speeds $v_e$ are shown in Tab.~\ref{tab:apollo-merging}. They reveal accidents involving the front of the ego vehicle hitting the side of the arriving vehicle (\textsf{Ae}).

For $v_e=0.0$, accidents occur when $x_a=90.0$ and $x_f\ge 90.0$ during the transition from caution to progress. These values are both larger than the critical ones $\hat{x}_a = 59.5$ and $\hat{x}_f=0.0$. Thus, safe progress is possible. However, in accident scenarios, the ego vehicle is stationary at the beginning and starts to merge onto the main road when the arriving vehicle is at the merging point, resulting in a collision on the side of the arriving vehicle.

For $v_e=5.0$ and $v_e=10.0$, there is no accident in the transition from caution to progress. In fact, the arriving vehicle decelerates cautiously when it detects that the ego vehicle is approaching the merging point. This allows the ego vehicle to progress safely without applying maximum acceleration.

For $v_e=15.0$, the ego vehicle can progress safely when $x_f < \hat x_f = 40.1$ and $x_a < \hat x_a = 103.3$ due to the caution of the arriving vehicle towards the ego vehicle. 
In addition, for $x_f \ge 5.0$ and $90.0 \le x_a \le 150.0$, the ego vehicle is cautious but can progress safely for fixed $x_f$ and decreasing $x_a$ in the interval $[65.0, 85.0]$. In both cautious and progressive scenarios, the ego vehicle stops at the boundary of the main road. For $90.0 \le x_a \le 155.0$, the arriving vehicle passes the merging point first by applying lateral offsets within the lane to avoid collision with the ego vehicle. Conversely, for $65.0 \le x_a \le 80.0$, the arriving vehicle is initially closer to the merging point and lacks sufficient time to apply lateral offsets. Consequently, it stops before the merging point and lets the ego vehicle proceed.

\clearpage

\setlength{\tabcolsep}{2pt}
\renewcommand{\arraystretch}{1}

\vspace{2em}

\begin{minipage}{\textwidth}
\begin{table}[H]
    \centering
    \caption{Apollo autopilot: Verdicts for lane change scenarios}
    \label{tab:apollo-lane_change}
\end{table}
\vspace{-20pt}
    \begin{center}
    Table \ref{tab:apollo-lane_change} (a) Apollo autopilot: Verdicts for lane change scenarios with $v_e=5.0$
    \scriptsize 
    \adjustbox{max width=\linewidth}{\begin{tabular}{|c|c|c|c|c|c|c|c|c|c|c|c|c|c|c|c|c|c|c|c|c|c|c|c|c|}
\hline
\diagbox{$\bm{xa}$}{$\bm{xf}$}
& 0.0 & 5.0 & \ct{6.1} & 10.0 & 15.0 & 20.0 & 25.0 & 30.0 & 35.0 & 40.0 & 45.0 & 50.0 & 55.0 & 60.0 & 80.0 & 120.0 & 160.0 & 165.0 & 170.0 & 180.0 & 200.0 & 240.0 & 280.0 & 320.0 \\
\hline
0.0 & - & - & - & - & - & - & - & - & - & - & - & - & - & \cs{\textsf{CS}} & \cs{\textsf{CS}} & \cs{\textsf{CS}} & \cs{\textsf{CS}} & \cs{\textsf{CS}} & \cs{\textsf{CS}} & \cs{\textsf{CS}} & \cs{\textsf{CS}} & \cs{\textsf{CS}} & \cs{\textsf{CS}} & \cs{\textsf{CS}} \\
\hline
40.0 & - & - & - & - & - & \cs{\textsf{CS}} & \cs{\textsf{CS}} & \cs{\textsf{CS}} & \cs{\textsf{CS}} & \cs{\textsf{CS}} & \cs{\textsf{CS}} & \cs{\textsf{CS}} & \cs{\textsf{CS}} & \cs{\textsf{CS}} & \cs{\textsf{CS}} & \cs{\textsf{CS}} & \cs{\textsf{CS}} & \cs{\textsf{CS}} & \cs{\textsf{CS}} & \cs{\textsf{CS}} & \cs{\textsf{CS}} & \cs{\textsf{CS}} & \cs{\textsf{CS}} & \cs{\textsf{CS}} \\
\hline
80.0 & \cs{\textsf{CS}} & \cs{\textsf{CS}} & \cs{\textsf{CS}} & \cs{\textsf{CS}} & \cs{\textsf{CS}} & \cs{\textsf{CS}} & \cs{\textsf{CS}} & \cs{\textsf{CS}} & \cs{\textsf{CS}} & \cs{\textsf{CS}} & \cs{\textsf{CS}} & \cs{\textsf{CS}} & \cs{\textsf{CS}} & \cs{\textsf{CS}} & \cs{\textsf{CS}} & \cs{\textsf{CS}} & \cs{\textsf{CS}} & \cs{\textsf{CS}} & \cs{\textsf{CS}} & \cs{\textsf{CS}} & \cs{\textsf{CS}} & \cs{\textsf{CS}} & \cs{\textsf{CS}} & \cs{\textsf{CS}} \\
\hline
90.0 & \cs{\textsf{CS}} & \cs{\textsf{CS}} & \cs{\textsf{CS}} & \cs{\textsf{CS}} & \cs{\textsf{CS}} & \cs{\textsf{CS}} & \cs{\textsf{CS}} & \cs{\textsf{CS}} & \cs{\textsf{CS}} & \cs{\textsf{CS}} & \cs{\textsf{CS}} & \cs{\textsf{CS}} & \cs{\textsf{CS}} & \cs{\textsf{CS}} & \cs{\textsf{CS}} & \cs{\textsf{CS}} & \cs{\textsf{CS}} & \cs{\textsf{CS}} & \cs{\textsf{CS}} & \cs{\textsf{CS}} & \cs{\textsf{CS}} & \cs{\textsf{CS}} & \cs{\textsf{CS}} & \cs{\textsf{CS}} \\
\hline
\end{tabular}}\end{center}
\end{minipage}
\vspace{10pt}
    \begin{center}
    Table \ref{tab:apollo-lane_change} (b) Apollo autopilot: Verdicts for lane change scenarios with $v_e=10.0$
    \scriptsize 
    \adjustbox{max width=\linewidth}
    {\begin{tabular}{|c|c|c|c|c|c|c|c|c|c|c|c|c|c|c|c|c|c|c|c|c|c|c|c|c|c|c|}
\hline
\diagbox{$\bm{xa}$}{$\bm{xf}$}
& 0.0 & 10.0 & 15.0 & \ct{17.2} & 20.0 & 25.0 & 30.0 & 35.0 & 40.0 & 45.0 & 50.0 & 55.0 & 60.0 & 65.0 & 70.0 & 75.0 & 80.0 & 85.0 & 90.0 & 100.0 & 120.0 & 160.0 & 200.0 & 240.0 & 280.0 & 320.0 \\
\hline
0.0 & - & - & - & - & - & - & - & - & - & - & - & - & \cs{\textsf{CS}} & \cs{\textsf{CS}} & \cs{\textsf{CS}} & \cs{\textsf{CS}} & \cs{\textsf{CS}} & \cs{\textsf{CS}} & \cs{\textsf{CS}} & \cs{\textsf{CS}} & \cs{\textsf{CS}} & \cs{\textsf{CS}} & \cs{\textsf{CS}} & \cs{\textsf{CS}} & \cs{\textsf{CS}} & \cs{\textsf{CS}} \\
\hline
40.0 & - & - & - & - & \cs{\textsf{CS}} & \cs{\textsf{CS}} & \cs{\textsf{CS}} & \cs{\textsf{CS}} & \cs{\textsf{CS}} & \cs{\textsf{CS}} & \cs{\textsf{CS}} & \cs{\textsf{CS}} & \cs{\textsf{CS}} & \cs{\textsf{CS}} & \cs{\textsf{CS}} & \cs{\textsf{CS}} & \cs{\textsf{CS}} & \cs{\textsf{CS}} & \cs{\textsf{CS}} & \cs{\textsf{CS}} & \cs{\textsf{CS}} & \cs{\textsf{CS}} & \cs{\textsf{CS}} & \cs{\textsf{CS}} & \cs{\textsf{CS}} & \cs{\textsf{CS}} \\
\hline
50.0 & - & \cs{\textsf{CS}} & \cs{\textsf{CS}} & \cs{\textsf{CS}} & \cs{\textsf{CS}} & \cs{\textsf{CS}} & \cs{\textsf{CS}} & \cs{\textsf{CS}} & \cs{\textsf{CS}} & \cs{\textsf{CS}} & \cs{\textsf{CS}} & \cs{\textsf{CS}} & \cs{\textsf{CS}} & \cs{\textsf{CS}} & \cs{\textsf{CS}} & \cs{\textsf{CS}} & \cs{\textsf{CS}} & \cs{\textsf{CS}} & \cs{\textsf{CS}} & \cs{\textsf{CS}} & \cs{\textsf{CS}} & \cs{\textsf{CS}} & \cs{\textsf{CS}} & \cs{\textsf{CS}} & \cs{\textsf{CS}}& \cs{\textsf{CS}} \\
\hline
\end{tabular}}\end{center}
\vspace{10pt}
    \begin{center}
    Table \ref{tab:apollo-lane_change} (c) Apollo autopilot: Verdicts for lane change scenarios with $v_e=15.0$
    \footnotesize
    \adjustbox{max width=\linewidth}{\begin{tabular}{|c|c|c|c|c|c|c|c|c|c|c|c|c|c|c|c|c|c|c|c|c|c|c|c|c|c|c|c|c|c|}
\hline
\diagbox{$\bm{xa}$}{$\bm{xf}$}
& 0.0 & 20.0 & 30.0 & \ct{31.7} & 35.0 & 40.0 & 45.0 & 50.0 & 55.0 & 60.0 & 65.0 & 70.0 & 80.0 & 85.0 & 90.0 & 95.0 & 100.0 & 120.0 & 160.0 & 180.0 & 185.0 & 190.0 & 200.0 & 240.0 & 280.0 & 300.0 & 310.0 & 315.0 & 320.0 \\
\hline
0.0 & - & - & - & - & - & - & - & - & - & \cs{\textsf{CS}} & \cs{\textsf{CS}} & \cs{\textsf{CS}} & \cs{\textsf{CS}} & \cs{\textsf{CS}} & \cs{\textsf{CS}} & \cs{\textsf{CS}} & \cs{\textsf{CS}} & \cs{\textsf{CS}} & \cs{\textsf{CS}} & \cs{\textsf{CS}} & \cs{\textsf{CS}} & \cs{\textsf{CS}} & \cs{\textsf{CS}} & \cs{\textsf{CS}} & \cs{\textsf{CS}} & \cs{\textsf{CS}} & \cs{\textsf{CS}} & \cs{\textsf{CS}} & \cs{\textsf{CS}} \\
\hline
40.0 & - & \cs{\textsf{CS}} & \cs{\textsf{CS}} & \cs{\textsf{CS}} & \cs{\textsf{CS}} & \cs{\textsf{CS}} & \cs{\textsf{CS}} & \cs{\textsf{CS}} & \cs{\textsf{CS}} & \cs{\textsf{CS}} & \cs{\textsf{CS}} & \cs{\textsf{CS}} & \cs{\textsf{CS}} & \cs{\textsf{CS}} & \cs{\textsf{CS}} & \cs{\textsf{CS}} & \cs{\textsf{CS}} & \cs{\textsf{CS}} & \cs{\textsf{CS}} & \cs{\textsf{CS}} & \cs{\textsf{CS}} & \cs{\textsf{CS}} & \cs{\textsf{CS}} & \cs{\textsf{CS}} & \cs{\textsf{CS}} & \cs{\textsf{CS}} & \cs{\textsf{CS}} & \cs{\textsf{CS}} & \cs{\textsf{CS}} \\
\hline
50.0 & - & \cs{\textsf{CS}} & \cs{\textsf{CS}} & \cs{\textsf{CS}} & \cs{\textsf{CS}} & \cs{\textsf{CS}} & \cs{\textsf{CS}} & \cs{\textsf{CS}} & \cs{\textsf{CS}} & \cs{\textsf{CS}} & \cs{\textsf{CS}} & \cs{\textsf{CS}} & \cs{\textsf{CS}} & \cs{\textsf{CS}} & \cs{\textsf{CS}} & \cs{\textsf{CS}} & \cs{\textsf{CS}} & \cs{\textsf{CS}} & \cs{\textsf{CS}} & \cs{\textsf{CS}} & \cs{\textsf{CS}} & \cs{\textsf{CS}} & \cs{\textsf{CS}} & \cs{\textsf{CS}} & \cs{\textsf{CS}} & \cs{\textsf{CS}} & \cs{\textsf{CS}} & \cs{\textsf{CS}} & \cs{\textsf{CS}} \\
\hline
\end{tabular}}\end{center}
\vspace{10pt}
    \begin{center}
    Table \ref{tab:apollo-lane_change} (d) Apollo autopilot: Verdicts for lane change scenarios with $v_e=20.0$
    \footnotesize 
    \adjustbox{max width=\linewidth}{
    {\begin{tabular}{|c|c|c|c|c|c|c|c|c|c|c|c|c|c|c|c|c|c|c|c|c|}
\hline
\diagbox{$\bm{xa}$}{$\bm{xf}$}
& 0.0 & 40.0 & \ct{50.0} & 55.0 & 60.0 & 70.0 & 75.0 & 80.0 & 85.0 & 90.0 & 100.0 & 120.0 & 160.0 & 200.0 & 205.0 & 210.0 & 220.0 & 240.0 & 280.0 & 320.0 \\
\hline
0.0 & - & - & - & - & \cs{\textsf{CS}} & \cs{\textsf{CS}} & \cs{\textsf{CS}} & \cs{\textsf{CS}} & \cs{\textsf{CS}} & \cs{\textsf{CS}} & \cs{\textsf{CS}} & \cs{\textsf{CS}} & \cs{\textsf{CS}} & \cs{\textsf{CS}} & \cs{\textsf{CS}} & \cs{\textsf{CS}} & \cs{\textsf{CS}} & \cs{\textsf{CS}} & \cs{\textsf{CS}} & \cs{\textsf{CS}} \\
\hline
40.0 & - & \cs{\textsf{CS}} & \cs{\textsf{CS}} & \cs{\textsf{CS}} & \cs{\textsf{CS}} & \cs{\textsf{CS}} & \cs{\textsf{CS}} & \cs{\textsf{CS}} & \cs{\textsf{CS}} & \cs{\textsf{CS}} & \cs{\textsf{CS}} & \cs{\textsf{CS}} & \cs{\textsf{CS}} & \cs{\textsf{CS}} & \cs{\textsf{CS}} & \cs{\textsf{CS}} & \cs{\textsf{CS}} & \cs{\textsf{CS}} & \cs{\textsf{CS}} & \cs{\textsf{CS}} \\
\hline
\ct{74.5} & \cs{\textsf{CS}} & \cs{\textsf{CS}} & \cs{\textsf{CS}} & \cs{\textsf{CS}} & \cs{\textsf{CS}} & \cs{\textsf{CS}} & \cs{\textsf{CS}} & \cs{\textsf{CS}} & \cs{\textsf{CS}} & \cs{\textsf{CS}} & \cs{\textsf{CS}} & \cs{\textsf{CS}} & \cs{\textsf{CS}} & \cs{\textsf{CS}} & \cs{\textsf{CS}} & \cs{\textsf{CS}} & \cs{\textsf{CS}} & \cs{\textsf{CS}} & \cs{\textsf{CS}} & \cs{\textsf{CS}} \\
\hline
80.0 & \cs{\textsf{CS}} & \cs{\textsf{CS}} & \cs{\textsf{CS}} & \cs{\textsf{CS}} & \cs{\textsf{CS}} & \cs{\textsf{CS}} & \cs{\textsf{CS}} & \cs{\textsf{CS}} & \cs{\textsf{CS}} & \cs{\textsf{CS}} & \cs{\textsf{CS}} & \cs{\textsf{CS}} & \cs{\textsf{CS}} & \cs{\textsf{CS}} & \cs{\textsf{CS}} & \cs{\textsf{CS}} & \cs{\textsf{CS}} & \cs{\textsf{CS}} & \cs{\textsf{CS}} & \cs{\textsf{CS}} \\
\hline
\end{tabular}}}\end{center}

\vspace{2em}
\paragraph{2) Apollo autopilot: Analysis for lane change scenarios}

Verdicts for lane change scenarios and different speeds $v_e$ are shown in Tab.~\ref{tab:apollo-lane_change}, revealing accidents, software failure, and blocking scenarios.

For $v_e = 5.0$, a wide range of parameters can trigger software failures. The failures all occur when the ego vehicle switches from the inner to the outer lane, mainly due to the autopilot's inability to find a feasible policy while moving. Safe progress is only possible when the front vehicle on the outer lane is a long way off, and the arriving vehicle has sufficient space to brake.

For $v_e=10.0$, the autopilot becomes more aggressive. It can progress safely even when $x_a < \hat x_a = 89.6$. However, accidents can occur during the transition from caution to progress as $x_a$ increases. The autopilot makes the wrong decision to progress, as accidents can be avoided if it applies the cautious safe policies generated for lower parameter values.  The autopilot may also suffer from software failures, but to a lesser extent than in cases where $v_e = 5.0$. In addition, a few cases in the transition from caution to progress show that the ego vehicle can stop when switching from the internal lane to the external lane and block the passage of the arriving vehicle.

For $v_e=15.0$ and $v_e=20.0$, the autopilot can only safely progress when both $x_a$ and $x_f$ are much larger than critical values. That is, for $v_e=15.0$, the ego vehicle progresses safely only if $\langle x_f, x_a \rangle \ge \langle 190.0, 180.0 \rangle$ with critical values $\langle \hat x_f, \hat x_a \rangle = \langle 31.7, 79.5 \rangle$; for $v_e=20.0$, the ego vehicle progresses safely only if $\langle x_f, x_a \rangle \ge \langle 210.0, 190.0 \rangle$ with critical values $\langle \hat x_f, \hat x_a \rangle = \langle 50.0, 74.5 \rangle$. Software failures occur during the transition from caution to progress with increasing $x_a$ and $x_f$. Furthermore, for $v_e=15.0$, the ego vehicle can decide to progress and has software failures for  $x_a$ close to its critical value. As the value of $x_a$ increases, the autopilot again decides to be cautious. 
In addition, the autopilot can lock up when switching from the internal lane to the external lane, blocking the passage of the arriving vehicle. 


\vspace{2em}
\begin{minipage}{\linewidth}
\begin{table}[H]
    \centering
        \caption{Apollo autopilot: Verdicts for crossing with yield sign scenarios}
    \label{tab:apollo-crossing}
\end{table}
\vspace{-20pt}
    \begin{center}
    Table \ref{tab:apollo-crossing} (a) Apollo autopilot: Verdicts for crossing with yield sign scenarios with $v_e=0.0$
    \footnotesize 
    \adjustbox{max width=\linewidth}{
    {\begin{tabular}{|c|c|c|c|c|c|c|c|c|c|c|c|c|c|c|c|c|}
\hline
\diagbox{$\bm{xa}$}{$\bm{xf}$}
& 0.0 & 5.0 & 10.0 & 15.0 & \ct{15.4} & 20.0 & 25.0 & 30.0 & 40.0 & 80.0 & 120.0 & 160.0 & 200.0 & 240.0 & 280.0 & 320.0 \\
\hline
0.0 & \cs{\textsf{CS}} & \cs{\textsf{CS}} & \cs{\textsf{CS}} & \cs{\textsf{CS}} & \cs{\textsf{CS}} & \cs{\textsf{CS}} & \cs{\textsf{CS}} & \cs{\textsf{CS}} & \cs{\textsf{CS}} & \cs{\textsf{CS}} & \cs{\textsf{CS}} & \cs{\textsf{CS}} & \cs{\textsf{CS}} & \cs{\textsf{CS}} & \cs{\textsf{CS}} & \cs{\textsf{CS}} \\
\hline
40.0 & \cs{\textsf{CS}} & \cs{\textsf{CS}} & \cs{\textsf{CS}} & \cs{\textsf{CS}} & \cs{\textsf{CS}} & \cs{\textsf{CS}} & \cs{\textsf{CS}} & \cs{\textsf{CS}} & \cs{\textsf{CS}} & \cs{\textsf{CS}} & \cs{\textsf{CS}} & \cs{\textsf{CS}} & \cs{\textsf{CS}} & \cs{\textsf{CS}} & \cs{\textsf{CS}} & \cs{\textsf{CS}} \\
\hline
80.0 & \cs{\textsf{CS}} & \cs{\textsf{CS}} & \cs{\textsf{CS}} & \cs{\textsf{CS}} & \cs{\textsf{CS}} & \cs{\textsf{CS}} & \cs{\textsf{CS}} & \cs{\textsf{CS}} & \cs{\textsf{CS}} & \cs{\textsf{CS}} & \cs{\textsf{CS}} & \cs{\textsf{CS}} & \cs{\textsf{CS}} & \cs{\textsf{CS}} & \cs{\textsf{CS}} & \cs{\textsf{CS}} \\
\hline
\ct{120.0} & \cu{\textsf{CU$p_1$}} & \cu{\textsf{CU$p_1$}} & \cu{\textsf{CU$p_1$}} & \cu{\textsf{CU$p_1$}} & \cu{\textsf{CU$p_1$}} & \cu{\textsf{CU$p_1$}} & \cu{\textsf{CU$p_1$}} & \cu{\textsf{CU$p_1$}} & \cu{\textsf{CU$p_1$}} & \cu{\textsf{CU$p_1$}} & \cu{\textsf{CU$p_1$}} & \cu{\textsf{CU$p_1$}} & \cu{\textsf{CU$p_1$}} & \cu{\textsf{CU$p_1$}} & \cu{\textsf{CU$p_1$}} & \cu{\textsf{CU$p_1$}} \\
\hline
\end{tabular}}}\end{center}
\end{minipage}
\vspace{10pt}
    \begin{center}
    Table \ref{tab:apollo-crossing} (b) Apollo autopilot: Verdicts for crossing with yield sign scenarios with $v_e=5.0$
    \footnotesize {
    \adjustbox{max width=\linewidth}{
    \begin{tabular}{|c|c|c|c|c|c|c|c|c|c|c|c|c|c|c|c|c|}
\hline
\diagbox{$\bm{xa}$}{$\bm{xf}$}
& 0.0 & 5.0 & 10.0 & 15.0 & 20.0 & \ct{20.2} & 25.0 & 30.0 & 40.0 & 80.0 & 120.0 & 160.0 & 200.0 & 240.0 & 280.0 & 320.0 \\
\hline
0.0 & \cs{\textsf{CS}} & \cs{\textsf{CS}} & \cs{\textsf{CS}} & \cs{\textsf{CS}} & \cs{\textsf{CS}} & \cs{\textsf{CS}} & \cs{\textsf{CS}} & \cs{\textsf{CS}} & \cs{\textsf{CS}} & \cs{\textsf{CS}} & \cs{\textsf{CS}} & \cs{\textsf{CS}} & \cs{\textsf{CS}} & \cs{\textsf{CS}} & \cs{\textsf{CS}} & \cs{\textsf{CS}} \\
\hline
40.0 & \cs{\textsf{CS}} & \cs{\textsf{CS}} & \cs{\textsf{CS}} & \cs{\textsf{CS}} & \cs{\textsf{CS}} & \cs{\textsf{CS}} & \cs{\textsf{CS}} & \cs{\textsf{CS}} & \cs{\textsf{CS}} & \cs{\textsf{CS}} & \cs{\textsf{CS}} & \cs{\textsf{CS}} & \cs{\textsf{CS}} & \cs{\textsf{CS}} & \cs{\textsf{CS}} & \cs{\textsf{CS}} \\
\hline
80.0 & \cs{\textsf{CS}} & \cs{\textsf{CS}} & \cs{\textsf{CS}} & \cs{\textsf{CS}} & \cs{\textsf{CS}} & \cs{\textsf{CS}} & \cs{\textsf{CS}} & \cs{\textsf{CS}} & \cs{\textsf{CS}} & \cs{\textsf{CS}} & \cs{\textsf{CS}} & \cs{\textsf{CS}} & \cs{\textsf{CS}} & \cs{\textsf{CS}} & \cs{\textsf{CS}} & \cs{\textsf{CS}} \\
\hline
\ct{84.8} & \cs{\textsf{CS}} & \cs{\textsf{CS}} & \cs{\textsf{CS}} & \cs{\textsf{CS}} & \cs{\textsf{CS}} & \cs{\textsf{CS}} & \cs{\textsf{CS}} & \cs{\textsf{CS}} & \cs{\textsf{CS}} & \cs{\textsf{CS}} & \cs{\textsf{CS}} & \cs{\textsf{CS}} & \cs{\textsf{CS}} & \cs{\textsf{CS}} & \cs{\textsf{CS}} & \cs{\textsf{CS}} \\
\hline
120.0 & \cu{\textsf{CU$p_1$}} & \cu{\textsf{CU$p_1$}} & \cu{\textsf{CU$p_1$}} & \cu{\textsf{CU$p_1$}} & \cu{\textsf{CU$p_1$}} & \cu{\textsf{CU$p_1$}} & \cu{\textsf{CU$p_1$}} & \cu{\textsf{CU$p_1$}} & \cu{\textsf{CU$p_1$}} & \cu{\textsf{CU$p_1$}} & \cu{\textsf{CU$p_1$}} & \cu{\textsf{CU$p_1$}} & \cu{\textsf{CU$p_1$}} & \cu{\textsf{CU$p_1$}} & \cu{\textsf{CU$p_1$}} & \cu{\textsf{CU$p_1$}} \\
\hline
\end{tabular}}}\end{center}
\vspace{10pt}
    \begin{center}
    Table \ref{tab:apollo-crossing} (c) Apollo autopilot: Verdicts for crossing with yield sign scenarios with $v_e=10.0$
    \footnotesize 
    \adjustbox{max width=\linewidth}{
    {\begin{tabular}{|c|c|c|c|c|c|c|c|c|c|c|c|c|c|}
\hline
\diagbox{$\bm{xa}$}{$\bm{xf}$}
& 0.0 & 20.0 & 25.0 & 30.0 & \ct{32.2} & 40.0 & 80.0 & 120.0 & 160.0 & 200.0 & 240.0 & 280.0 & 320.0 \\
\hline
0.0 & \cs{\textsf{CS}} & \cs{\textsf{CS}} & \cs{\textsf{CS}} & \cs{\textsf{CS}} & \cs{\textsf{CS}} & \cs{\textsf{CS}} & \cs{\textsf{CS}} & \cs{\textsf{CS}} & \cs{\textsf{CS}} & \cs{\textsf{CS}} & \cs{\textsf{CS}} & \cs{\textsf{CS}} & \cs{\textsf{CS}} \\
\hline
40.0 & \cs{\textsf{CS}} & \cs{\textsf{CS}} & \cs{\textsf{CS}} & \cs{\textsf{CS}} & \cs{\textsf{CS}} & \cs{\textsf{CS}} & \cs{\textsf{CS}} & \cs{\textsf{CS}} & \cs{\textsf{CS}} & \cs{\textsf{CS}} & \cs{\textsf{CS}} & \cs{\textsf{CS}} & \cs{\textsf{CS}} \\
\hline
\ct{73.9} & \cs{\textsf{CS}} & \cs{\textsf{CS}} & \cs{\textsf{CS}} & \cs{\textsf{CS}} & \cs{\textsf{CS}} & \cs{\textsf{CS}} & \cs{\textsf{CS}} & \cs{\textsf{CS}} & \cs{\textsf{CS}} & \cs{\textsf{CS}} & \cs{\textsf{CS}} & \cs{\textsf{CS}} & \cs{\textsf{CS}} \\
\hline
80.0 & \cs{\textsf{CS}} & \cs{\textsf{CS}} & \cs{\textsf{CS}} & \cs{\textsf{CS}} & \cs{\textsf{CS}} & \cs{\textsf{CS}} & \cs{\textsf{CS}} & \cs{\textsf{CS}} & \cs{\textsf{CS}} & \cs{\textsf{CS}} & \cs{\textsf{CS}} & \cs{\textsf{CS}} & \cs{\textsf{CS}} \\
\hline
120.0 & \cs{\textsf{CS}} & \cs{\textsf{CS}} & \cs{\textsf{CS}} & \cs{\textsf{CS}} & \cs{\textsf{CS}} & \cs{\textsf{CS}} & \cs{\textsf{CS}} & \cs{\textsf{CS}} & \cs{\textsf{CS}} & \cs{\textsf{CS}} & \cs{\textsf{CS}} & \cs{\textsf{CS}} & \cs{\textsf{CS}} \\
\hline
160.0 & \cu{\textsf{CU$p_1$}} & \cu{\textsf{CU$p_1$}} & \cu{\textsf{CU$p_1$}} & \cu{\textsf{CU$p_1$}} & \cu{\textsf{CU$p_1$}} & \cu{\textsf{CU$p_1$}} & \cu{\textsf{CU$p_1$}} & \cu{\textsf{CU$p_1$}} & \cu{\textsf{CU$p_1$}} & \cu{\textsf{CU$p_1$}} & \cu{\textsf{CU$p_1$}} & \cu{\textsf{CU$p_1$}} & \cu{\textsf{CU$p_1$}} \\
\hline
\end{tabular}}}\end{center}
    \hspace{\fill}
\vspace{10pt}
    \begin{center}
    Table \ref{tab:apollo-crossing} (d) Apollo autopilot: Verdicts for crossing with yield sign scenarios with $v_e=15.0$
    \footnotesize 
    \adjustbox{max width=\linewidth}{
    {\begin{tabular}{|c|c|c|c|c|c|c|c|c|c|c|c|}
\hline
\diagbox{$\bm{xa}$}{$\bm{xf}$}
& 0.0 & 20.0 & 40.0 & \ct{49.8} & 80.0 & 120.0 & 160.0 & 200.0 & 240.0 & 280.0 & 320.0 \\
\hline
0.0 & \cs{\textsf{CS}} & \cs{\textsf{CS}} & \cs{\textsf{CS}} & \cs{\textsf{CS}} & \cs{\textsf{CS}} & \cs{\textsf{CS}} & \cs{\textsf{CS}} & \cs{\textsf{CS}} & \cs{\textsf{CS}} & \cs{\textsf{CS}} & \cs{\textsf{CS}} \\
\hline
40.0 & \cs{\textsf{CS}} & \cs{\textsf{CS}} & \cs{\textsf{CS}} & \cs{\textsf{CS}} & \cs{\textsf{CS}} & \cs{\textsf{CS}} & \cs{\textsf{CS}} & \cs{\textsf{CS}} & \cs{\textsf{CS}} & \cs{\textsf{CS}} & \cs{\textsf{CS}} \\
\hline
\ct{71.5} & \cs{\textsf{CS}} & \cs{\textsf{CS}} & \cs{\textsf{CS}} & \cs{\textsf{CS}} & \cs{\textsf{CS}} & \cs{\textsf{CS}} & \cs{\textsf{CS}} & \cs{\textsf{CS}} & \cs{\textsf{CS}} & \cs{\textsf{CS}} & \cs{\textsf{CS}} \\
\hline
80.0 & \cs{\textsf{CS}} & \cs{\textsf{CS}} & \cs{\textsf{CS}} & \cs{\textsf{CS}} & \cs{\textsf{CS}} & \cs{\textsf{CS}} & \cs{\textsf{CS}} & \cs{\textsf{CS}} & \cs{\textsf{CS}} & \cs{\textsf{CS}} & \cs{\textsf{CS}} \\
\hline
120.0 & \cs{\textsf{CS}} & \cs{\textsf{CS}} & \cs{\textsf{CS}} & \cs{\textsf{CS}} & \cs{\textsf{CS}} & \cs{\textsf{CS}} & \cs{\textsf{CS}} & \cs{\textsf{CS}} & \cs{\textsf{CS}} & \cs{\textsf{CS}} & \cs{\textsf{CS}} \\
\hline
160.0 & \cu{\textsf{CU$p_1$}} & \cu{\textsf{CU$p_1$}} & \cu{\textsf{CU$p_1$}} & \cu{\textsf{CU$p_1$}} & \cu{\textsf{CU$p_1$}} & \cu{\textsf{CU$p_1$}} & \cu{\textsf{CU$p_1$}} & \cu{\textsf{CU$p_1$}} & \cu{\textsf{CU$p_1$}} & \cu{\textsf{CU$p_1$}} & \cu{\textsf{CU$p_1$}} \\
\hline
200.0 & \cu{\textsf{CU$p_1$}} & \cu{\textsf{CU$p_1$}} & \cu{\textsf{CU$p_1$}} & \cu{\textsf{CU$p_1$}} & \cu{\textsf{CU$p_1$}} & \cu{\textsf{CU$p_1$}} & \cu{\textsf{CU$p_1$}} & \cu{\textsf{CU$p_1$}} & \cu{\textsf{CU$p_1$}} & \cu{\textsf{CU$p_1$}} & \cu{\textsf{CU$p_1$}} \\
\hline
\end{tabular}}}\end{center}

\vspace{2em}
\paragraph{3) Apollo autopilot: Analysis for crossing with yield sign scenarios}

Verdicts for crossing with yield sign scenarios and different speeds $v_e$ are shown in Tab.~\ref{tab:apollo-crossing}. They reveal violations of safety properties $p_1$ (Two vehicles must not be in the critical zone at the same time) and $p_2$ (The ego vehicle must not stop inside the critical zone) when transitioning from caution to progress, as well as  overly cautious polices.

For all values of $v_e$, we observed overcautious policies as the ego vehicle only progresses when $x_a$ is much greater than its critical value, or does not progress within the tested range. For example, with $v_e=0.0$, the ego vehicle progresses only when $x_a > 240.0$, which is much larger than $\hat x_a=120.0$,  and for $v_e=15.0$, the ego vehicle does not progress for all values of $x_a$. 
Despite its over-caution, a wide range of parameters can lead to property violations. The ego vehicle can stop or drive at low speed inside the critical zone to wait for the arriving vehicle to pass which can lead to unsafe caution violating $p_1$ (The ego vehicle must not stop inside the critical zone) and $p_2$ (The ego vehicle must not stop inside the critical zone). When the ego vehicle reaches the conflict point first, it can also stop inside the critical zone, violating $p_2$ and the arriving vehicle may reach the critical zone during the ego vehicle's stop and violate $p_1$.

\clearpage
\begin{table}[H]
    \caption{Apollo autopilot: Verdicts for crossing with traffic light scenarios}
    \label{tab:apollo-traffic_light}
\end{table}
\vspace{-20pt}
    \begin{center}
    Table \ref{tab:apollo-traffic_light} (a) Apollo autopilot: Verdicts for crossing  with traffic light scenarios with $v_e=0.0$
    \small 
    {\begin{tabular}{|c|c|c|c|c|c|c|c|c|}
\hline
\multicolumn{9}{|c|}{$\bm{xf}$}\\
\hline
0.0 & 40.0 & 80.0 & 120.0 & 160.0 & 200.0 & 240.0 & 280.0 & 320.0 \\
\hline
\pu{\textsf{PU$p_4$}} & \pu{\textsf{PU$p_4$}} & \pu{\textsf{PU$p_4$}} & \pu{\textsf{PU$p_4$}} & \pu{\textsf{PU$p_4$}} & \pu{\textsf{PU$p_4$}} & \pu{\textsf{PU$p_4$}} & \pu{\textsf{PU$p_4$}} & \pu{\textsf{PU$p_4$}} \\
\hline
\end{tabular}
}\end{center}
\vspace{10pt}
    \begin{center}
    Table \ref{tab:apollo-traffic_light} (b) Apollo autopilot: Verdicts for crossing with traffic light scenarios with $v_e=5.0$
    \small {
    \begin{tabular}{|c|c|c|c|c|c|c|c|c|c|}
\hline
\multicolumn{10}{|c|}{$\bm{xf}$}\\
\hline
0.0 & \ct{20.2} & 40.0 & 80.0 & 120.0 & 160.0 & 200.0 & 240.0 & 280.0 & 320.0 \\
\hline
\cs{\textsf{CS}} & \cs{\textsf{CS}} & \cs{\textsf{CS}} & \cs{\textsf{CS}} & \cs{\textsf{CS}} & \cs{\textsf{CS}} & \cs{\textsf{CS}} & \cs{\textsf{CS}} & \cs{\textsf{CS}} & \cs{\textsf{CS}} \\
\hline
\end{tabular}
}\end{center}
\vspace{10pt}
    \begin{center}
    Table \ref{tab:apollo-traffic_light} (c) Apollo autopilot: Verdicts for crossing  with traffic light scenarios with $v_e=10.0$
    \small 
    {\begin{tabular}{|c|c|c|c|c|c|c|c|c|c|c|c|c|c|c|}
\hline
\multicolumn{15}{|c|}{$\bm{xf}$}\\
\hline
0.0 & 10.0 & 15.0 & 20.0 & 25.0 & 30.0 & \ct{32.2} & 40.0 & 80.0 & 120.0 & 160.0 & 200.0 & 240.0 & 280.0 & 320.0 \\
\hline
\cs{\textsf{CS}} & \cs{\textsf{CS}} & \pu{\textsf{PU$p_4$}} & \pu{\textsf{PU$p_4$}} & \ps{\textsf{PS}} & \ps{\textsf{PS}} & \ps{\textsf{PS}} & \ps{\textsf{PS}} & \ps{\textsf{PS}} & \ps{\textsf{PS}} & \ps{\textsf{PS}} & \ps{\textsf{PS}} & \ps{\textsf{PS}} & \ps{\textsf{PS}} & \ps{\textsf{PS}} \\
\hline
\end{tabular}}\end{center}
\vspace{10pt}
    \begin{center}
    Table \ref{tab:apollo-traffic_light} (d) Apollo autopilot: Verdicts for crossing with traffic light scenarios with $v_e=15.0$
    \small 
    {\begin{tabular}{|c|c|c|c|c|c|c|c|c|c|c|c|c|}
\hline
\multicolumn{13}{|c|}{$\bm{xf}$}\\
\hline
0.0 & 20.0 & 25.0 & 30.0 & 40.0 & \ct{49.8} & 80.0 & 120.0 & 160.0 & 200.0 & 240.0 & 280.0 & 320.0 \\
\hline
\pu{\textsf{PU$p_4$}} & \pu{\textsf{PU$p_4$}} & \ps{\textsf{PS}} & \ps{\textsf{PS}} & \ps{\textsf{PS}} & \ps{\textsf{PS}} & \ps{\textsf{PS}} & \ps{\textsf{PS}} & \ps{\textsf{PS}} & \ps{\textsf{PS}} & \ps{\textsf{PS}} & \ps{\textsf{PS}} & \ps{\textsf{PS}} \\
\hline
\end{tabular}}\end{center}
\vspace{10pt}
    \begin{center}
    Table \ref{tab:apollo-traffic_light} (e) Apollo autopilot: Verdicts for crossing with traffic light scenarios with $v_e=20.0$
    \small 
    {\begin{tabular}{|c|c|c|c|c|c|c|c|c|c|c|c|c|c|}
\hline
\multicolumn{14}{|c|}{$\bm{xf}$}\\
\hline
0.0 & 20.0 & 25.0 & 30.0 & 35.0 & 40.0 & \ct{59.5} & 80.0 & 120.0 & 160.0 & 200.0 & 240.0 & 280.0 & 320.0 \\
\hline
\pu{\textsf{PU$p_3p_4$}} & \pu{\textsf{PU$p_3p_4$}} & \pu{\textsf{PU$p_3p_4$}} & \pu{\textsf{PU$p_3$}} & \ps{\textsf{PS}} & \ps{\textsf{PS}} & \ps{\textsf{PS}} & \ps{\textsf{PS}} & \ps{\textsf{PS}} & \ps{\textsf{PS}} & \ps{\textsf{PS}} & \ps{\textsf{PS}} & \ps{\textsf{PS}} & \ps{\textsf{PS}} \\
\hline
\end{tabular}}\end{center}

\vspace{2em}
\paragraph{4) Apollo autopilot: Analysis for crossing with traffic light scenarios}

Verdicts for crossing with traffic light scenarios and different speeds $v_e$ are shown in Tab.~\ref{tab:apollo-traffic_light}. We found violations of safety properties $p_3$ (The ego vehicle must not enter the critical zone when the light is red) and $p_4$ (The ego vehicle must not be in the critical zone when the side light is green). Violations are caused by the autopilot's failure to take into account the time constants governing traffic light state changes.

For $v_0=0.0$, there is no safe progress policy according to the theory, as the ego vehicle cannot cross the critical zone before the side light turns green. However, the ego vehicle always decides to progress, which violates $p_4$ (The ego vehicle must not be in the critical zone when the side light is green). 

For $v_e=5.0$, the ego vehicle is cautious for all $x_f$ values. However, for values $x_f \ge \hat{x}_f =20.2$, it is too cautious as there is a feasible progress policy. 

For $v_e=10.0$, the ego vehicle is cautious when $x_f \le 10.0$ and decides to move forward otherwise.
However, it moves forward when $x_f$ is smaller than $\hat{x}_f=33.0$ and fails to exit the critical zone within the 2-second time limit imposed by the all-red phase.

For $v_e=15.0$, the ego vehicle always decides to progress. However, for $x_f < \hat x_f = 49.8$, the ego vehicle may not progress safely by applying sufficient acceleration, as the front vehicle imposes a speed constraint. As a result, it violates property $p_4$ (The ego vehicle must not be in the critical zone when the side light is green) when $x_f \le 20.0$.

For $v_e=20.0$, the ego vehicle can violate $p_4$ for the same reason as for  $v_e=15.0$ when $x_f \le 25.0 < \hat x_f = 59.5$. Moreover, it can also fail to reach the critical zone within the 3-second time limit imposed by the yellow-light phase when $x_f \le 30.0$, violating $p_3$ (The ego vehicle must not enter the critical zone when the light is red).

\vspace{3em}
\subsection{Testing the Autoware Autopilot}

\subsubsection{Autoware autopilot: Estimating the A/D functions}

The Autoware autopilot system uses A/D functions
similar to those of the Apollo autopilot. The jerk is in the range [-5.0 m/s$^3$, 1.0 m/s$^3$], the maximum acceleration rate is 1.0 m/s$^2$ and the maximum deceleration rate is 5.0 m/s$^3$. From the analytical specification of the A/D functions, we have obtained the values given in Tab.~\ref{tab:autoware-braking} and Tab.~\ref{tab:autoware-av}.

\begin{table}[H]
    \centering
    \caption{$B(v)$ for Autoware autopilot}
    \label{tab:autoware-braking}
    \vspace{-10pt}
    \begin{tabular}{|c|c|c|c|c|c|}
    \hline
        $\bm{v}$ & 0.0 & 5.0 & 10.0 & 15.0 & 20.0 \\
    \hline
        $\bm{B(v)}$ & 0.0 & 4.8 & 14.8 & 29.8 & 49.8 \\
    \hline
    \end{tabular}
\end{table}

\begin{table}[H]
    \caption{$AV(v, x)$, $AT(v, x)$ for Autoware autopilot}
    \label{tab:autoware-av}
    \vspace{-10pt}
        \begin{tabular}{|c|c|c|c|c|c|c|c|}
        \hline
        \diagbox[width=3em,height=2em]{~$\bm v$}{$\bm x$~} 
             & 0.0 & 10.0 & 20.0 & 30.0 & 40.0 & 50.0 & 60.0 \\
        \hline
0.0& 0.0, 0.0& 4.4, 5.0& 6.2, 6.8& 7.6, 8.2& 8.8, 9.4& 9.9, 10.5& 10.9, 11.5 \\\hline
5.0& 5.0, 0.0& 6.3, 1.7& 7.6, 3.2& 8.8, 4.4& 9.9, 5.5& 10.9, 6.5& 11.7, 7.3 \\\hline
10.0& 10.0, 0.0& 10.4, 1.0& 11.4, 1.8& 12.1, 2.7& 12.9, 3.5& 13.7, 4.3& 14.4, 5.0 \\\hline
15.0& 15.0, 0.0& 15.2, 0.7& 15.7, 1.3& 16.5, 1.9& 16.9, 2.5& 17.5, 3.1& 18.1, 3.7 \\\hline
        \end{tabular}
\end{table}

\subsubsection{Autoware autopilot: Results and analysis}

The test results for the Carla autopilot are shown in Tab. \ref{tab:autoware-merging} to Tab. \ref{tab:autoware-traffic_light} where the critical values $\hat{x}_a$ and $\hat{x}_f$ are indicated in red.

\begin{table}[H]
    \centering
    \caption{Autoware autopilot: Verdicts for merging scenarios}
    \label{tab:autoware-merging}
\end{table}
\vspace{-20pt}
\begin{center}
    Table \ref{tab:autoware-merging} (a) Autoware autopilot: Verdicts for merging scenarios with $v_e=0.0$
    \footnotesize 
    \adjustbox{max width=\textwidth}{\begin{tabular}{|c|c|c|c|c|c|c|c|c|c|c|c|c|c|}
\hline
\diagbox{$\bm{xa}$}{$\bm{xf}$}
& \ct{0.0} & 5.0 & 10.0 & 15.0 & 20.0 & 40.0 & 80.0 & 120.0 & 160.0 & 200.0 & 240.0 & 280.0 & 320.0 \\
\hline
0.0 & - & - & - & - & - & - & \cs{\textsf{CS}} & \cs{\textsf{CS}} & \cs{\textsf{CS}} & \cs{\textsf{CS}} & \cs{\textsf{CS}} & \cs{\textsf{CS}} & \cs{\textsf{CS}} \\
\hline
40.0 & - & - & - & - & - & \cs{\textsf{CS}} & \cs{\textsf{CS}} & \cs{\textsf{CS}} & \cs{\textsf{CS}} & \cs{\textsf{CS}} & \cs{\textsf{CS}} & \cs{\textsf{CS}} & \cs{\textsf{CS}} \\
\hline
\ct{60.3} & \cs{\textsf{CS}} & \cs{\textsf{CS}} & \cs{\textsf{CS}} & \cs{\textsf{CS}} & \cs{\textsf{CS}} & \cs{\textsf{CS}} & \cs{\textsf{CS}} & \cs{\textsf{CS}} & \cs{\textsf{CS}} & \cs{\textsf{CS}} & \cs{\textsf{CS}} & \cs{\textsf{CS}} & \cs{\textsf{CS}} \\
\hline
80.0 & \cs{\textsf{CS}} & \cs{\textsf{CS}} & \cs{\textsf{CS}} & \cs{\textsf{CS}} & \cs{\textsf{CS}} & \cs{\textsf{CS}} & \cs{\textsf{CS}} & \cs{\textsf{CS}} & \cs{\textsf{CS}} & \cs{\textsf{CS}} & \cs{\textsf{CS}} & \cs{\textsf{CS}} & \cs{\textsf{CS}} \\
\hline
120.0 & \cs{\textsf{CS}} & \cs{\textsf{CS}} & \cs{\textsf{CS}} & \cs{\textsf{CS}} & \cs{\textsf{CS}} & \cs{\textsf{CS}} & \cs{\textsf{CS}} & \cs{\textsf{CS}} & \cs{\textsf{CS}} & \cs{\textsf{CS}} & \cs{\textsf{CS}} & \cs{\textsf{CS}} & \cs{\textsf{CS}} \\
\hline
160.0 & \cs{\textsf{CS}} & \cs{\textsf{CS}} & \cs{\textsf{CS}} & \cs{\textsf{CS}} & \cs{\textsf{CS}} & \cs{\textsf{CS}} & \cs{\textsf{CS}} & \cs{\textsf{CS}} & \cs{\textsf{CS}} & \cs{\textsf{CS}} & \cs{\textsf{CS}} & \cs{\textsf{CS}} & \cs{\textsf{CS}} \\
\hline
200.0 & \cs{\textsf{CS}} & \cs{\textsf{CS}} & \cs{\textsf{CS}} & \cs{\textsf{CS}} & \cs{\textsf{CS}} & \cs{\textsf{CS}} & \cs{\textsf{CS}} & \cs{\textsf{CS}} & \cs{\textsf{CS}} & \cs{\textsf{CS}} & \cs{\textsf{CS}} & \cs{\textsf{CS}} & \cs{\textsf{CS}} \\
\hline
\end{tabular}}
\end{center}
\vspace{10pt}
\begin{center}
    Table \ref{tab:autoware-merging} (b) Autoware autopilot: Verdicts for merging scenarios with $v_e=5.0$
    \footnotesize 
    \adjustbox{max width=\textwidth}{\begin{tabular}{|c|c|c|c|c|c|c|c|c|c|c|c|c|c|}
\hline
\diagbox{$\bm{xa}$}{$\bm{xf}$}
& 0.0 & 5.0 & \ct{5.3} & 10.0 & 20.0 & 40.0 & 80.0 & 120.0 & 160.0 & 200.0 & 240.0 & 280.0 & 320.0 \\
\hline
0.0 & - & - & - & - & - & - & \cs{\textsf{CS}} & \cs{\textsf{CS}} & \cs{\textsf{CS}} & \cs{\textsf{CS}} & \cs{\textsf{CS}} & \cs{\textsf{CS}} & \cs{\textsf{CS}} \\
\hline
40.0 & - & - & - & - & - & \cs{\textsf{CS}} & \cs{\textsf{CS}} & \cs{\textsf{CS}} & \cs{\textsf{CS}} & \cs{\textsf{CS}} & \cs{\textsf{CS}} & \cs{\textsf{CS}} & \cs{\textsf{CS}} \\
\hline
80.0 & \cs{\textsf{CS}} & \cs{\textsf{CS}} & \cs{\textsf{CS}} & \cs{\textsf{CS}} & \cs{\textsf{CS}} & \cs{\textsf{CS}} & \cs{\textsf{CS}} & \cs{\textsf{CS}} & \cs{\textsf{CS}} & \cs{\textsf{CS}} & \cs{\textsf{CS}} & \cs{\textsf{CS}} & \cs{\textsf{CS}} \\
\hline
\ct{80.7} & \cs{\textsf{CS}} & \cs{\textsf{CS}} & \cs{\textsf{CS}} & \cs{\textsf{CS}} & \cs{\textsf{CS}} & \cs{\textsf{CS}} & \cs{\textsf{CS}} & \cs{\textsf{CS}} & \cs{\textsf{CS}} & \cs{\textsf{CS}} & \cs{\textsf{CS}} & \cs{\textsf{CS}} & \cs{\textsf{CS}} \\
\hline
120.0 & \cs{\textsf{CS}} & \cs{\textsf{CS}} & \cs{\textsf{CS}} & \cs{\textsf{CS}} & \cs{\textsf{CS}} & \cs{\textsf{CS}} & \cs{\textsf{CS}} & \cs{\textsf{CS}} & \cs{\textsf{CS}} & \cs{\textsf{CS}} & \cs{\textsf{CS}} & \cs{\textsf{CS}} & \cs{\textsf{CS}} \\
\hline
140.0 & \cs{\textsf{CS}} & \cs{\textsf{CS}} & \cs{\textsf{CS}} & \cs{\textsf{CS}} & \cs{\textsf{CS}} & \cs{\textsf{CS}} & \cs{\textsf{CS}} & \cs{\textsf{CS}} & \cs{\textsf{CS}} & \cs{\textsf{CS}} & \cs{\textsf{CS}} & \cs{\textsf{CS}} & \cs{\textsf{CS}} \\
\hline
145.0 & \cs{\textsf{CS}} & \cs{\textsf{CS}} & \cs{\textsf{CS}} & \ps{\textsf{PS}} & \ps{\textsf{PS}} & \ps{\textsf{PS}} & \ps{\textsf{PS}} & \ps{\textsf{PS}} & \ps{\textsf{PS}} & \ps{\textsf{PS}} & \ps{\textsf{PS}} & \ps{\textsf{PS}} & \ps{\textsf{PS}} \\
\hline
\end{tabular}}
\end{center}
\vspace{10pt}
\begin{center}
    Table \ref{tab:autoware-merging} (c) Autoware autopilot: Verdicts for merging scenarios with $v_e=10.0$
    \footnotesize 
    \adjustbox{max width=\textwidth}{\begin{tabular}{|c|c|c|c|c|c|c|c|c|c|c|c|c|c|c|c|c|c|c|c|c|}
\hline
\diagbox{$\bm{xa}$}{$\bm{xf}$}
& 0.0 & 5.0 & 10.0 & 15.0 & \ct{16.9} & 20.0 & 40.0 & 60.0 & 65.0 & 70.0 & 80.0 & 85.0 & 90.0 & 100.0 & 120.0 & 160.0 & 200.0 & 240.0 & 280.0 & 320.0 \\
\hline
0.0 & - & - & - & - & - & - & - & - & \cs{\textsf{CS}} & \cs{\textsf{CS}} & \cs{\textsf{CS}} & \cs{\textsf{CS}} & \cs{\textsf{CS}} & \cs{\textsf{CS}} & \cs{\textsf{CS}} & \cs{\textsf{CS}} & \cs{\textsf{CS}} & \cs{\textsf{CS}} & \cs{\textsf{CS}} & \cs{\textsf{CS}} \\
\hline
40.0 & - & - & - & - & - & - & \cs{\textsf{CS}} & \cs{\textsf{CS}} & \cs{\textsf{CS}} & \cs{\textsf{CS}} & \cs{\textsf{CS}} & \cs{\textsf{CS}} & \cs{\textsf{CS}} & \cs{\textsf{CS}} & \cs{\textsf{CS}} & \cs{\textsf{CS}} & \cs{\textsf{CS}} & \cs{\textsf{CS}} & \cs{\textsf{CS}} & \cs{\textsf{CS}} \\
\hline
80.0 & \cs{\textsf{CS}} & \cs{\textsf{CS}} & \cs{\textsf{CS}} & \cs{\textsf{CS}} & \cs{\textsf{CS}} & \cs{\textsf{CS}} & \cs{\textsf{CS}} & \cs{\textsf{CS}} & \cs{\textsf{CS}} & \cs{\textsf{CS}} & \cs{\textsf{CS}} & \cs{\textsf{CS}} & \cs{\textsf{CS}} & \cs{\textsf{CS}} & \cs{\textsf{CS}} & \cs{\textsf{CS}} & \cs{\textsf{CS}} & \cs{\textsf{CS}} & \cs{\textsf{CS}} & \cs{\textsf{CS}} \\
\hline
85.0 & \cs{\textsf{CS}} & \cs{\textsf{CS}} & \cs{\textsf{CS}} & \cs{\textsf{CS}} & \cs{\textsf{CS}} & \cs{\textsf{CS}} & \cs{\textsf{CS}} & \cs{\textsf{CS}} & \cs{\textsf{CS}} & \cs{\textsf{CS}} & \cs{\textsf{CS}} & \cs{\textsf{CS}} & \cs{\textsf{CS}} & \cs{\textsf{CS}} & \cs{\textsf{CS}} & \cs{\textsf{CS}} & \cs{\textsf{CS}} & \cs{\textsf{CS}} & \cs{\textsf{CS}} & \cs{\textsf{CS}} \\
\hline
90.0 & \cs{\textsf{CS}} & \cs{\textsf{CS}} & \ac{\textsf{Aa}} & \ac{\textsf{Aa}} & \ac{\textsf{Aa}} & \ac{\textsf{Aa}} & \ac{\textsf{Aa}} & \ac{\textsf{Aa}} & \ac{\textsf{Aa}} & \ac{\textsf{Aa}} & \ac{\textsf{Aa}} & \ac{\textsf{Aa}} & \ac{\textsf{Aa}} & \ac{\textsf{Aa}} & \ac{\textsf{Aa}} & \ac{\textsf{Aa}} & \ac{\textsf{Aa}} & \ac{\textsf{Aa}} & \ac{\textsf{Aa}} & \ac{\textsf{Aa}} \\
\hline
\end{tabular}}
\end{center}
\vspace{10pt}
\begin{center}
    Table \ref{tab:autoware-merging} (d) Autoware autopilot: Verdicts for merging scenarios with $v_e=15.0$
    \footnotesize 
    \adjustbox{max width=\textwidth}{
    \begin{tabular}{|c|c|c|c|c|c|c|c|c|c|c|c|c|c|c|c|c|c|c|c|}
\hline
\diagbox{$\bm{xa}$}{$\bm{xf}$}
& 0.0 & 5.0 & 10.0 & 15.0 & 20.0 & 25.0 & 30.0 & \ct{35.1} & 40.0 & 45.0 & 50.0 & 60.0 & 80.0 & 120.0 & 160.0 & 200.0 & 240.0 & 280.0 & 320.0 \\
\hline
0.0 & - & - & - & - & - & - & - & - & - & - & - & - & \cs{\textsf{CS}} & \cs{\textsf{CS}} & \cs{\textsf{CS}} & \cs{\textsf{CS}} & \cs{\textsf{CS}} & \cs{\textsf{CS}} & \cs{\textsf{CS}} \\
\hline
40.0 & - & - & - & - & - & \cs{\textsf{CS}} & \cs{\textsf{CS}} & \cs{\textsf{CS}} & \cs{\textsf{CS}} & \cs{\textsf{CS}} & \cs{\textsf{CS}} & \cs{\textsf{CS}} & \cs{\textsf{CS}} & \cs{\textsf{CS}} & \cs{\textsf{CS}} & \cs{\textsf{CS}} & \cs{\textsf{CS}} & \cs{\textsf{CS}} & \cs{\textsf{CS}} \\
\hline
80.0 & \cs{\textsf{CS}} & \cs{\textsf{CS}} & \cs{\textsf{CS}} & \cs{\textsf{CS}} & \cs{\textsf{CS}} & \cs{\textsf{CS}} & \cs{\textsf{CS}} & \cs{\textsf{CS}} & \cs{\textsf{CS}} & \cs{\textsf{CS}} & \cs{\textsf{CS}} & \cs{\textsf{CS}} & \cs{\textsf{CS}} & \cs{\textsf{CS}} & \cs{\textsf{CS}} & \cs{\textsf{CS}} & \cs{\textsf{CS}} & \cs{\textsf{CS}} & \cs{\textsf{CS}} \\
\hline
85.0 & \cs{\textsf{CS}} & \cs{\textsf{CS}} & \cs{\textsf{CS}} & \cs{\textsf{CS}} & \cs{\textsf{CS}} & \cs{\textsf{CS}} & \cs{\textsf{CS}} & \cs{\textsf{CS}} & \cs{\textsf{CS}} & \cs{\textsf{CS}} & \cs{\textsf{CS}} & \cs{\textsf{CS}} & \cs{\textsf{CS}} & \cs{\textsf{CS}} & \cs{\textsf{CS}} & \cs{\textsf{CS}} & \cs{\textsf{CS}} & \cs{\textsf{CS}} & \cs{\textsf{CS}} \\
\hline
90.0 & \cs{\textsf{CS}} & \cs{\textsf{CS}} & \ac{\textsf{Aa}} & \ac{\textsf{Aa}} & \ac{\textsf{Aa}} & \ac{\textsf{Aa}} & \ac{\textsf{Aa}} & \ac{\textsf{Aa}} & \ac{\textsf{Aa}} & \ac{\textsf{Aa}} & \ac{\textsf{Aa}} & \ac{\textsf{Aa}} & \ac{\textsf{Aa}} & \ac{\textsf{Aa}} & \ac{\textsf{Aa}} & \ac{\textsf{Aa}} & \ac{\textsf{Aa}} & \ac{\textsf{Aa}} & \ac{\textsf{Aa}} \\
\hline
\end{tabular}}
\end{center}

\vspace{2em}
\paragraph{1) Autoware autopilot: Analysis for merging scenarios}

Verdicts for merging scenarios and different speeds $v_e$ are presented in Tab. \ref{tab:autoware-merging}, revealing accidents in which the ego vehicle hits the arriving vehicle (\textsf{Ae}) and the arriving vehicle hits the ego vehicle (\textsf{Aa}).

For $v_e=0.0$, accidents occur during the transition from caution to progress when {$x_f \ge 10.0 > \hat x_f = 0.0$} and $x_a \ge 205.0 > \hat x_a = 60.3$. In those cases, the ego vehicle can progress safely by applying adequate acceleration. However, in accident scenarios, the ego vehicle is stationary at the beginning and starts to move onto the main road when the arriving vehicle is at the merging point, resulting in a collision on the side of the arriving vehicle.

For $v_e=5.0$, there is no accident in the transition from caution to progress. However, the ego vehicle only progresses when {$x_f \ge 10.0 > \hat x_f = 5.3$ and $x_a > 140.0 > \hat x_a = 80.7$. The difference between $x_a$ and $\hat x_a$ is greater than 59.3, indicating its over-cautiousness.}

{For $v_e=10.0$, accidents occur when $x_f \ge 10.0$ and $90.0 \le x_a \le 100.0$, where the range of $x_f$ includes $\hat x_f=16.9$ and the range of $x_a$ includes $\hat x_a=91.6$. For $\langle x_a, x_f \rangle \ge \langle \hat x_f, \hat x_a \rangle$, the ego vehicle should accelerate sufficiently to progress safely. Otherwise, it should be cautious. However, for $10.0 \le x_f \le 60.0$, the ego vehicle stops at the merging point, leaving insufficient space for the arriving vehicle to brake. Finally, the arriving vehicle hits the ego vehicle's side. For $x_f \ge 65.0$, the ego vehicle decelerates to pass the merging point and is hit in the back by the arriving vehicle.}

For $v_e=15.0$, a wider range of $x_a$ around $\hat x_a=101.9$ results in accidents where the ego vehicle does not apply sufficient acceleration as for $v_e=10.0$ and is hit by the arriving vehicle on its side or on its back.

\vspace{2em}
\begin{minipage}{\linewidth}
\begin{table}[H]
    \centering
    \caption{Autoware autopilot: Verdicts for lane change scenarios}
    \label{tab:autoware-lane_change}
\end{table}
\vspace{-20pt}
\begin{center}
    Table \ref{tab:autoware-lane_change} (a) Autoware autopilot: Verdicts for lane change scenarios with $v_e=5.0$
    \footnotesize 
    \adjustbox{max width=\textwidth}{\begin{tabular}{|c|c|c|c|c|c|c|c|c|c|c|}
\hline
\diagbox{$\bm{xa}$}{$\bm{xf}$}
& 0.0 & \ct{4.8} & 40.0 & 80.0 & 120.0 & 160.0 & 200.0 & 240.0 & 280.0 & 320.0 \\
\hline
0.0 & - & - & - & \cs{\textsf{CS}} & \cs{\textsf{CS}} & \cs{\textsf{CS}} & \cs{\textsf{CS}} & \cs{\textsf{CS}} & \cs{\textsf{CS}} & \cs{\textsf{CS}} \\
\hline
40.0 & - & - & \cs{\textsf{CS}} & \cs{\textsf{CS}} & \cs{\textsf{CS}} & \cs{\textsf{CS}} & \cs{\textsf{CS}} & \cs{\textsf{CS}} & \cs{\textsf{CS}} & \cs{\textsf{CS}} \\
\hline
80.0 & \cs{\textsf{CS}} & \cs{\textsf{CS}} & \cs{\textsf{CS}} & \cs{\textsf{CS}} & \cs{\textsf{CS}} & \cs{\textsf{CS}} & \cs{\textsf{CS}} & \cs{\textsf{CS}} & \cs{\textsf{CS}} & \cs{\textsf{CS}} \\
\hline
120.0 & \cs{\textsf{CS}} & \cs{\textsf{CS}} & \cs{\textsf{CS}} & \cs{\textsf{CS}} & \cs{\textsf{CS}} & \cs{\textsf{CS}} & \cs{\textsf{CS}} & \cs{\textsf{CS}} & \cs{\textsf{CS}} & \cs{\textsf{CS}} \\
\hline
\ct{120.4} & \cs{\textsf{CS}} & \cs{\textsf{CS}} & \cs{\textsf{CS}} & \cs{\textsf{CS}} & \cs{\textsf{CS}} & \cs{\textsf{CS}} & \cs{\textsf{CS}} & \cs{\textsf{CS}} & \cs{\textsf{CS}} & \cs{\textsf{CS}} \\
\hline
160.0 & \cs{\textsf{CS}} & \cs{\textsf{CS}} & \cs{\textsf{CS}} & \cs{\textsf{CS}} & \cs{\textsf{CS}} & \cs{\textsf{CS}} & \cs{\textsf{CS}} & \cs{\textsf{CS}} & \cs{\textsf{CS}} & \cs{\textsf{CS}} \\
\hline
200.0 & \cs{\textsf{CS}} & \cs{\textsf{CS}} & \cs{\textsf{CS}} & \cs{\textsf{CS}} & \cs{\textsf{CS}} & \cs{\textsf{CS}} & \cs{\textsf{CS}} & \cs{\textsf{CS}} & \cs{\textsf{CS}} & \cs{\textsf{CS}} \\
\hline
\end{tabular}}
\end{center}
\end{minipage}
\vspace{10pt}
\begin{center}
    Table \ref{tab:autoware-lane_change} (b) Autoware autopilot: Verdicts for lane change scenarios with $v_e=10.0$
    \footnotesize 
    \adjustbox{max width=\textwidth}{\begin{tabular}{|c|c|c|c|c|c|c|c|c|c|c|}
\hline
\diagbox{$\bm{xa}$}{$\bm{xf}$}
& 0.0 & \ct{14.8} & 40.0 & 80.0 & 120.0 & 160.0 & 200.0 & 240.0 & 280.0 & 320.0 \\
\hline
0.0 & - & - & - & \cs{\textsf{CS}} & \cs{\textsf{CS}} & \cs{\textsf{CS}} & \cs{\textsf{CS}} & \cs{\textsf{CS}} & \cs{\textsf{CS}} & \cs{\textsf{CS}} \\
\hline
40.0 & - & - & \cs{\textsf{CS}} & \cs{\textsf{CS}} & \cs{\textsf{CS}} & \cs{\textsf{CS}} & \cs{\textsf{CS}} & \cs{\textsf{CS}} & \cs{\textsf{CS}} & \cs{\textsf{CS}} \\
\hline
80.0 & \cs{\textsf{CS}} & \cs{\textsf{CS}} & \cs{\textsf{CS}} & \cs{\textsf{CS}} & \cs{\textsf{CS}} & \cs{\textsf{CS}} & \cs{\textsf{CS}} & \cs{\textsf{CS}} & \cs{\textsf{CS}} & \cs{\textsf{CS}} \\
\hline
\ct{90.3} & \cs{\textsf{CS}} & \cs{\textsf{CS}} & \cs{\textsf{CS}} & \cs{\textsf{CS}} & \cs{\textsf{CS}} & \cs{\textsf{CS}} & \cs{\textsf{CS}} & \cs{\textsf{CS}} & \cs{\textsf{CS}} & \cs{\textsf{CS}} \\
\hline
120.0 & \cs{\textsf{CS}} & \cs{\textsf{CS}} & \cs{\textsf{CS}} & \cs{\textsf{CS}} & \cs{\textsf{CS}} & \cs{\textsf{CS}} & \cs{\textsf{CS}} & \cs{\textsf{CS}} & \cs{\textsf{CS}} & \cs{\textsf{CS}} \\
\hline
160.0 & \cs{\textsf{CS}} & \cs{\textsf{CS}} & \cs{\textsf{CS}} & \cs{\textsf{CS}} & \cs{\textsf{CS}} & \cs{\textsf{CS}} & \cs{\textsf{CS}} & \cs{\textsf{CS}} & \cs{\textsf{CS}} & \cs{\textsf{CS}} \\
\hline
200.0 & \cs{\textsf{CS}} & \cs{\textsf{CS}} & \cs{\textsf{CS}} & \cs{\textsf{CS}} & \cs{\textsf{CS}} & \cs{\textsf{CS}} & \cs{\textsf{CS}} & \cs{\textsf{CS}} & \cs{\textsf{CS}} & \cs{\textsf{CS}} \\
\hline
\end{tabular}}
\end{center}
\vspace{10pt}
\begin{center}
    Table \ref{tab:autoware-lane_change} (c) Autoware autopilot: Verdicts for lane change scenarios with $v_e=15.0$
    \footnotesize 
    \adjustbox{max width=\textwidth}{\begin{tabular}{|c|c|c|c|c|c|c|c|c|c|c|}
\hline
\diagbox{$\bm{xa}$}{$\bm{xf}$}
& 0.0 & \ct{29.8} & 40.0 & 80.0 & 120.0 & 160.0 & 200.0 & 240.0 & 280.0 & 320.0 \\
\hline
0.0 & - & - & - & \cs{\textsf{CS}} & \cs{\textsf{CS}} & \cs{\textsf{CS}} & \cs{\textsf{CS}} & \cs{\textsf{CS}} & \cs{\textsf{CS}} & \cs{\textsf{CS}} \\
\hline
40.0 & - & - & \cs{\textsf{CS}} & \cs{\textsf{CS}} & \cs{\textsf{CS}} & \cs{\textsf{CS}} & \cs{\textsf{CS}} & \cs{\textsf{CS}} & \cs{\textsf{CS}} & \cs{\textsf{CS}} \\
\hline
80.0 & \cs{\textsf{CS}} & \cs{\textsf{CS}} & \cs{\textsf{CS}} & \cs{\textsf{CS}} & \cs{\textsf{CS}} & \cs{\textsf{CS}} & \cs{\textsf{CS}} & \cs{\textsf{CS}} & \cs{\textsf{CS}} & \cs{\textsf{CS}} \\
\hline
\ct{80.3} & \cs{\textsf{CS}} & \cs{\textsf{CS}} & \cs{\textsf{CS}} & \cs{\textsf{CS}} & \cs{\textsf{CS}} & \cs{\textsf{CS}} & \cs{\textsf{CS}} & \cs{\textsf{CS}} & \cs{\textsf{CS}} & \cs{\textsf{CS}} \\
\hline
120.0 & \cs{\textsf{CS}} & \cs{\textsf{CS}} & \cs{\textsf{CS}} & \cs{\textsf{CS}} & \cs{\textsf{CS}} & \cs{\textsf{CS}} & \cs{\textsf{CS}} & \cs{\textsf{CS}} & \cs{\textsf{CS}} & \cs{\textsf{CS}} \\
\hline
160.0 & \cs{\textsf{CS}} & \cs{\textsf{CS}} & \cs{\textsf{CS}} & \cs{\textsf{CS}} & \cs{\textsf{CS}} & \cs{\textsf{CS}} & \cs{\textsf{CS}} & \cs{\textsf{CS}} & \cs{\textsf{CS}} & \cs{\textsf{CS}} \\
\hline
200.0 & \cs{\textsf{CS}} & \cs{\textsf{CS}} & \cs{\textsf{CS}} & \cs{\textsf{CS}} & \cs{\textsf{CS}} & \cs{\textsf{CS}} & \cs{\textsf{CS}} & \cs{\textsf{CS}} & \cs{\textsf{CS}} & \cs{\textsf{CS}} \\
\hline
\end{tabular}}
\end{center}
\vspace{10pt}
\begin{center}
    Table \ref{tab:autoware-lane_change} (d) Autoware autopilot: Verdicts for lane change scenarios with $v_e=20.0$
    \footnotesize 
    \adjustbox{max width=\textwidth}{
    \begin{tabular}{|c|c|c|c|c|c|c|c|c|c|c|}
\hline
\diagbox{$\bm{xa}$}{$\bm{xf}$}
& 0.0 & 40.0 & \ct{49.8} & 80.0 & 120.0 & 160.0 & 200.0 & 240.0 & 280.0 & 320.0 \\
\hline
0.0 & - & - & - & \cs{\textsf{CS}} & \cs{\textsf{CS}} & \cs{\textsf{CS}} & \cs{\textsf{CS}} & \cs{\textsf{CS}} & \cs{\textsf{CS}} & \cs{\textsf{CS}} \\
\hline
40.0 & - & \cs{\textsf{CS}} & \cs{\textsf{CS}} & \cs{\textsf{CS}} & \cs{\textsf{CS}} & \cs{\textsf{CS}} & \cs{\textsf{CS}} & \cs{\textsf{CS}} & \cs{\textsf{CS}} & \cs{\textsf{CS}} \\
\hline
\ct{75.3} & \cs{\textsf{CS}} & \cs{\textsf{CS}} & \cs{\textsf{CS}} & \cs{\textsf{CS}} & \cs{\textsf{CS}} & \cs{\textsf{CS}} & \cs{\textsf{CS}} & \cs{\textsf{CS}} & \cs{\textsf{CS}} & \cs{\textsf{CS}} \\
\hline
80.0 & \cs{\textsf{CS}} & \cs{\textsf{CS}} & \cs{\textsf{CS}} & \cs{\textsf{CS}} & \cs{\textsf{CS}} & \cs{\textsf{CS}} & \cs{\textsf{CS}} & \cs{\textsf{CS}} & \cs{\textsf{CS}} & \cs{\textsf{CS}} \\
\hline
120.0 & \cs{\textsf{CS}} & \cs{\textsf{CS}} & \cs{\textsf{CS}} & \cs{\textsf{CS}} & \cs{\textsf{CS}} & \cs{\textsf{CS}} & \cs{\textsf{CS}} & \cs{\textsf{CS}} & \cs{\textsf{CS}} & \cs{\textsf{CS}} \\
\hline
160.0 & \cs{\textsf{CS}} & \cs{\textsf{CS}} & \cs{\textsf{CS}} & \cs{\textsf{CS}} & \cs{\textsf{CS}} & \cs{\textsf{CS}} & \cs{\textsf{CS}} & \cs{\textsf{CS}} & \cs{\textsf{CS}} & \cs{\textsf{CS}} \\
\hline
200.0 & \cs{\textsf{CS}} & \cs{\textsf{CS}} & \cs{\textsf{CS}} & \cs{\textsf{CS}} & \cs{\textsf{CS}} & \cs{\textsf{CS}} & \cs{\textsf{CS}} & \cs{\textsf{CS}} & \cs{\textsf{CS}} & \cs{\textsf{CS}} \\
\hline
\end{tabular}}
\end{center}

\vspace{2em}
\paragraph{2) Autoware autopilot: Analysis for lane change scenarios}

Verdicts for lane change scenarios and different speeds $v_e$ are presented in Tab. \ref{tab:autoware-lane_change}. It is surprising that for all configurations we tested, the ego vehicle is cautious, indicating its over-cautiousness. 

Autoware’s over-cautiousness can be explained by analyzing its lane change policy. We found that the lane change angle is a function of its speed. When the speed is different from 0, the angle is so small that performing a lane change would collide with the front vehicle on the inner lane even if it is located 54.3 meters forward. The lane change angle is larger if it decides to decelerate to 0 speed. However, Autoware considers the safe value of $x_a$, the distance of the arriving vehicle, to be $v_a * 3 + v_a^2 / 2 - v_e^2/2$, where $v_e$ and $v_a$ are speeds of the ego vehicle and the arriving vehicle, respectively. In the test cases, we have $v_a = vl = 22.2$. Consequently, even if the ego vehicle stops, which is abnormal for such a maneuver, the safe distance $x_a$ is calculated at 313.02, which is close to the upper limit of the values we tested.

\clearpage

\vspace{3em}
\begin{minipage}{\linewidth}
\begin{table}[H]
    \centering
    \caption{Autoware autopilot: Verdicts for crossing with yield sign scenarios}
    \label{tab:autoware-crossing}
\end{table}
\vspace{-20pt}
\begin{center}
    Table \ref{tab:autoware-crossing} (a) Autoware autopilot: Verdicts for crossing with yield sign scenarios with $v_e=0.0$
    \footnotesize 
    \adjustbox{max width=\textwidth}{\begin{tabular}{|c|c|c|c|c|c|c|c|c|c|c|c|c|c|}
\hline
\diagbox{$\bm{xa}$}{$\bm{xf}$}
& 0.0 & 5.0 & \ct{7.9} & 10.0 & 20.0 & 40.0 & 80.0 & 120.0 & 160.0 & 200.0 & 240.0 & 280.0 & 320.0 \\
\hline
0.0 & \cs{\textsf{CS}} & \cs{\textsf{CS}} & \cs{\textsf{CS}} & \cs{\textsf{CS}} & \cs{\textsf{CS}} & \cs{\textsf{CS}} & \cs{\textsf{CS}} & \cs{\textsf{CS}} & \cs{\textsf{CS}} & \cs{\textsf{CS}} & \cs{\textsf{CS}} & \cs{\textsf{CS}} & \cs{\textsf{CS}} \\
\hline
40.0 & \cu{\textsf{CU$p_1p_2$}} & \cu{\textsf{CU$p_1$}} & \cu{\textsf{CU$p_1$}} & \cu{\textsf{CU$p_1$}} & \cu{\textsf{CU$p_1$}} & \cu{\textsf{CU$p_1$}} & \cu{\textsf{CU$p_1$}} & \cu{\textsf{CU$p_1$}} & \cu{\textsf{CU$p_1$}} & \cu{\textsf{CU$p_1$}} & \cu{\textsf{CU$p_1$}} & \cu{\textsf{CU$p_1$}} & \cu{\textsf{CU$p_1$}} \\
\hline
80.0 & \cu{\textsf{CU$p_1$}} & \cu{\textsf{CU$p_1$}} & \cu{\textsf{CU$p_1$}} & \cu{\textsf{CU$p_1$}} & \cu{\textsf{CU$p_1$}} & \cu{\textsf{CU$p_1$}} & \cu{\textsf{CU$p_1$}} & \cu{\textsf{CU$p_1$}} & \cu{\textsf{CU$p_1$}} & \cu{\textsf{CU$p_1$}} & \cu{\textsf{CU$p_1$}} & \cu{\textsf{CU$p_1$}} & \cu{\textsf{CU$p_1$}} \\
\hline
120.0 & \cu{\textsf{CU$p_1$}} & \cu{\textsf{CU$p_1$}} & \cu{\textsf{CU$p_1$}} & \cu{\textsf{CU$p_1$}} & \cu{\textsf{CU$p_1$}} & \cu{\textsf{CU$p_1$}} & \cu{\textsf{CU$p_1$}} & \cu{\textsf{CU$p_1$}} & \cu{\textsf{CU$p_1$}} & \cu{\textsf{CU$p_1$}} & \cu{\textsf{CU$p_1$}} & \cu{\textsf{CU$p_1$}} & \cu{\textsf{CU$p_1$}} \\
\hline
160.0 & \cu{\textsf{CU$p_1$}} & \cu{\textsf{CU$p_1$}} & \cu{\textsf{CU$p_1$}} & \cu{\textsf{CU$p_1$}} & \cu{\textsf{CU$p_1$}} & \cu{\textsf{CU$p_1$}} & \cu{\textsf{CU$p_1$}} & \cu{\textsf{CU$p_1$}} & \cu{\textsf{CU$p_1$}} & \cu{\textsf{CU$p_1$}} & \cu{\textsf{CU$p_1$}} & \cu{\textsf{CU$p_1$}} & \cu{\textsf{CU$p_1$}} \\
\hline
\end{tabular}}
\end{center}
\end{minipage}
\vspace{10pt}
\begin{center}
    Table \ref{tab:autoware-crossing} (b) Autoware autopilot: Verdicts crossing with yield sign scenarios with $v_e=5.0$
    \footnotesize 
    \adjustbox{max width=\textwidth}{\begin{tabular}{|c|c|c|c|c|c|c|c|c|c|c|c|c|c|}
\hline
\diagbox{$\bm{xa}$}{$\bm{xf}$}
& 0.0 & 5.0 & 10.0 & \ct{11.7} & 20.0 & 40.0 & 80.0 & 120.0 & 160.0 & 200.0 & 240.0 & 280.0 & 320.0 \\
\hline
0.0 & \cs{\textsf{CS}} & \cs{\textsf{CS}} & \cs{\textsf{CS}} & \cs{\textsf{CS}} & \cs{\textsf{CS}} & \cs{\textsf{CS}} & \cs{\textsf{CS}} & \cs{\textsf{CS}} & \cs{\textsf{CS}} & \cs{\textsf{CS}} & \cs{\textsf{CS}} & \cs{\textsf{CS}} & \cs{\textsf{CS}} \\
\hline
40.0 & \cu{\textsf{CU$p_1p_2$}} & \cu{\textsf{CU$p_1p_2$}} & \cu{\textsf{CU$p_1p_2$}} & \cu{\textsf{CU$p_1p_2$}} & \cu{\textsf{CU$p_1p_2$}} & \cu{\textsf{CU$p_1p_2$}} & \cu{\textsf{CU$p_1p_2$}} & \cu{\textsf{CU$p_1p_2$}} & \cu{\textsf{CU$p_1p_2$}} & \cu{\textsf{CU$p_1p_2$}} & \cu{\textsf{CU$p_1p_2$}} & \cu{\textsf{CU$p_1p_2$}} & \cu{\textsf{CU$p_1$}} \\
\hline
80.0 & \cu{\textsf{CU$p_1p_2$}} & \cu{\textsf{CU$p_1p_2$}} & \cu{\textsf{CU$p_1p_2$}} & \cu{\textsf{CU$p_1p_2$}} & \cu{\textsf{CU$p_1p_2$}} & \cu{\textsf{CU$p_1p_2$}} & \cu{\textsf{CU$p_1p_2$}} & \cu{\textsf{CU$p_1p_2$}} & \cu{\textsf{CU$p_1p_2$}} & \cu{\textsf{CU$p_1p_2$}} & \cu{\textsf{CU$p_1p_2$}} & \cu{\textsf{CU$p_1p_2$}} & \cu{\textsf{CU$p_1p_2$}} \\
\hline
\ct{95.6} & \cu{\textsf{CU$p_1p_2$}} & \cu{\textsf{CU$p_1p_2$}} & \cu{\textsf{CU$p_1p_2$}} & \cu{\textsf{CU$p_1p_2$}} & \cu{\textsf{CU$p_1p_2$}} & \cu{\textsf{CU$p_1p_2$}} & \cu{\textsf{CU$p_1p_2$}} & \cu{\textsf{CU$p_1p_2$}} & \cu{\textsf{CU$p_1p_2$}} & \cu{\textsf{CU$p_1p_2$}} & \cu{\textsf{CU$p_1p_2$}} & \cu{\textsf{CU$p_1p_2$}} & \cu{\textsf{CU$p_1p_2$}} \\
\hline
120.0 & \cu{\textsf{CU$p_1p_2$}} & \cu{\textsf{CU$p_1p_2$}} & \cu{\textsf{CU$p_1p_2$}} & \cu{\textsf{CU$p_1p_2$}} & \cu{\textsf{CU$p_1p_2$}} & \cu{\textsf{CU$p_1p_2$}} & \cu{\textsf{CU$p_1p_2$}} & \cu{\textsf{CU$p_1p_2$}} & \cu{\textsf{CU$p_1p_2$}} & \cu{\textsf{CU$p_1p_2$}} & \cu{\textsf{CU$p_1p_2$}} & \cu{\textsf{CU$p_1p_2$}} & \cu{\textsf{CU$p_1p_2$}} \\
\hline
\end{tabular}}
\end{center}
\vspace{10pt}
\begin{center}
    Table \ref{tab:autoware-crossing} (c) Autoware autopilot: Verdicts for crossing with yield sign scenarios with $v_e=10.0$
    \footnotesize 
    \adjustbox{max width=\textwidth}{\begin{tabular}{|c|c|c|c|c|c|c|c|c|c|c|c|c|c|c|}
\hline
\diagbox{$\bm{xa}$}{$\bm{xf}$}
& 0.0 & 5.0 & 10.0 & 15.0 & 20.0 & \ct{22.7} & 40.0 & 80.0 & 120.0 & 160.0 & 200.0 & 240.0 & 280.0 & 320.0 \\
\hline
0.0 & \cu{\textsf{CU$p_2$}} & \cs{\textsf{CS}} & \cs{\textsf{CS}} & \cs{\textsf{CS}} & \cs{\textsf{CS}} & \cs{\textsf{CS}} & \cs{\textsf{CS}} & \cs{\textsf{CS}} & \cs{\textsf{CS}} & \cs{\textsf{CS}} & \cs{\textsf{CS}} & \cs{\textsf{CS}} & \cs{\textsf{CS}} & \cs{\textsf{CS}} \\
\hline
40.0 & \cu{\textsf{CU$p_1$}} & \cu{\textsf{CU$p_1$}} & \cu{\textsf{CU$p_1$}} & \cu{\textsf{CU$p_1$}} & \cu{\textsf{CU$p_1$}} & \cu{\textsf{CU$p_1$}} & \cu{\textsf{CU$p_1p_2$}} & \cu{\textsf{CU$p_1p_2$}} & \cu{\textsf{CU$p_1p_2$}} & \cu{\textsf{CU$p_1p_2$}} & \cu{\textsf{CU$p_1p_2$}} & \cu{\textsf{CU$p_1p_2$}} & \cu{\textsf{CU$p_1p_2$}} & \cu{\textsf{CU$p_1p_2$}} \\
\hline
\ct{76.5} & \cu{\textsf{CU$p_1$}} & \cu{\textsf{CU$p_1$}} & \cu{\textsf{CU$p_1$}} & \cu{\textsf{CU$p_1$}} & \cu{\textsf{CU$p_1$}} & \cu{\textsf{CU$p_1p_2$}} & \cu{\textsf{CU$p_1$}} & \cu{\textsf{CU$p_1p_2$}} & \cu{\textsf{CU$p_1$}} & \cu{\textsf{CU$p_1$}} & \cu{\textsf{CU$p_1$}} & \cu{\textsf{CU$p_1$}} & \cu{\textsf{CU$p_1$}} & \cu{\textsf{CU$p_1$}} \\
\hline
80.0 & \cu{\textsf{CU$p_1$}} & \cu{\textsf{CU$p_1$}} & \cu{\textsf{CU$p_1p_2$}} & \cu{\textsf{CU$p_1$}} & \cu{\textsf{CU$p_1$}} & \cu{\textsf{CU$p_1$}} & \cu{\textsf{CU$p_1$}} & \cu{\textsf{CU$p_1p_2$}} & \cu{\textsf{CU$p_1$}} & \cu{\textsf{CU$p_1$}} & \cu{\textsf{CU$p_1$}} & \cu{\textsf{CU$p_1$}} & \cu{\textsf{CU$p_1$}} & \cu{\textsf{CU$p_1$}} \\
\hline
100.0 & \cu{\textsf{CU$p_1p_2$}} & \cu{\textsf{CU$p_1p_2$}} & \cu{\textsf{CU$p_1$}} & \cu{\textsf{CU$p_1p_2$}} & \cu{\textsf{CU$p_1p_2$}} & \cu{\textsf{CU$p_1p_2$}} & \cu{\textsf{CU$p_1p_2$}} & \cu{\textsf{CU$p_1p_2$}} & \cu{\textsf{CU$p_1p_2$}} & \cu{\textsf{CU$p_1p_2$}} & \cu{\textsf{CU$p_1p_2$}} & \cu{\textsf{CU$p_1p_2$}} & \cu{\textsf{CU$p_1p_2$}} & \cu{\textsf{CU$p_1p_2$}} \\
\hline
\end{tabular}}
\end{center}
\vspace{10pt}
\begin{center}
    Table \ref{tab:autoware-crossing} (d) Autoware autopilot: Verdicts for crossing with yield sign scenarios with $v_e=15.0$
    \scriptsize 
    \adjustbox{max width=\textwidth}{
    \begin{tabular}{|c|c|c|c|c|c|c|c|c|c|c|c|c|c|}
\hline
\diagbox{$\bm{xa}$}{$\bm{xf}$}
& 0.0 & 5.0 & 10.0 & 15.0 & 20.0 & \ct{40.0} & 80.0 & 120.0 & 160.0 & 200.0 & 240.0 & 280.0 & 320.0 \\
\hline
0.0 & \cu{\textsf{CU$p_2$}} & \cs{\textsf{CS}} & \cs{\textsf{CS}} & \cs{\textsf{CS}} & \cs{\textsf{CS}} & \cs{\textsf{CS}} & \cs{\textsf{CS}} & \cs{\textsf{CS}} & \cs{\textsf{CS}} & \cs{\textsf{CS}} & \cs{\textsf{CS}} & \cs{\textsf{CS}} & \cs{\textsf{CS}} \\
\hline
40.0 & \cu{\textsf{CU$p_2$}} & \cs{\textsf{CS}} & \cs{\textsf{CS}} & \cs{\textsf{CS}} & \cs{\textsf{CS}} & \cs{\textsf{CS}} & \cs{\textsf{CS}} & \cs{\textsf{CS}} & \cs{\textsf{CS}} & \cs{\textsf{CS}} & \cs{\textsf{CS}} & \cs{\textsf{CS}} & \cs{\textsf{CS}} \\
\hline
\ct{73.8} & \cu{\textsf{CU$p_1p_2$}} & \cu{\textsf{CU$p_1p_2$}} & \cu{\textsf{CU$p_1p_2$}} & \cu{\textsf{CU$p_1p_2$}} & \cu{\textsf{CU$p_1p_2$}} & \cu{\textsf{CU$p_1p_2$}} & \cu{\textsf{CU$p_1p_2$}} & \cu{\textsf{CU$p_1p_2$}} & \cu{\textsf{CU$p_1p_2$}} & \cu{\textsf{CU$p_1p_2$}} & \cu{\textsf{CU$p_1p_2$}} & \cu{\textsf{CU$p_1p_2$}} & \cu{\textsf{CU$p_1p_2$}} \\
\hline
80.0 & \cu{\textsf{CU$p_1p_2$}} & \cu{\textsf{CU$p_1p_2$}} & \cu{\textsf{CU$p_1p_2$}} & \cu{\textsf{CU$p_1p_2$}} & \cu{\textsf{CU$p_1p_2$}} & \cu{\textsf{CU$p_1p_2$}} & \cu{\textsf{CU$p_1p_2$}} & \cu{\textsf{CU$p_1p_2$}} & \cu{\textsf{CU$p_1p_2$}} & \cu{\textsf{CU$p_1p_2$}} & \cu{\textsf{CU$p_1p_2$}} & \cu{\textsf{CU$p_1p_2$}} & \cu{\textsf{CU$p_1p_2$}} \\
\hline
85.0 & \cu{\textsf{CU$p_1p_2$}} & \cu{\textsf{CU$p_1p_2$}} & \cu{\textsf{CU$p_1p_2$}} & \cu{\textsf{CU$p_1p_2$}} & \cu{\textsf{CU$p_1p_2$}} & \cu{\textsf{CU$p_1p_2$}} & \cu{\textsf{CU$p_1p_2$}} & \cu{\textsf{CU$p_1p_2$}} & \cu{\textsf{CU$p_1p_2$}} & \cu{\textsf{CU$p_1p_2$}} & \cu{\textsf{CU$p_1p_2$}} & \cu{\textsf{CU$p_1p_2$}} & \cu{\textsf{CU$p_1p_2$}} \\
\hline
\end{tabular}}
\end{center}

\vspace{2em}
\paragraph{3) Autoware autopilot: Analysis for crossing with yield sign scenarios}

Verdicts for crossing with yield sign scenarios and different speeds $v_e$ are presented in Tab. \ref{tab:autoware-crossing}, revealing accidents in which the arriving vehicle hits the ego vehicle (\textsf{Aa}) as well as violations of $p_1$ (Two vehicles must not be in the critical zone at the same time) and $p_2$ (The ego vehicle must not stop inside the critical zone).

For $v_e=0.0$, the ego vehicle can only be safely cautious when $x_a = 0.0$. All values of $x_f$ in a wide range of $x_a$ values around $\hat x_a = 165.0$ lead to unsafe caution with property violations. The ego vehicle stops or drives at low speed inside the critical zone to wait for the arriving vehicle to pass, which violates $p_1$ (Two vehicles must not be in the critical zone at the same time) and $p_2$ (The ego vehicle must not stop inside the critical zone). For $x_f \le 10.0$ and $x_a \ge 220.0$ during the transition from caution to progress, the ego vehicle passes the conflict point first but does not leave the critical zone before the arriving vehicle enters it, violating $p_1$. Additionally, for $\langle x_f, x_a \rangle = \langle 0.0, 320.0 \rangle$, the ego vehicle stops inside the critical zone near the exit, which violates $p_2$ (the ego vehicle must not stop inside the critical zone).

For $v_e = 5.0$, $v_e=10.0$, and $v_e=15.0$, we also observe unsafe caution and progress as for $v_e=0.0$. Moreover, for $v_e=10.0$ and $v_e=15.0$, accidents, in which the ego vehicle is hit by the arriving vehicle (\textsf{Aa}), occur during the transition from caution to progress when $x_a > \hat x_a$ but $x_f < \hat x_f$. In accident scenarios, the ego vehicle decides to progress at the beginning but decides to brake after a period of acceleration and finally stops at the conflict point, leaving insufficient space for the arriving vehicle to brake.

\begin{minipage}{\linewidth}
\begin{table}[H]
    \caption{Autoware autopilot: Verdicts for crossing with traffic light scenarios}
    \label{tab:autoware-traffic_light}
\end{table}
\vspace{-20pt}
    \begin{center}
    Table \ref{tab:autoware-traffic_light} (a) Autoware autopilot: Verdicts for crossing with traffic light scenarios with $v_e=0.0$
    \small 
    {\begin{tabular}{|c|c|c|c|c|c|c|c|c|c|}
\hline
\multicolumn{9}{|c|}{$\bm{xf}$}\\
\hline
0.0 & 40.0 & 80.0 & 120.0 & 160.0 & 200.0 & 240.0 & 280.0 & 320.0 \\
\hline
\cs{\textsf{CS}} & \cs{\textsf{CS}} & \cs{\textsf{CS}} & \cs{\textsf{CS}} & \cs{\textsf{CS}} & \cs{\textsf{CS}} & \cs{\textsf{CS}} & \cs{\textsf{CS}} & \cs{\textsf{CS}} \\
\hline
\end{tabular}
}\end{center}
\end{minipage}
\vspace{10pt}
    \begin{center}
    Table \ref{tab:autoware-traffic_light} (b) Autoware autopilot: Verdicts for crossing with traffic light scenarios with $v_e=5.0$
    \small {
    \begin{tabular}{|c|c|c|c|c|c|c|c|c|c|c|c|c|c|}
\hline
\multicolumn{13}{|c|}{$\bm{xf}$}\\
\hline
0.0 & \ct{11.7} & 40.0 & 45.0 & 50.0 & 60.0 & 80.0 & 120.0 & 160.0 & 200.0 & 240.0 & 280.0 & 320.0 \\
\hline
\pu{\textsf{PU$p_4$}} & \pu{\textsf{PU$p_4$}} & \pu{\textsf{PU$p_4$}} & \pu{\textsf{PU$p_4$}} & \ps{\textsf{PS}} & \ps{\textsf{PS}} & \ps{\textsf{PS}} & \ps{\textsf{PS}} & \ps{\textsf{PS}} & \ps{\textsf{PS}} & \ps{\textsf{PS}} & \ps{\textsf{PS}} & \ps{\textsf{PS}} \\
\hline
\end{tabular}}\end{center}
\vspace{10pt}
    \begin{center}
    Table \ref{tab:autoware-traffic_light} (c) Autoware autopilot: Verdicts for crossing with traffic light scenarios with $v_e=10.0$
    \small 
    {\begin{tabular}{|c|c|c|c|c|c|c|c|c|c|c|c|c|}
\hline
\multicolumn{13}{|c|}{$\bm{xf}$}\\
\hline
 0.0 & 10.0 & 15.0 & 20.0 & \ct{22.7} & 40.0 & 80.0 & 120.0 & 160.0 & 200.0 & 240.0 & 280.0 & 320.0 \\
\hline
  \pu{\textsf{PU$p_4$}} & \pu{\textsf{PU$p_4$}} & \ps{\textsf{PS}} & \ps{\textsf{PS}} & \ps{\textsf{PS}} & \ps{\textsf{PS}} & \ps{\textsf{PS}} & \ps{\textsf{PS}} & \ps{\textsf{PS}} & \ps{\textsf{PS}} & \ps{\textsf{PS}} & \ps{\textsf{PS}} & \ps{\textsf{PS}} \\
\hline
\end{tabular}}\end{center}
\vspace{10pt}
    \begin{center}
    Table \ref{tab:autoware-traffic_light} (d) Autoware autopilot: Verdicts for crossing with traffic light scenarios with $v_e=15.0$
    \small 
    {\begin{tabular}{|c|c|c|c|c|c|c|c|c|c|c|c|}
\hline
\multicolumn{12}{|c|}{$\bm{xf}$}\\
\hline
0.0 & 5.0 & 10.0 & 20.0 & \ct{40.0} & 80.0 & 120.0 & 160.0 & 200.0 & 240.0 & 280.0 & 320.0 \\
\hline
\pu{\textsf{PU$p_4$}} & \pu{\textsf{PU$p_4$}} & \ps{\textsf{PS}} & \ps{\textsf{PS}} & \ps{\textsf{PS}} & \ps{\textsf{PS}} & \ps{\textsf{PS}} & \ps{\textsf{PS}} & \ps{\textsf{PS}} & \ps{\textsf{PS}} & \ps{\textsf{PS}} & \ps{\textsf{PS}} \\
\hline
\end{tabular}}\end{center}
\vspace{10pt}
    \begin{center}
    Table \ref{tab:autoware-traffic_light} (e) Autoware autopilot: Verdicts for crossing with traffic light scenarios with $v_e=20.0$
    \small 
    {\begin{tabular}{|c|c|c|c|c|c|c|c|c|c|c|c|c|}
\hline
\multicolumn{13}{|c|}{$\bm{xf}$}\\
\hline
0.0 & 5.0 & 10.0 & 20.0 & 40.0 & \ct{60.3} & 80.0 & 120.0 & 160.0 & 200.0 & 240.0 & 280.0 & 320.0 \\
\hline
\pu{\textsf{PU$p_3p_4$}} & \pu{\textsf{PU$p_3p_4$}} & \ps{PS} & \ps{PS}& \ps{PS}& \ps{PS}& \ps{PS}& \ps{PS}& \ps{PS}& \ps{PS}& \ps{PS}& \ps{PS}& \ps{PS} \\
\hline
\end{tabular}
}\end{center}

\vspace{2em}
\paragraph{4) Autoware autopilot: Analysis for crossing with traffic light scenarios}

{
Verdicts for crossing with traffic light scenarios and different speeds $v_e$ are presented in Tab. \ref{tab:autoware-traffic_light}, revealing violations of $p_3$ (The ego vehicle must not enter the critical zone when the light is red) and $p_4$ (The ego vehicle must not be in the critical zone when the side light is green).

For $v_e=0.0$, and for all $x_f$, there is theoretically no safe progress policy as the ego vehicle cannot cross the critical zone before the side light turns green. The ego vehicle succeeds in being safely cautious.

For $v_e=5.0$, the ego vehicle can theoretically make safe progress when $x_f \ge \hat x_f=11.7$. However, for $x_f \le 45.0$, it fails to apply the maximal acceleration, and violates $p_4$ (The ego vehicle must not be in the critical zone when the side light is green).

For $v_e=10.0$, the ego vehicle can progress safely by applying sufficient acceleration when $x_f\ge \hat x_f=22.7$. However, it may decide to progress when $x_f \le 10.0$, but it fails to reach a sufficiently high speed due to the presence of the front vehicle. Consequently, it fails to exit the critical zone within the 2-second time limit imposed by the all-red phase, violating $p_4$ (The ego vehicle must not be in the critical zone when the side light is green).

For $v_e=15.0$, as for $v_e=10.0$, the ego vehicle decides to progress when $x_f \le 5.0 < \hat x_f=40.0$, and fails to exit the critical zone before the side light turns green, violating $p_4$.

For $v_e = 20.0$, the ego vehicle also decides to progress when $x_f \le 5.0 < \hat x_f = 40.0$. As a result, it fails both to enter the critical zone within the 3-second time limit imposed by the yellow light, violating $p_3$, and to exit the critical zone within the 2-second time limit imposed by the all-red phase, violating $p_4$.


\clearpage
\subsection{Testing the Carla Autopilot}

\subsubsection{Carla autopilot: Estimating the A/D functions}

For the Carla autopilot, there is no specification of the A/D functions. We estimate these functions experimentally and provide their values for the arguments used in our experiments. We estimate these functions experimentally and provide their values for the arguments used in our experiments. The functions are monotonic. However, they are not realistic because the average deceleration rates calculated from Tab. \ref{tab:carla-braking} are all greater than 7 m/s$^2$ and reach 15.6 m/s$^2$ for $v = 5.0$ m/s. Furthermore, according to Tab. \ref{tab:carla-av}, the average acceleration rate can reach 4.7 m/s$^2$.

\begin{table}[H]
    \centering
    \caption{$B(v)$ for Carla autopilot}
    \label{tab:carla-braking}
    \vspace{-10pt}
    \begin{tabular}{|c|c|c|c|c|c|}
    \hline
        $\bm{v}$ & 0.0 & 5.0 & 10.0 & 15.0 & 20.0 \\
    \hline
        $\bm{B(v)}$ & 0.0 & 0.8 & 6.8 & 15.8 & 26.0 \\
    \hline
    \end{tabular}
\end{table}

\begin{table}[H]
    \centering
    \caption{$AV(v, x)$, $AT(v, x)$ for Carla autopilot}
    \vspace{-10pt}
    \label{tab:carla-av}
        \begin{tabular}{|c|c|c|c|c|c|c|c|}
        \hline
        \diagbox[width=3em,height=2em]{~$\bm v$}{$\bm x$~} 
             & 0.0 & 10.0 & 20.0 & 30.0 & 40.0 & 50.0 & 60.0 \\
             \hline
             0.0 &   0.0, 0.0& 10.3, 2.2& 14.2, 3.0& 16.9, 3.7& 19.1, 4.2& 21.0, 4.7& 22.2, 5.2\\
             \hline
             5.0 &   5.0, 0.0& 11.0, 1.4& 14.7, 2.2& 17.2, 2.8& 19.4, 3.4& 21.3, 3.9& 22.2, 4.3\\
             \hline
             10.0 &  10.0, 0.0& 13.6, 1.0&  16.4, 1.6& 18.9, 2.2& 20.8, 2.7& 22.2, 3.2& 22.2, 3.6\\
             \hline
             15.0 &  15.0, 0.0& 17.0, 0.7& 19.3, 1.3& 21.1, 1.8& 22.2, 2.2& 22.2, 2.7& 22.2, 3.1\\
             \hline
        \end{tabular}
\end{table}

\subsubsection{Carla autopilot: Results and analysis}

The test results for the Carla autopilot are shown in Tab. \ref{tab:carla-merging} to Tab. \ref{tab:carla-traffic_light} where the critical values $\hat{x}_a$ and $\hat{x}_f$ are indicated in red.

\begin{table}[H]
    \centering
    \caption{Carla autopilot: Verdicts for merging scenarios}
    \label{tab:carla-merging}
\end{table}
\vspace{-20pt}
\begin{center}
    Table \ref{tab:carla-merging} (a) Carla autopilot: Verdicts for merging scenarios with $v_e=0.0$ \\
    \scriptsize 
    \adjustbox{max width=\textwidth}{\begin{tabular}{|c|c|c|c|c|c|c|c|c|c|c|c|c|}
\hline
\diagbox{$\bm{xa}$}{$\bm{xf}$}
& \ct{0.0} & 5.0 & 10.0 & 20.0 & 40.0 & 80.0 & 120.0 & 160.0 & 200.0 & 240.0 & 280.0 & 320.0 \\
\hline
0.0 & - & - & - & - & \cs{\textsf{CS}} & \cs{\textsf{CS}} & \cs{\textsf{CS}} & \cs{\textsf{CS}} & \cs{\textsf{CS}} & \cs{\textsf{CS}} & \cs{\textsf{CS}} & \cs{\textsf{CS}} \\
\hline
10.0 & - & - & - & - & \cs{\textsf{CS}} & \cs{\textsf{CS}} & \cs{\textsf{CS}} & \cs{\textsf{CS}} & \cs{\textsf{CS}} & \cs{\textsf{CS}} & \cs{\textsf{CS}} & \cs{\textsf{CS}} \\
\hline
12.5 & - & - & - & \cs{\textsf{CS}} & \cs{\textsf{CS}} & \cs{\textsf{CS}} & \cs{\textsf{CS}} & \cs{\textsf{CS}} & \cs{\textsf{CS}} & \cs{\textsf{CS}} & \cs{\textsf{CS}} & \cs{\textsf{CS}} \\
\hline
13.1 & - & - & - & \cs{\textsf{CS}} & \cs{\textsf{CS}} & \cs{\textsf{CS}} & \cs{\textsf{CS}} & \cs{\textsf{CS}} & \cs{\textsf{CS}} & \cs{\textsf{CS}} & \cs{\textsf{CS}} & \cs{\textsf{CS}} \\
\hline
13.8 & - & - & - & \ac{\textsf{Ae}} & \cs{\textsf{CS}} & \cs{\textsf{CS}} & \cs{\textsf{CS}} & \cs{\textsf{CS}} & \cs{\textsf{CS}} & \cs{\textsf{CS}} & \cs{\textsf{CS}} & \cs{\textsf{CS}} \\
\hline
14.4 & - & - & - & \ac{\textsf{Ae}} & \ac{\textsf{Ae}} & \ac{\textsf{Ae}} & \ac{\textsf{Ae}} & \ac{\textsf{Ae}} & \ac{\textsf{Ae}} & \ac{\textsf{Ae}} & \ac{\textsf{Ae}} & \ac{\textsf{Ae}} \\
\hline
15.0 & - & - & - & \ac{\textsf{Ae}} & \ac{\textsf{Ae}} & \ac{\textsf{Ae}} & \ac{\textsf{Ae}} & \ac{\textsf{Ae}} & \ac{\textsf{Ae}} & \ac{\textsf{Ae}} & \ac{\textsf{Ae}} & \ac{\textsf{Ae}} \\
\hline
\end{tabular}}
\end{center}
\vspace{10pt}
\begin{center}
    Table \ref{tab:carla-merging} (b) Carla autopilot: Verdicts for merging scenarios with $v_e=5.0$
    \footnotesize 
    \adjustbox{max width=\textwidth}{\input{latex-table/carla-merging-5.0}}
\end{center}
\vspace{10pt}
\begin{center}
    Table \ref{tab:carla-merging} (c) Carla autopilot: Verdicts for merging scenarios with $v_e=10.0$
    \footnotesize 
    \adjustbox{max width=\textwidth}{\input{latex-table/carla-merging-10.0}}
\end{center}
\vspace{10pt}
\begin{center}
    Table \ref{tab:carla-merging} (d) Carla autopilot: Verdicts for merging scenarios with $v_e=15.0$ 
    \footnotesize 
    \adjustbox{max width=\textwidth}{
    \input{latex-table/carla-merging-15.0}}
\end{center}

\vspace{2em}
\paragraph{1) Carla autopilot: Analysis for merging scenarios}

Verdicts for merging scenarios and different speeds $v_e$ are presented in Tab.~\ref{tab:carla-merging}, revealing accidents in which the ego vehicle hits the arriving vehicle (\textsf{Ae}) and the arriving vehicle hits the ego vehicle (\textsf{Aa}).

For $v_e=0.0$ and $v_e=5.0$, accidents occur during the transition from caution to progression for $x_a < \hat{x}_a$, showing the autopilot's inability to estimate the danger induced by the arriving vehicle.

For $v_e=10.0$ and $v_e=15.0$, the autopilot proceeds safely at moderate speed when the arriving vehicle cautiously decelerates to avoid collision. 

\clearpage
\begin{minipage}{\linewidth}
\begin{table}[H]
    \centering
    \caption{Carla autopilot: Verdicts for lane change scenarios}
    \label{tab:carla-lane_change}
\end{table}
\vspace{-20pt}
\begin{center}
    Table \ref{tab:carla-lane_change} (a) Carla autopilot: Verdicts for lane change scenarios with $v_e=5.0$
    \scriptsize 
    \adjustbox{max width=\textwidth}{\input{latex-table/carla-lane_change-5.0}}
\end{center}
\end{minipage}
\vspace{10pt}
\begin{center}
    Table \ref{tab:carla-lane_change} (b) Carla autopilot: Verdicts for lane change scenarios with $v_e=10.0$
    \scriptsize 
    \adjustbox{max width=\textwidth}{\input{latex-table/carla-lane_change-10.0}}
\end{center}
\vspace{10pt}
\newpage
\begin{center}
    Table \ref{tab:carla-lane_change} (c) Carla autopilot: Verdicts for lane change scenarios with $v_e=15.0$\\
    \footnotesize 
    \adjustbox{max width=\textwidth}{
    \input{latex-table/carla-lane_change-15.0}}
\end{center}
\vspace{10pt}
\begin{center}
    Table \ref{tab:carla-lane_change} (d) Carla autopilot: Verdicts for lane change scenarios with $v_e=20.0$\\
    \footnotesize 
    \adjustbox{max width=\textwidth}{\input{latex-table/carla-lane_change-20.0}}
\end{center}
\vspace{2em}
\paragraph{2) Carla autopilot: Analysis for lane change scenarios}

Verdicts for lane change scenarios and different speeds $v_e$ are presented in Tab.~\ref{tab:carla-lane_change}, revealing accidents in which the ego vehicle hits the arriving vehicle (\textsf{Ae}) and the arriving vehicle hits the ego vehicle (\textsf{Aa}), as well as over-caution of the ego vehicle.

For $v_e=5.0$, accidents occur during the transition from caution to progress when $x_a$ is around 20.0, which is less than $\hat x_a = 31.5$, showing the autopilot's inability to estimate the danger induced by the arriving vehicle.

For $v_e=10.0$, accidents also happen during the transition from caution to progress around $x_a=15.0$, which is less than  $\hat x_a=37.1$. This also indicates the bad decision made by the ego vehicle.

For $v_e=15.0$ and $v_e=20.0$, the ego vehicle does not change lanes, regardless of the configuration, indicating that it is over-cautious. It also goes against common sense that a vehicle confronted with a stationary obstacle at high speed would not change lanes to make progress. 

\clearpage
\begin{minipage}{\linewidth}
\begin{table}[H]
    \centering
    \caption{Carla autopilot: Verdicts for crossing with yield sign scenarios}
    \label{tab:carla-crossing}
\end{table}
\vspace{-20pt}
\begin{center}
    Table \ref{tab:carla-crossing} (a) Carla autopilot: Verdicts for crossing with yield sign scenarios with $v_e=0.0$
    \scriptsize 
    \adjustbox{max width=\textwidth}{\input{latex-table/carla-crossing-0.0}}
\end{center}
\end{minipage}
\vspace{10pt}
\begin{center}
    Table \ref{tab:carla-crossing} (b) Carla autopilot: Verdicts for crossing with yield sign scenarios with $v_e=5.0$
    \footnotesize 
    \adjustbox{max width=\textwidth}{\input{latex-table/carla-crossing-5.0}}
\end{center}
\vspace{10pt}
\begin{center}
    Table \ref{tab:carla-crossing} (c) Carla autopilot: Verdicts for crossing with yield sign scenarios with $v_e=10.0$
    \footnotesize 
    \adjustbox{max width=\textwidth}{\input{latex-table/carla-crossing-10.0}}
\end{center}
\vspace{10pt}
\begin{center}
    Table \ref{tab:carla-crossing} (d) Carla autopilot: Verdicts for crossing with yield sign scenarios with $v_e=15.0$
    \footnotesize 
    \adjustbox{max width=\textwidth}{
    \input{latex-table/carla-crossing-15.0}}
\end{center}

\vspace{2em}
\paragraph{3) Carla autopilot: Analysis for crossing with yield sign scenarios}

Verdicts for crossing with yield sign scenarios and different speeds $v_e$ are shown in Tab.~\ref{tab:carla-crossing}, revealing accidents and violations of $p_1$ (Two vehicles must not be in the critical zone at the same time) and $p_2$ (The ego vehicle must not stop inside the critical zone).

For all four values of $v_e$, the ego autopilot is safely cautious for $x_f = 0.0$, and dangerously cautious for $x_f \ge 5.0$ even if $x_a=0.0$. {Moreover, for $x_f \ge 5.0$, the transition occurs only if $x_a < \hat x_a$. Accidents occur during the transition for $v_e = 0.0$ and $v_e=5.0$.} This indicates that the autopilot cannot properly manage the risk induced by the arriving vehicle.

\clearpage
\begin{table}[H]
    \centering
    \caption{Carla autopilot: Verdicts for crossing with traffic light scenarios}
    \label{tab:carla-traffic_light}
\end{table}
\vspace{-20pt}
\begin{center}
    Table \ref{tab:carla-traffic_light} (a) Carla autopilot: Verdicts for crossing with traffic light scenarios with $v_e=0.0$
    \small 
    {\begin{tabular}{|c|c|c|c|c|c|c|c|c|c|c|c|c|}
\hline
\multicolumn{13}{|c|}{$\bm{xf}$}\\
\hline
0.0 & 5.0 & 10.0 & \ct{16.3} & 20.0 & 40.0 & 80.0 & 120.0 & 160.0 & 200.0 & 240.0 & 280.0 & 320.0 \\
\hline
\cs{\textsf{CS}} & \ps{\textsf{PS}} & \ps{\textsf{PS}} & \ps{\textsf{PS}} & \ps{\textsf{PS}} & \ps{\textsf{PS}} & \ps{\textsf{PS}} & \ps{\textsf{PS}} & \ps{\textsf{PS}} & \ps{\textsf{PS}} & \ps{\textsf{PS}} & \ps{\textsf{PS}} & \ps{\textsf{PS}} \\
\hline
\end{tabular}}
\end{center}
\vspace{10pt}
\begin{center}
    Table \ref{tab:carla-traffic_light} (b) Carla autopilot: Verdicts for crossing with traffic light scenarios with $v_e=5.0$
    \small 
    {\begin{tabular}{|c|c|c|c|c|c|c|c|c|c|c|c|c|}
\hline
\multicolumn{13}{|c|}{$\bm{xf}$}\\
\hline
0.0 & 5.0 & 10.0 & \ct{17.6} & 20.0 & 40.0 & 80.0 & 120.0 & 160.0 & 200.0 & 240.0 & 280.0 & 320.0 \\
\hline
\cs{\textsf{CS}} & \ps{\textsf{PS}} & \ps{\textsf{PS}} & \ps{\textsf{PS}} & \ps{\textsf{PS}} & \ps{\textsf{PS}} & \ps{\textsf{PS}} & \ps{\textsf{PS}} & \ps{\textsf{PS}} & \ps{\textsf{PS}} & \ps{\textsf{PS}} & \ps{\textsf{PS}} & \ps{\textsf{PS}} \\
\hline
\end{tabular}}
\end{center}
\vspace{10pt}
\begin{center}
    Table \ref{tab:carla-traffic_light} (c) Carla autopilot: Verdicts for crossing with traffic light scenarios with $v_e=10.0$
    \small 
    {\begin{tabular}{|c|c|c|c|c|c|c|c|c|c|}
\hline
\multicolumn{10}{|c|}{$\bm{xf}$}\\
\hline
0.0 & \ct{24.1} & 40.0 & 80.0 & 120.0 & 160.0 & 200.0 & 240.0 & 280.0 & 320.0 \\
\hline
\cs{\textsf{CS}} & \cs{\textsf{CS}} & \cs{\textsf{CS}} & \cs{\textsf{CS}} & \cs{\textsf{CS}} & \cs{\textsf{CS}} & \cs{\textsf{CS}} & \cs{\textsf{CS}} & \cs{\textsf{CS}} & \cs{\textsf{CS}} \\
\hline
\end{tabular}
}
\end{center}
\vspace{10pt}
\begin{center}
    Table \ref{tab:carla-traffic_light} (d) Carla autopilot: Verdicts for crossing with traffic light scenarios with $v_e=15.0$
    \small 
    {\begin{tabular}{|c|c|c|c|c|c|c|c|c|c|}
\hline
\multicolumn{10}{|c|}{$\bm{xf}$}\\
\hline
0.0 & \ct{31.5} & 40.0 & 80.0 & 120.0 & 160.0 & 200.0 & 240.0 & 280.0 & 320.0 \\
\hline
\cs{\textsf{CS}} & \cs{\textsf{CS}} & \cs{\textsf{CS}} & \cs{\textsf{CS}} & \cs{\textsf{CS}} & \cs{\textsf{CS}} & \cs{\textsf{CS}} & \cs{\textsf{CS}} & \cs{\textsf{CS}} & \cs{\textsf{CS}} \\
\hline
\end{tabular}
}
\end{center}
\vspace{10pt}
\begin{center}
    Table \ref{tab:carla-traffic_light} (e) Carla autopilot: Verdicts for crossing with traffic light scenarios with $v_e=20.0$
    \small 
    {\begin{tabular}{|c|c|c|c|c|c|c|c|c|c|}
\hline
\multicolumn{10}{|c|}{$\bm{xf}$}\\
\hline
0.0 & \ct{31.5} & 40.0 & 80.0 & 120.0 & 160.0 & 200.0 & 240.0 & 280.0 & 320.0 \\
\hline
\cs{\textsf{CS}} & \cs{\textsf{CS}} & \cs{\textsf{CS}} & \cs{\textsf{CS}} & \cs{\textsf{CS}} & \cs{\textsf{CS}} & \cs{\textsf{CS}} & \cs{\textsf{CS}} & \cs{\textsf{CS}} & \cs{\textsf{CS}} \\
\hline
\end{tabular}
}
\end{center}

\vspace{2em}
\paragraph{4) Carla autopilot: Analysis for crossing with traffic light scenarios}

Verdicts for crossing with traffic light scenarios and different speeds $v_e$ are presented in Tab.~\ref{tab:carla-traffic_light}, where no safety problems were found.  However, the autopilot adopts a very cautious policy.

For $ v_e=0.0$ and $ v_e=5.0$, the ego vehicle has $ x_e=0.0$ and $ x_e=0.8$, respectively. As it is close to the critical zone, it has enough time to progress without violating the time limit given by the yellow-light and all-red phases. 

For $v_e=10.0$, $v_e=15.0$, and $v_e=20.0$, the ego vehicle is always cautious within the range of $x_f$ values tested,  even though there are theoretically feasible progress policies.

\vspace{3em}
\subsection{Testing the LGSVL Autopilot}

\subsubsection{LGSVL autopilot: Estimating the A/D functions}

LGSVL has speed-updating policies specified by differential equations. 
The equations are $dv / dt = -4v$ for deceleration and $dv / dt = (3* \min(4, t_a) / 5 + 1) * (vl - v)$ for acceleration, where $t_a$ is the time elapsed since the start of acceleration, and $vl$ is the speed limit defined by the context. By solving the differential equations, we can get $a(v, T, t)$ and $b(v, T, t)$ to obtain the A/D functions. 

However, these functions are not realistic.  As shown in Tab.~\ref{tab:lgsvl-braking}, reducing speed from 20.0 m/s to 0.0 m/s gives an average deceleration rate of 40 m/s$^2$. The acceleration functions shown in Tab.~\ref{tab:lgsvl-av} also indicate that the vehicle can accelerate from 0.0 to 16.6 m/s within 1.1 seconds, giving the average acceleration of 15.1 m/s$^2$.

In addition, we found violations of monotonicity of the speed acceleration function. As shown by the values marked in bold in Tab.~\ref{tab:lgsvl-av}, the speed reached after acceleration can decrease as the initial speed increases.

\begin{table}[H]
    \centering
    \caption{$B(v)$ for LGSVL autopilot}
    \label{tab:lgsvl-braking}
    \vspace{-10pt}
    \begin{tabular}{|c|c|c|c|c|c|}
    \hline
        $\bm v$ & 0.0 & 5.0 & 10.0 & 15.0 & 20.0 \\
    \hline
        $\bm{B(v)}$ & 0.0 & 1.3 & 2.5 & 3.8 & 5.0 \\
    \hline
    \end{tabular}
\end{table}

\begin{table}[H]
    \centering
    \caption{$AV(v, x)$ for LGSVL autopilot}
    \label{tab:lgsvl-av}
    \vspace{-10pt}
        \begin{tabular}{|c|c|c|c|c|c|c|c|}
        \hline
        \diagbox[width=3em,height=2em]{~$\bm v$}{$\bm x$~} 
             & 0.0 & 10.0 & 20.0 & 30.0 & 40.0 & 50.0 & 60.0 \\
        \hline
        0.0 & 0.0, 0.0& \textbf{16.6}, 1.1& \textbf{20.1}, 1.6& 21.4, 2.1& 22.0, 2.5& 22.1, 3.0& 22.2, 3.4\\
        \hline
        5.0 & 5.0, 0.0& \textbf{16.5}, 0.9& \textbf{19.9}, 1.4& \textbf{21.4}, 1.9& 21.9, 2.4& 22.1, 2.8& 22.2, 3.3\\
        \hline
        10.0 & 10.0, 0.0& 17.1, 0.7& 20.0, 1.3& \textbf{21.3}, 1.7& 21.9, 2.2& 22.1, 2.7& 22.2, 3.1\\
        \hline
        15.0 & 15.0, 0.0& 18.6, 0.6& 20.5, 1.1& 21.5, 1.6& 21.9, 2.0& 22.1, 2.5& 22.2, 2.9\\
        \hline
        \end{tabular}
\end{table}

\subsubsection{LGSVL autopilot: Results and analysis}

The test results for the LGSVL autopilot are shown in Tab. \ref{tab:lgsvl-merging} to Tab. \ref{tab:lgsvl-traffic_light} where the critical values $\hat{x}_a$ and $\hat{x}_f$ are indicated in red.

\vspace{2em}

\vspace{-5pt}
\begin{minipage}{\linewidth}
\begin{table}[H]
    \centering
    \caption{LGSVL autopilot: Verdicts for merging scenarios}
    \label{tab:lgsvl-merging}
\end{table}
\vspace{-20pt}
\begin{center}
    Table \ref{tab:lgsvl-merging} (a) LGSVL autopilot: Verdicts for merging scenarios with $v_e=0.0$
    \scriptsize 
    \adjustbox{max width=\textwidth}{\input{latex-table/lgsvl-merging-0.0}}
\end{center}
\end{minipage}
\vspace{10pt}
\begin{center}
    Table \ref{tab:lgsvl-merging} (b) LGSVL autopilot: Verdicts for merging scenarios with $v_e=5.0$
    \footnotesize 
    \adjustbox{max width=\textwidth}{\input{latex-table/lgsvl-merging-5.0}}
\end{center}
\vspace{10pt}
\begin{center}
    Table \ref{tab:lgsvl-merging} (c) LGSVL autopilot: Verdicts for merging scenarios with $v_e=10.0$
    \scriptsize 
    \adjustbox{max width=\textwidth}{\input{latex-table/lgsvl-merging-10.0}}
\end{center}
\vspace{10pt}
\begin{center}
    Table \ref{tab:lgsvl-merging} (d) LGSVL autopilot: Verdicts for merging scenarios with $v_e=15.0$
    \scriptsize 
    \adjustbox{max width=\textwidth}{
    \input{latex-table/lgsvl-merging-15.0}}
\end{center}

\vspace{2em}
\paragraph{1) LGSVL autopilot: Analysis for merging scenarios}

Verdicts for merging scenarios and different speeds $v_e$ are shown in Tab.~\ref{tab:lgsvl-merging}, revealing accidents involving both the ego vehicle hitting the arriving vehicle (\textsf{Ae}) and the arriving vehicle hitting the ego vehicle (\textsf{Aa}).

For all values of $v_e$,  accidents can happen for $x_a$ close to its critical value and $x_f \ge 15.0$. Accidents occur in two cases
1) the ego vehicle enters the main road and collides with the arriving vehicle at the merging point (\textsf{Ae}); 2) the ego vehicle enters the main road and there is not enough space for the arriving vehicle to stop and collides with the ego vehicle (\textsf{Aa}).

In addition, for all values of $v_e$, accidents can occur when $x_f$ is around 5.0, regardless of the value of $x_a$. In these accidents, the ego vehicle stops at the edge of the main road, and the  vehicle arriving from a distance may collide with the ego vehicle  which has stopped due to its poor estimate of the clearance.

\clearpage

\begin{minipage}{\linewidth}
\begin{table}[H]
    \centering
    \caption{LGSVL autopilot: Verdicts for lane change scenarios}
    \label{tab:lgsvl-lane_change}
\end{table}
\vspace{-20pt}
\begin{center}
    Table \ref{tab:lgsvl-lane_change} (a) LGSVL autopilot: Verdicts for lane change scenarios with $v_e=5.0$
    \scriptsize 
    \adjustbox{max width=\textwidth}{\input{latex-table/lgsvl-lane_change-5.0}}
\end{center}
\end{minipage}
\vspace{10pt}
\begin{center}
    Table \ref{tab:lgsvl-lane_change} (b) LGSVL autopilot: Verdicts for lane change scenarios with $v_e=10.0$
    \scriptsize 
    \adjustbox{max width=\textwidth}{\input{latex-table/lgsvl-lane_change-10.0}}
\end{center}
\vspace{10pt}
\begin{center}
    Table \ref{tab:lgsvl-lane_change} (c) LGSVL autopilot: Verdicts for lane change scenarios with $v_e=15.0$
    \scriptsize 
    \adjustbox{max width=\textwidth}{\input{latex-table/lgsvl-lane_change-15.0}}
\end{center}
\vspace{10pt}
\begin{center}
    Table \ref{tab:lgsvl-lane_change} (d) LGSVL autopilot: Verdicts for lane change scenarios with $v_e=20.0$
    \scriptsize 
    \adjustbox{max width=\textwidth}{
    \input{latex-table/lgsvl-lane_change-20.0}}
\end{center}
\vspace{2em}
\clearpage
\paragraph{2) LGSVL autopilot: Analysis for lane change}

Verdicts for lane change scenarios and different speeds $v_e$ are shown in Tab.~\ref{tab:lgsvl-lane_change}. They reveal accidents involving both the ego vehicle hitting the arriving vehicle (\textsf{Ae}) and the arriving vehicle hitting the ego vehicle (\textsf{Aa}).

For all values of $v_e$, accidents occur during the transition from caution to progress. 

For $v_e=5.0$, the transition occurs at $x_a \le 35.0 < \hat x_a = 65.7$; and for $v_e=10.0$, the transition occurs for $x_a \le 34.4 < \hat x_a=35.6$, which explains the autopilot's wrong decision.

For $v_e=15.0$, the transition occurs at $x_a = 32.5 > \hat x_a = 25.6$; and for $v_e=20.0$, the transition occurs at $x_a=30.0 > \hat x_a = 20.6$. {Accidents occur when the ego vehicle decelerates to change lanes, taking longer to reach the outside lane and leaving insufficient space to stop. They can be avoided if the ego vehicle maintains its initial speed.}

\vspace{2em}
\begin{table}[H]
    \centering
    \caption{LGSVL autopilot: Verdicts for crossing with yield sign scenarios}
    \label{tab:lgsvl-crossing}
\end{table}
\vspace{-20pt}
\begin{center}
    Table \ref{tab:lgsvl-crossing} (a) LGSVL autopilot: Verdicts for crossing with yield sign scenarios with $v_e=0.0$
    \footnotesize 
    \adjustbox{max width=\textwidth}{\input{latex-table/lgsvl-crossing-0.0}}
\end{center}
\vspace{10pt}
\begin{center}
    Table \ref{tab:lgsvl-crossing} (b) LGSVL autopilot: Verdicts for crossing with yield sign scenarios with $v_e=5.0$
    \footnotesize 
    \adjustbox{max width=\textwidth}{\input{latex-table/lgsvl-crossing-5.0}}
\end{center}
\vspace{10pt}
\begin{center}
    Table \ref{tab:lgsvl-crossing} (c) LGSVL autopilot: Verdicts for crossing with yield sign scenarios with $v_e=10.0$
    \footnotesize 
    \adjustbox{max width=\textwidth}{\input{latex-table/lgsvl-crossing-10.0}}
\end{center}
\vspace{10pt}
\begin{center}
    Table \ref{tab:lgsvl-crossing} (d) LGSVL autopilot: Verdicts for crossing with yield sign scenarios with $v_e=15.0$
    \footnotesize 
    \adjustbox{max width=\textwidth}{
    \input{latex-table/lgsvl-crossing-15.0}}
\end{center}

\vspace{2em}
\paragraph{3) LGSVL autopilot: Analysis for crossing with yield sign scenarios}

Verdicts for crossing with yield sign scenarios and different speeds $v_e$ are shown in Tab.~\ref{tab:lgsvl-crossing}, revealing accidents and violations of $p_1$ (Two vehicles must not be in the critical zone at the same time).

For all four values of $v_e$, we found that the transition from caution to progress occurs for $x_a$ increasing from its critical value. However, accidents and violations of rule $p_1$ (Two vehicles must not be in the critical zone at the same time) can occur during this transition, indicating a poor control policy. Moreover, the autopilot fails to apply safe cautious policies for all test cases with $\langle x_f, x_a \rangle = \langle 0.0, 0.0 \rangle$.

\begin{table}[H]
    \centering
    \caption{LGSVL autopilot: Verdicts for crossing with traffic light scenarios}
    \label{tab:lgsvl-traffic_light}
\end{table}
\vspace{-20pt}
\begin{center}
    Table \ref{tab:lgsvl-traffic_light} (a) LGSVL autopilot: Verdicts for crossing with traffic light scenarios with $v_e=0.0$
    \small 
    {\begin{tabular}{|c|c|c|c|c|c|c|c|c|c|}
\hline
\multicolumn{10}{|c|}{$\bm{xf}$}\\
\hline
0.0 & \ct{5.2} & 40.0 & 80.0 & 120.0 & 160.0 & 200.0 & 240.0 & 280.0 & 320.0 \\
\hline
\cu{\textsf{CU$p_2p_4$}} & \cu{\textsf{CU$p_2p_4$}} & \cu{\textsf{CU$p_2p_4$}} & \cu{\textsf{CU$p_2p_4$}} & \cu{\textsf{CU$p_2p_4$}} & \cu{\textsf{CU$p_2p_4$}} & \cu{\textsf{CU$p_2p_4$}} & \cu{\textsf{CU$p_2p_4$}} & \cu{\textsf{CU$p_2p_4$}} & \cu{\textsf{CU$p_2p_4$}} \\
\hline
\end{tabular}}
\end{center}
\vspace{10pt}
\begin{center}
    Table \ref{tab:lgsvl-traffic_light} (b) LGSVL autopilot: Verdicts for crossing with traffic light scenarios with $v_e=5.0$
    \small 
    {\input{latex-table/lgsvl-traffic_light-5.0}}
\end{center}
\vspace{10pt}
\begin{center}
    Table \ref{tab:lgsvl-traffic_light} (c) LGSVL autopilot: Verdicts for crossing with traffic light scenarios with $v_e=10.0$
    \small 
    {\input{latex-table/lgsvl-traffic_light-10.0}}
\end{center}
\vspace{10pt}
\begin{center}
    Table \ref{tab:lgsvl-traffic_light} (d) LGSVL autopilot: Verdicts for crossing with traffic light scenarios with $v_e=15.0$
    \small 
    {\input{latex-table/lgsvl-traffic_light-15.0}}
\end{center}
\vspace{10pt}
\begin{center}
    Table \ref{tab:lgsvl-traffic_light} (e) LGSVL autopilot: Verdicts for crossing with traffic light scenarios with $v_e=20.0$
    \small 
    {\input{latex-table/lgsvl-traffic_light-20.0}}
\end{center}

\vspace{2em}
\paragraph{4) LGSVL autopilot: Analysis for crossing with traffic light scenarios}

Verdicts for crossing with traffic light scenarios and different speeds $v_e$ are shown in Tab.~\ref{tab:lgsvl-traffic_light}. They reveal accidents and violations of $p_2$ (The ego vehicle must not stop inside the critical zone) and $p_4$ (The ego vehicle must not be in the critical zone when the side light is green).

For  $v_e=0.0$, $v_e=5.0$, $v_e=10.0$, and $v_e=15.0$, the ego vehicle decides to progress when confronted with a green light at the start and begins to be cautious when it detects a yellow light at the next time step. However, there is not enough space to stop before the entrance, so it stops inside the critical zone to wait for the green light, thus violating $p_2$ (The ego vehicle must not stop inside the critical zone) and $p_4$ (The ego vehicle must not be in the critical zone when the side light is green) for all values of $x_f$.

For  $v_e=20.0$, the ego vehicle also decides to progress initially and be cautious then. However, after it stops inside the critical zone, it decides to progress again. Thus, it violates $p_2$ (The ego vehicle must not stop inside the critical zone). {In addition, when $x_f \le \hat x_f = 5.5$, as the ego vehicle cannot accelerate to a high speed, it exceeds the time limit of 2-second all-red phases violating $p_4$ (The ego must not be in the critical zone when the side light is green).}

\clearpage
\subsection{Summary of results}

\subsubsection{Macroscopic analysis of the verdicts }

Figures~ \ref{fig:apollo-result} to \ref{fig:lgsvl-result} show the partition of verdicts for each autopilot. The results confirm that 
problematic scenarios mainly occur during the transition from safe caution to safe progress as the values of $x_f$ and $x_a$ increase. 
In particular, we observe irrational policies in terms of safety or  performance:

\vspace{-2pt}
\begin{enumerate}
    \item Irrational safety violations occur when we have safe progress for given values $x_a$ and $x_f$ , and problematic situations for larger values, or we have safe caution for given values $x_a$ and $x_f$, and problematic situations for smaller values. We observed the latter case for the Apollo autopilot in lane change scenarios with $v_e=15.0$.
    
    \item Irrational performance degradation occurs when we have safe progress for given values of $x_a$ and $x_f$, and cautious policy for larger values. We observed this for the Apollo autopilot in merging scenarios with $v_e=15.0$.
\end{enumerate}

 \vspace{-2pt}
Such anomalies compromise autopilot testability, making systematic defect finding extremely difficult.

\vspace{5pt}

In the following cases, the autopilots remain uniformly cautious for all test cases generated in the $x_a$ and $x_f$ range under consideration, while progress is theoretically possible as $x_a$ and $x_f$ increase.
\vspace{-2pt}
\begin{enumerate} 
    \item The Apollo autopilot in crossing with traffic light scenarios where $v_e=5.0$.
    \item The Autoware autopilot in lane change scenarios where $v_e\in \{5.0, 10.0, 15.0, 20.0\}$.
    \item The Carla autopilot in lane change scenarios where $v_e\in \{15.0, 20.0\}$.
    \item The Carla autopilot in crossing with traffic light scenarios where $v_e\in \{10.0, 15.0, 20.0\}$.
\end{enumerate}

\vspace{5pt}

In the following cases, the autopilots uniformly fail in the caution phase for the test cases generated in the $x_f$ and $x_a$ range under consideration, showing inherent defects in compliance with safety rules.
\vspace{-2pt}
\begin{enumerate}
\item  The Apollo autopilot in crossing with traffic light scenarios where $v_e\in\{0.0, 15.0, 20.0\}$, either it progresses unsafely or it fails to progress. 
\item  The Autoware autopilot in crossing with traffic light scenarios where $v_e \in \{5.0, 10.0, 15.0, 20.0\}$, it either progresses unsafely or it does not apply sufficient acceleration. 
\item The LGSVL autopilot in crossing with yield sign scenarios where $v_e\in \{0.0, 5.0, 10.0, 15.0\}$, if it decides to be cautious, it always enters the critical zone to wait for the arriving vehicle.
\item  The LGSVL autopilot in crossing with traffic light scenarios where $v_e \in \{0.0, 5.0, 10.0, 15.0, 20.0\}$, is unsafely cautious for $v_e\in \{0.0, 5.0, 10.0, 15.0\}$ and progresses  unsafely for $v_e=20.0$.
\end{enumerate}

\vspace{5pt}

In addition, the Apollo autopilot in crossing with yield sign scenarios where $v_e\in \{10.0, 15.0\}$, after the phase of safe caution, it fails to make safe progress in the range of the considered values of $x_f$ and $x_a$.

Our test method provides good coverage of the vista types considered. Specifically, for merging, lane change, and crossing with yield sign, there is a regularity in the patterns observed, as $v_e$ increases. Except for irrational, over-cautious, and lacking safe caution or progress patterns, the progress zone is approximately a convex area whose position changes with the values of $v_e$. Furthermore, as $v_e$ increases, some problem situations may disappear, while others may emerge.

The results for crossing with traffic light show similarity in different ways between the Apollo and Autoware autopilots, and between the Carla and LGSVL autopilots.

When $v_e=0.0$, Apollo and Autoware autopilots do not have feasible progress policies for all values of $x_f$. 
Furthermore, for the other speed values, we observe similar patterns in the transitions from unsafe progress to safe progress.

For Carla and LGSVL autopilots, for all values of $v_e$, the theory predicts feasible progress policies for $x_f \ge \hat x_f$. For the Carla autopilot, we observe no safety violations.  
For $v_e \in \{0.0, 5.0\}$, the vehicle progresses safely for all $x_f$ values except $x_f=0$, 
and for $v_e \in \{10.0, 15.0, 20.0\}$, the vehicle is safely cautious for all $x_f$ values. For the LGSVL autopilot, the vehicle is always unsafely cautious when $v_e \in \{0.0, 5.0, 10.0, 15.0\}$ and always progresses unsafely when $v_e = 20.0$.

\clearpage

\begin{figure}[H]
    \centering
    \includegraphics[width=\linewidth]{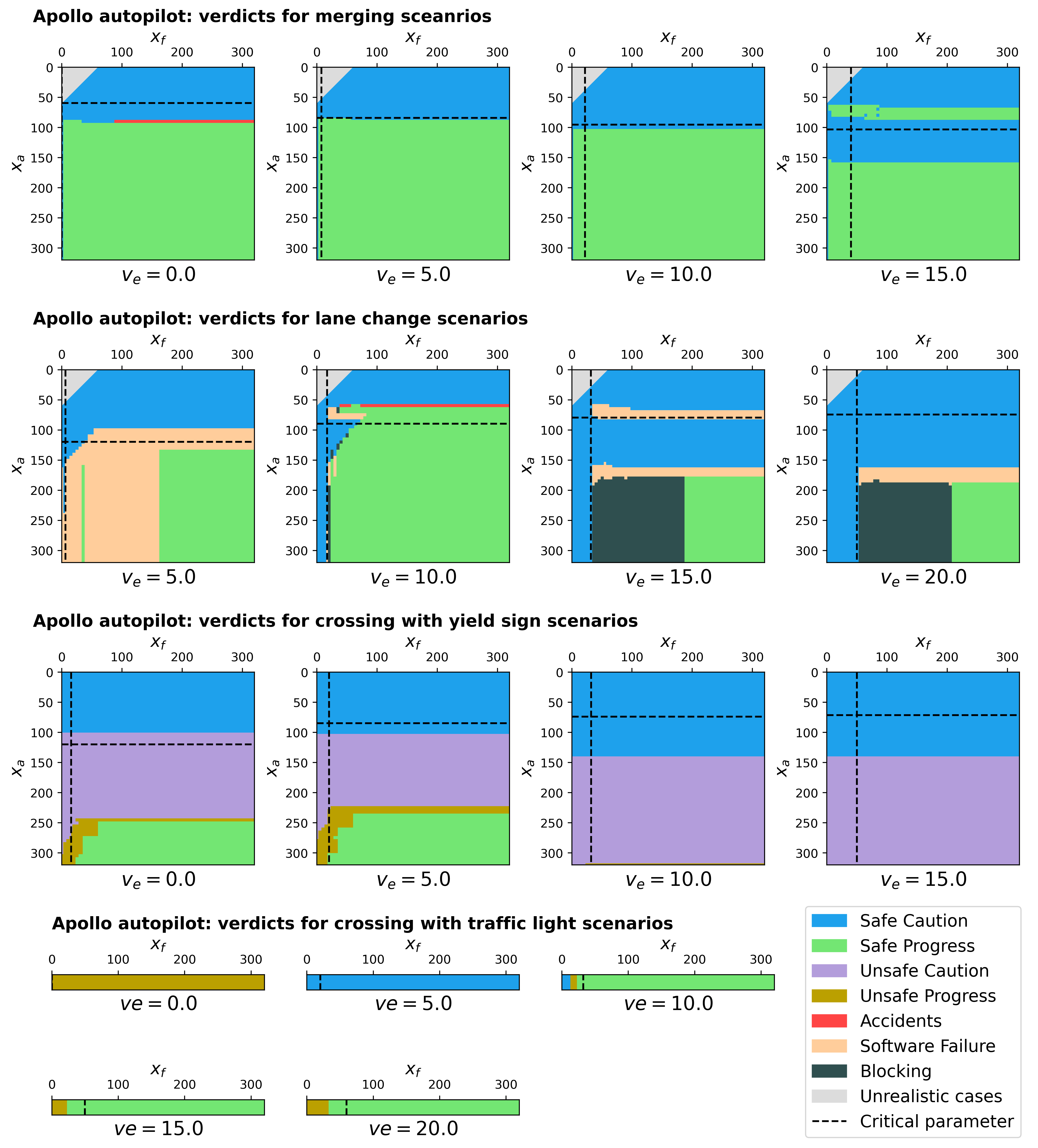}
    \caption{Macroscopic analysis of verdicts  for Apollo autopilot}
    \label{fig:apollo-result}
\end{figure}

\setlength{\tabcolsep}{10pt}
\renewcommand{\arraystretch}{1}

\begin{figure}[H]
    \centering
    \includegraphics[width=\linewidth]{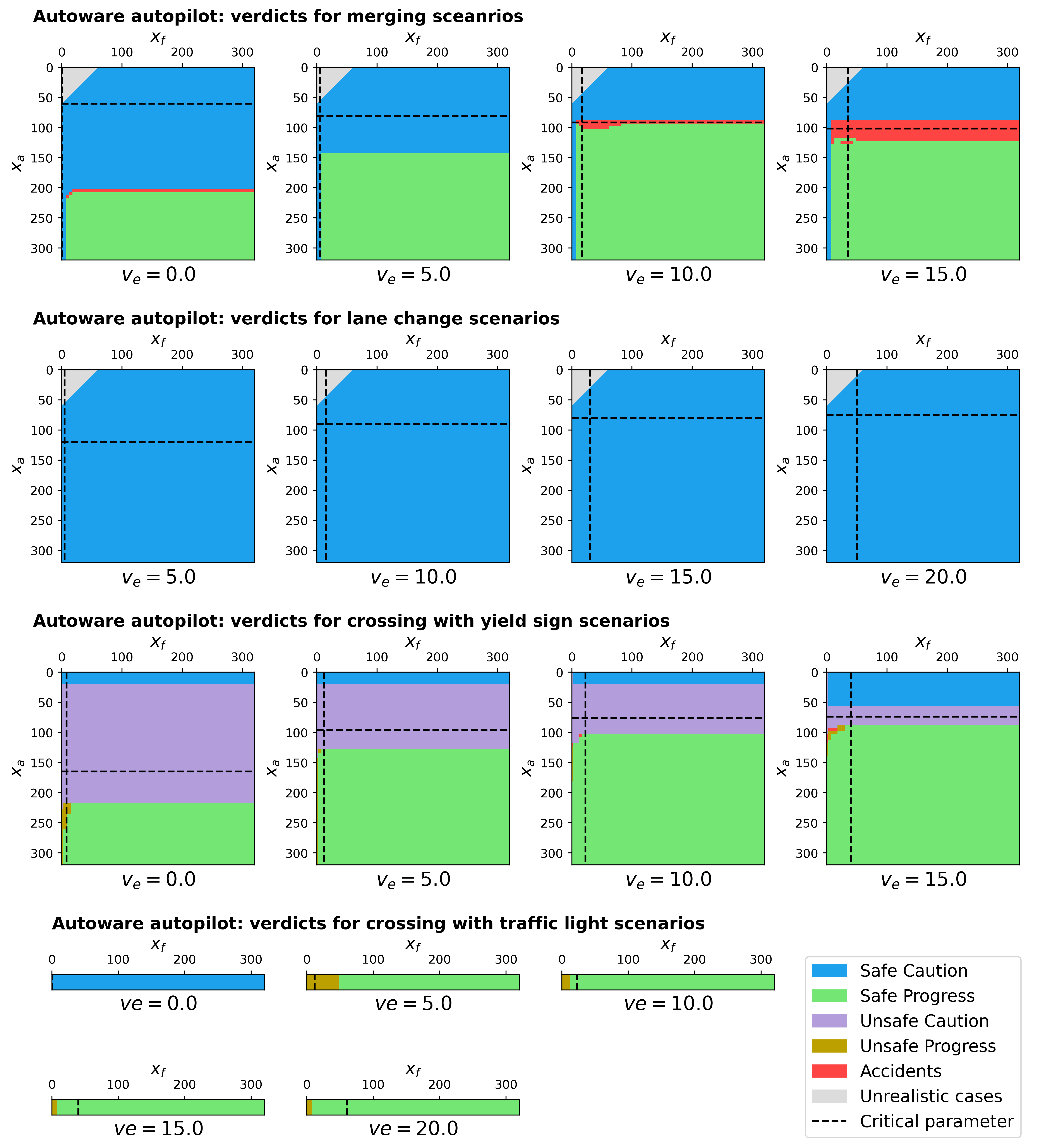}
    \caption{Macroscopic analysis of verdicts  for Autoware autopilot}
    \label{fig:autoware-result}
\end{figure}
\vspace{-15pt}
\begin{figure}[H]
    \centering
    \includegraphics[width=\linewidth]{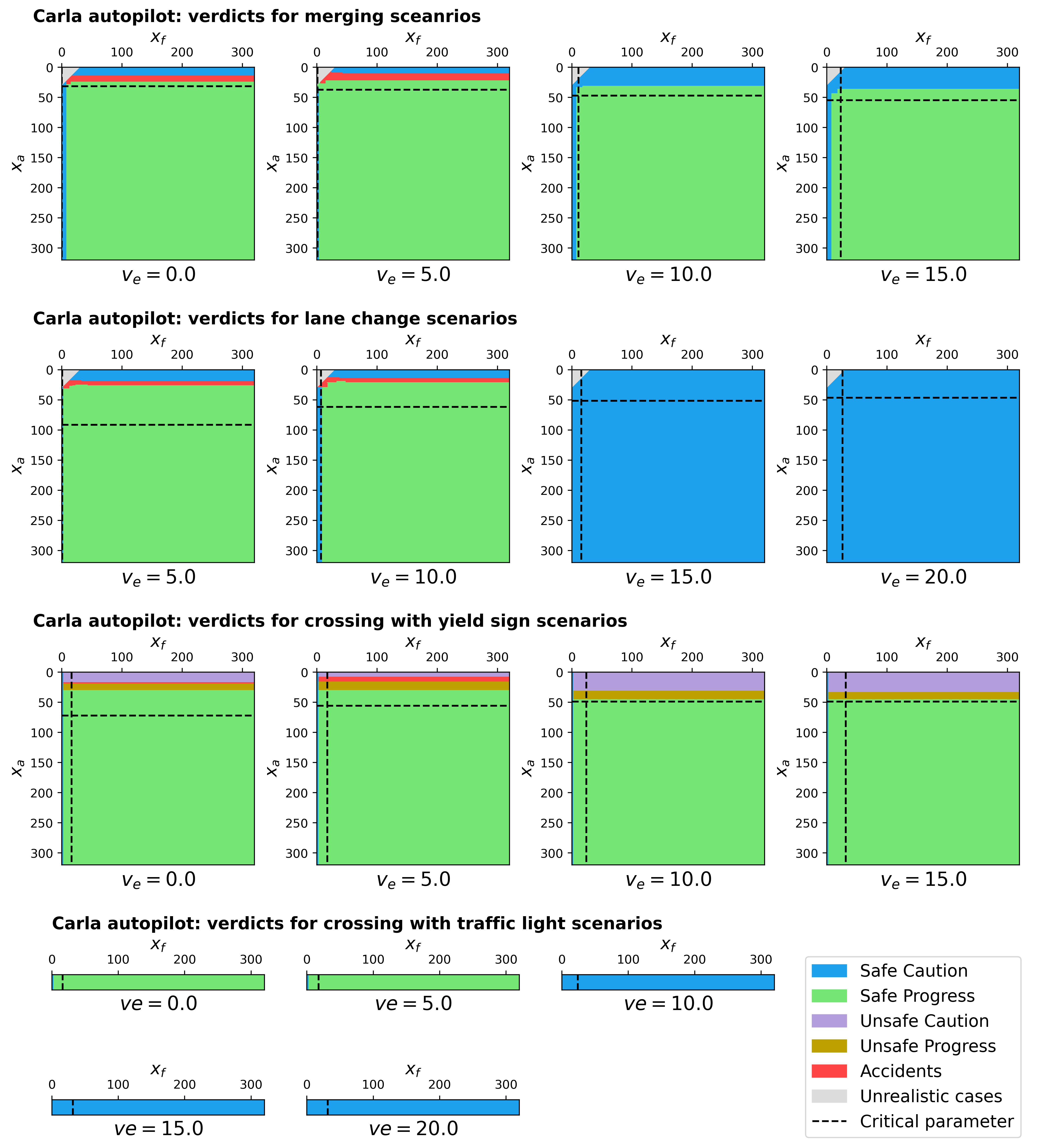}
    \caption{Macroscopic analysis of verdicts  for Carla autopilot}
    \label{fig:carla-result}
\end{figure}
\begin{figure}[H]
    \centering
    \includegraphics[width=\linewidth]{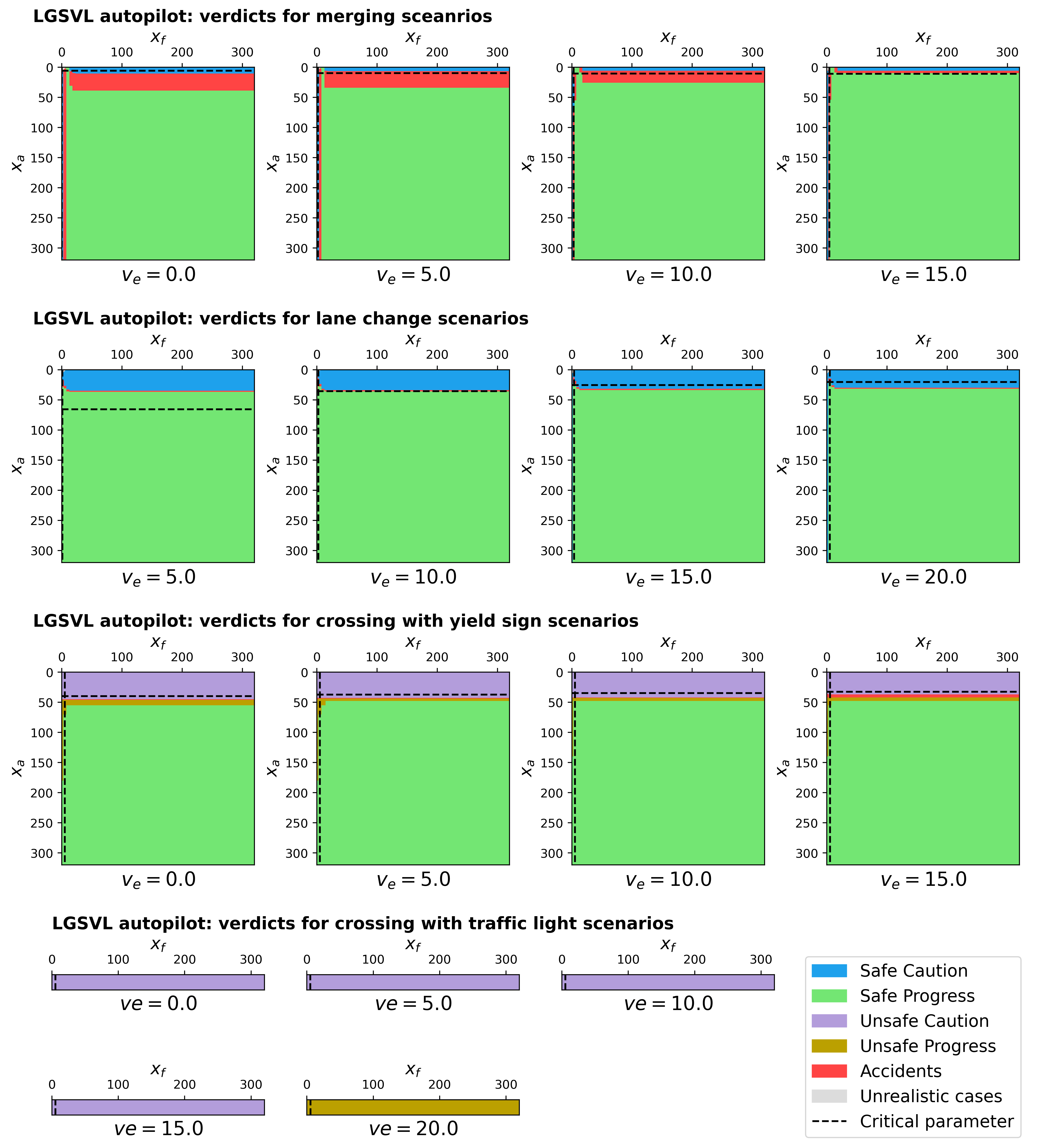}
    \caption{Macroscopic analysis of verdicts  for LGSVL autopilot
    }
    \label{fig:lgsvl-result}
\end{figure}

\clearpage
\subsubsection{Effectiveness of the test method}

Our test method is proving highly effective in uncovering autopilot defects. Tables \ref{tab:sum-merging} to \ref{tab:sum-traffic_light} summarize the number and frequency of 
the different types of verdicts for each vista and autopilot.
It reveals safety problems for all autopilots and all vista types, with the exception of {the Autoware autopilot when changing lanes and the Carla autopilot when crossing traffic lights.} For a total of 14506 test cases, we have detected 3962 various types of defects, which corresponds to 27.31\% of the test cases. 

Defects of merging scenarios include accidents: 781 accidents for a total of 4687 test cases, which corresponds to 16.66\% of the test cases. The LGSVL autopilot has the highest number of accidents at 495 (31.55\% of its test cases), while the Apollo autopilot shows the lowest number with 8 (0.80\% of its test cases) accidents.

Defects of lane change scenarios include accidents, software failures, and blocking: 1282 defects for a total of 5967 test cases, which corresponds to 21.48\% of the test cases. In particular, there are 1070 accidents and software failures, resulting in a rate of 17.93\% among all test cases. The Carla and LGSVL autopilots have high accident numbers, with 209 (17.55\% of its test cases) and 179 (10.49\% of its test cases) accidents, respectively. The Apollo autopilot is the only one with software failures and blocking defects. The number of software failures is 667 (24.81\% of its test cases), and blocking is 212 (7.89\% of its test cases). No defects were observed for Autoware autopilot, as it is cautious in all test cases.

Defects of crossing with yield sign scenarios include accidents and violations of safety properties $p_1$ (Two vehicles must not be in the critical zone at the same time) and $p_2$ (The ego vehicle must not stop inside the critical zone): 1819 defects for a total of 3622 test cases, which corresponds to 50.22\% of the test cases. All four autopilots exhibit high amount of property violations, where Apollo has 500 violations (55.37\% of its test cases), Autoware has 321 violations (37.81\% of its test cases), Carla has 348 violations (38.24\% of its test cases), and LGSVL has 429 violations (44.69\% of its test cases). For accidents, Apollo autopilot has none, Autoware autopilot has 8 (0.94\% of its test cases), while Carla and LGSVL have more accidents, with 84 (9.23\% of its test cases) and 129 (13.44\% of its test cases) accidents, respectively.

Defects of crossing with traffic light scenarios include accidents and violations of safety properties $p_2$ (The ego vehicle must not stop inside the critical zone), $p_3$ (The ego vehicle must not enter the critical zone when the light is red.), and $p_4$ (The ego vehicle must not be in the critical zone when a side light is green): 80 defects for a total of 230 test cases, which corresponds to 34.78\% of the test cases. Apollo and Autoware autopilots have defects for respecting the time limits imposed by the yellow-light phase and all-red phase, violating $p_3$ and $p_4$. They have defects of such violations at 17 (27.87\% of its test cases) and 10 (16.67\% of its test cases), respectively. There is no defect detected for the Carla autopilot due to its over-cautiousness. The LGSVL autopilot exhibits a unique defect of violating $p_2$ (The ego vehicle must not stop inside the critical zone) across all test cases. It also violates the time limit imposed by the all-red phase in 3 (5.66\% of its test cases) test cases, violating $p_4$.

Our global analysis for the four autopilots shows that scenarios for crossings with yield signs present the highest percentage of defects (50.22\%), followed by scenarios for crossings with traffic lights (34.78\%). Lane change and merging scenarios have lower defect rates with 21.48\%  and 16.66\%, respectively. When considering only accidents, merging scenarios have a significantly higher accident rate (16.66\%), compared with lane change scenarios (6.75\%) and scenarios for crossings with yield signs (6.10\%).

\begin{table}[H]
    \centering
    \caption{Statistics on verdicts  for merging scenarios}
    \vspace{-10pt}
    \label{tab:sum-merging}
    \begin{tabular}{|c|r@{\ \ }l|r@{\ \ }l|r@{\ \ }l|r@{\ \ }l|}
\hline
\multirow{2}{*}{Verdict} & \multicolumn{8}{c|}{Autopilot} \\
\cline{2-9}
         &  \multicolumn{2}{c|}{Apollo}  &  \multicolumn{2}{c|}
         {Autoware} &  \multicolumn{2}{c|}{Carla}  &  \multicolumn{2}{c|}{LGSVL} \\\hline

CS & 432 & (43.11\%) & 364 & (37.18\%) & 304 & (26.74\%) & 313 & (19.95\%) \\\hline
PS & 562 & (56.09\%) & 439 & (44.84\%) & 731 & (64.29\%) & 761 & (48.50\%) \\\hline

Ae & 8 & (0.80\%) & 12 & (1.23\%) & 42 & (3.69\%) & 126 & (8.03\%) \\\hline
Aa & 0 & (0.00\%) & 164 & (16.75\%) & 60 & (5.28\%) & 369 & (23.52\%) \\\hline

\textbf{Total} & \multicolumn{2}{c|}{1002} & \multicolumn{2}{c|}{979} & \multicolumn{2}{c|}{1137} & \multicolumn{2}{c|}{1569} \\\hline
\end{tabular}
\end{table}

\begin{table}[H]
    \centering
    \caption{Statistics on verdicts  for lane change scenarios}
    \vspace{-10pt}
    \label{tab:sum-lane_change}
    \begin{tabular}{|c|r@{\ \ }l|r@{\ \ }l|r@{\ \ }l|r@{\ \ }l|}
\hline
\multirow{2}{*}{Verdict} & \multicolumn{8}{c|}{Autopilot} \\
\cline{2-9}
         &  \multicolumn{2}{c|}{Apollo}  &  \multicolumn{2}{c|}
         {Autoware} &  \multicolumn{2}{c|}{Carla}  &  \multicolumn{2}{c|}{LGSVL} \\\hline

CS & 1067 & (39.69\%) & 381 & (100.00\%) & 371 & (31.15\%) & 887 & (51.96\%) \\\hline
PS & 727 & (27.05\%) & 0 & (0.00\%) & 611 & (51.30\%) & 641 & (37.55\%) \\\hline

Blk & 212 & (7.89\%) & 0 & (0.00\%) & 0 & (0.00\%) & 0 & (0.00\%) \\\hline

Ae & 0 & (0.00\%) & 0 & (0.00\%) & 76 & (6.38\%) & 92 & (5.39\%) \\\hline
Aa & 15 & (0.56\%) & 0 & (0.00\%) & 133 & (11.17\%) & 87 & (5.10\%) \\\hline

Fsw & 667 & (24.81\%) & 0 & (0.00\%) & 0 & (0.00\%) & 0 & (0.00\%) \\\hline

\textbf{Total} & \multicolumn{2}{c|}{2688} & \multicolumn{2}{c|}{381} & \multicolumn{2}{c|}{1191} & \multicolumn{2}{c|}{1707} \\\hline
\end{tabular}
\end{table}

\begin{table}[H]
\centering
  \caption{Statistics on verdicts  for crossing with yield sign scenarios}
  \vspace{-10pt}
  \label{tab:sum-crossing}
  \begin{tabular}{|c|r@{\ \ }l|r@{\ \ }l|r@{\ \ }l|r@{\ \ }l|}
\hline
\multirow{2}{*}{Verdict} & \multicolumn{8}{c|}{Autopilot} \\
\cline{2-9}
         &  \multicolumn{2}{c|}{Apollo}  &  \multicolumn{2}{c|}
         {Autoware} &  \multicolumn{2}{c|}{Carla}  &  \multicolumn{2}{c|}{LGSVL} \\\hline

CS & 232 & (25.69\%) & 63 & (7.42\%) & 70 & (7.69\%) & 0 & (0.00\%) \\\hline
PS & 171 & (18.94\%) & 457 & (53.83\%) & 408 & (44.84\%) & 402 & (41.88\%) \\\hline

CU$p_1$ & 176 & (19.49\%) & 110 & (12.96\%) & 60 & (6.59\%) & 270 & (28.12\%) \\\hline
CU$p_2$ & 0 & (0.00\%) & 3 & (0.35\%) & 12 & (1.32\%) & 0 & (0.00\%) \\\hline
CU$p_1p_2$ & 161 & (17.83\%) & 173 & (20.38\%) & 96 & (10.55\%) & 0 & (0.00\%) \\\hline

PU$p_1$ & 2 & (0.22\%) & 30 & (3.53\%) & 72 & (7.91\%) & 159 & (16.56\%) \\\hline
PU$p_2$ & 8 & (0.89\%) & 3 & (0.35\%) & 10 & (1.10\%) & 0 & (0.00\%) \\\hline
PU$p_1p_2$ & 153 & (16.94\%) & 2 & (0.24\%) & 98 & (10.77\%) & 0 & (0.00\%) \\\hline

Ae & 0 & (0.00\%) & 0 & (0.00\%) & 48 & (5.27\%) & 53 & (5.52\%) \\\hline
Aa & 0 & (0.00\%) & 8 & (0.94\%) & 36 & (3.96\%) & 76 & (7.92\%) \\\hline
\textbf{Total} & \multicolumn{2}{c|}{903} & \multicolumn{2}{c|}{849} & \multicolumn{2}{c|}{910} & \multicolumn{2}{c|}{960} \\\hline
\end{tabular}
\end{table}

\begin{table}[H]
\centering
  \caption{Statistics on verdicts  for crossing with traffic light scenarios}
  \vspace{-10pt}
  \label{tab:sum-traffic_light}
  \begin{tabular}{|c|r@{\ \ }l|r@{\ \ }l|r@{\ \ }l|r@{\ \ }l|}
\hline
\multirow{2}{*}{Verdict} & \multicolumn{8}{c|}{Autopilot} \\
\cline{2-9}
         &  \multicolumn{2}{c|}{Apollo}  &  \multicolumn{2}{c|}
         {Autoware} &  \multicolumn{2}{c|}{Carla}  &  \multicolumn{2}{c|}{LGSVL} \\\hline

CS & 12 & (19.67\%) & 9 & (15.00\%) & 32 & (57.14\%) & 0 & (0.00\%) \\\hline
PS & 32 & (52.46\%) & 41 & (68.33\%) & 24 & (42.86\%) & 0 & (0.00\%) \\\hline

CU$p_1p_2$ & 0 & (0.00\%) & 0 & (0.00\%) & 0 & (0.00\%) & 40 & (75.47\%) \\\hline
PU$p_2$ & 0 & (0.00\%) & 0 & (0.00\%) & 0 & (0.00\%) & 10 & (18.87\%) \\\hline
PU$p_3$ & 1 & (1.64\%) & 0 & (0.00\%) & 0 & (0.00\%) & 0 & (0.00\%) \\\hline
PU$p_4$ & 13 & (21.31\%) & 8 & (13.33\%) & 0 & (0.00\%) & 0 & (0.00\%) \\\hline
PU$p_2p_4$ & 0 & (0.00\%) & 0 & (0.00\%) & 0 & (0.00\%) & 3 & (5.66\%) \\\hline
PU$p_3p_4$ & 3 & (4.92\%) & 2 & (3.33\%) & 0 & (0.00\%) & 0 & (0.00\%) \\\hline

\textbf{Total} & \multicolumn{2}{c|}{61} & \multicolumn{2}{c|}{60} & \multicolumn{2}{c|}{56} & \multicolumn{2}{c|}{53} \\\hline
\end{tabular}
\end{table}

\subsubsection{Open access and experimentation }

We provide the Simulator, including Maps, Bridge, and the configured Autopilots and Runtimes at \url{https://github.com/LIIHWF/autopilot-compositional-testing.git}. The experiments presented in this paper can be reproduced by following the instructions provided online. 

\section{Discussion}\label{se:discussion}
The paper confirms that much remains to be done to develop rigorous validation methods that systematically cover the wide variety of situations encountered by autonomous vehicles. It may be objected that industrial autopilots built with much greater care are undoubtedly more trustworthy than open-source autopilots. However, our results corroborate real-life observations and confirm that autonomous driving systems still have a long way to go before offering acceptable safety guarantees~\cite{cum2023}.

The test method presented shows that the four autopilots examined, and very probably others, have their Achilles heel: the transition from cautious control to progress for the most critical situations. 
These situations can be characterized by finely-tuned parameter combinations that are difficult to generate by random simulation. In this respect, it differs from most work which focuses on simple scenarios, typically freeway driving.  In addition, most simulators allow only one autopilot, which limits the possibility of creating dangerous situations by controlling several vehicles.

Analysis of the dynamics of some autopilots reveals a lack of realism, as they assume excessive acceleration or deceleration rates, which can make it easier to manage critical situations. In this case, the test results are of limited value, as the system under test is not realistic.		

Our analysis shows the importance of clearly defined A/D functions for accurate prediction and validation. Without precise, mathematically explicit definitions, it is virtually impossible for the autopilot to predict vehicle behavior. It is also very difficult to apply rigorous test methods. Compliance with the principle of rationality facilitates the application of worst-case reasoning, which is the basis of risk analysis techniques. 

The effectiveness of our method for an ADS depends on the testability of its autopilots. 
Our analysis shows the need for testable autopilots, free from anomalies that complicate or even make impossible guaranteed coverage. These anomalies may be inherent to the operation of neural networks such as adversarial examples~\cite{kurakin2018adversarial}, but they may simply be the result of sloppy design. Significantly, most of the problems detected concern the transition from caution to  progress  when the degree of criticality of the configurations is relaxed. An obvious safe solution would have been to avoid faults by remaining cautious until a safe progression is possible.

The results presented suggest that the complexity of the problem can be controlled by decomposing complex scenarios into sequences of simple types of scenarios for a limited number of traffic patterns and configurations of a small number of vehicles. Simulating billions of miles~\cite{etherington2019waymo} without specifying how they relate to and cover real-life situations is not a convincing argument for safety. Safety critical scenarios are rare and the probability of discovering the situations identified may be low for a car in simulation but may become non-negligible for very large number of cars in real-life situations.

This is an initial work in which we have considered a number of significant scenario types, without seeking to be exhaustive. Other types of intersections remain to be analyzed, such as four-way intersections and traffic circles. We believe that our test method, backed up by convincing experimental results, provides a systematic basis for rigorous compositional testing of autopilots. Compositionality is also supported by the intuitive idea that driving ability boils down to the combination of skills required to perform a relatively small number of operations.

The results obtained suggest further developments toward a rigorous test method for ADS validation, based on test coverage criteria, if the following conditions are jointly met. 

The first condition is to guarantee compositionality: having validated a system for a limited number of vista types and the resulting scenarios, guarantee validation for any scenario. The results presented in \cite{bozga2024safe} show that this condition can be met under certain reasonable assumptions about the autopilot behavior.

The second condition concerns the reproducibility of the results, and requires that the autopilot's behavior, although dependent on multiple parameters, does not exhibit non-determinism at runtime, which would make real-time constraint validation problematic. 

The third condition concerns the rationality of control policies, which must allow for worst-case reasoning, thus considerably reducing the complexity of the space of test cases to be explored.  Achieving this property depends on the decision algorithm, and is greatly facilitated for additive A/D functions.

All these results and their further development by realizing the above conditions argue in favor of traditional design techniques for autopilot decision-making, limiting the role of AI components to perception alone.

\section*{Acknowledgments}
We would like to thank Feng Liu and Zifan Zeng of Huawei for their insightful comments on early drafts of the article and their independent contribution to the experimental results confirming the validity of the theory.  

Joseph Sifakis thanks Marius Bozga for his essential contribution on theoretical aspects of the caution/progress paradigm, and on the compositionality  of safe policies for the different types of vistas, which are the subject of ongoing collaboration.

\newpage
\normalem 
\bibliographystyle{ACM-Reference-Format}
\bibliography{reference}


\begin{thebibliography}{35}


\ifx \showCODEN    \undefined \def \showCODEN     #1{\unskip}     \fi
\ifx \showDOI      \undefined \def \showDOI       #1{#1}\fi
\ifx \showISBNx    \undefined \def \showISBNx     #1{\unskip}     \fi
\ifx \showISBNxiii \undefined \def \showISBNxiii  #1{\unskip}     \fi
\ifx \showISSN     \undefined \def \showISSN      #1{\unskip}     \fi
\ifx \showLCCN     \undefined \def \showLCCN      #1{\unskip}     \fi
\ifx \shownote     \undefined \def \shownote      #1{#1}          \fi
\ifx \showarticletitle \undefined \def \showarticletitle #1{#1}   \fi
\ifx \showURL      \undefined \def \showURL       {\relax}        \fi
\providecommand\bibfield[2]{#2}
\providecommand\bibinfo[2]{#2}
\providecommand\natexlab[1]{#1}
\providecommand\showeprint[2][]{arXiv:#2}

\bibitem[Bozga and Sifakis(2022)]%
        {bozga2022specification}
\bibfield{author}{\bibinfo{person}{Marius Bozga} {and} \bibinfo{person}{Joseph Sifakis}.} \bibinfo{year}{2022}\natexlab{}.
\newblock \showarticletitle{Specification and validation of autonomous driving systems: A multilevel semantic framework}.
\newblock In \bibinfo{booktitle}{\emph{Principles of Systems Design: Essays Dedicated to Thomas A. Henzinger on the Occasion of His 60th Birthday}}. \bibinfo{publisher}{Springer}, \bibinfo{pages}{85--106}.
\newblock


\bibitem[Bozga and Sifakis(2024)]%
        {bozga2024safe}
\bibfield{author}{\bibinfo{person}{Marius Bozga} {and} \bibinfo{person}{Joseph Sifakis}.} \bibinfo{year}{2024}\natexlab{}.
\newblock \bibinfo{title}{Safe by Design Autonomous Driving Systems}.
\newblock
\newblock
\showeprint[arxiv]{2405.11995}~[cs.MA]


\bibitem[Cummings(2023)]%
        {cum2023}
\bibfield{author}{\bibinfo{person}{Mary~L. Cummings}.} \bibinfo{year}{2023}\natexlab{}.
\newblock \showarticletitle{What Self-Driving Cars Tell Us About {AI} Risks}.
\newblock \bibinfo{journal}{\emph{IEEE Spectrum}} (\bibinfo{year}{2023}).
\newblock
\urldef\tempurl%
\url{https://spectrum.ieee.org/self-driving-cars-2662494269}
\showURL{%
\tempurl}


\bibitem[Denise et~al\mbox{.}(2004)]%
        {denise2004generic}
\bibfield{author}{\bibinfo{person}{Alain Denise}, \bibinfo{person}{M-C Gaudel}, {and} \bibinfo{person}{S-D Gouraud}.} \bibinfo{year}{2004}\natexlab{}.
\newblock \showarticletitle{A generic method for statistical testing}. In \bibinfo{booktitle}{\emph{15th International Symposium on Software Reliability Engineering}}. IEEE, \bibinfo{pages}{25--34}.
\newblock


\bibitem[Ding et~al\mbox{.}(2023)]%
        {ding2023survey}
\bibfield{author}{\bibinfo{person}{Wenhao Ding}, \bibinfo{person}{Chejian Xu}, \bibinfo{person}{Mansur Arief}, \bibinfo{person}{Haohong Lin}, \bibinfo{person}{Bo Li}, {and} \bibinfo{person}{Ding Zhao}.} \bibinfo{year}{2023}\natexlab{}.
\newblock \showarticletitle{A survey on safety-critical driving scenario generation—A methodological perspective}.
\newblock \bibinfo{journal}{\emph{IEEE Transactions on Intelligent Transportation Systems}} (\bibinfo{year}{2023}).
\newblock


\bibitem[Djoudi et~al\mbox{.}(2020)]%
        {djoudi2020simulation}
\bibfield{author}{\bibinfo{person}{Adel Djoudi}, \bibinfo{person}{Loic Coquelin}, {and} \bibinfo{person}{R{\'e}mi Regnier}.} \bibinfo{year}{2020}\natexlab{}.
\newblock \showarticletitle{A simulation-based framework for functional testing of automated driving controllers}. In \bibinfo{booktitle}{\emph{2020 IEEE 23rd International Conference on intelligent transportation systems (ITSC)}}. IEEE, \bibinfo{pages}{1--6}.
\newblock


\bibitem[Dosovitskiy et~al\mbox{.}(2017)]%
        {dosovitskiy2017carla}
\bibfield{author}{\bibinfo{person}{Alexey Dosovitskiy}, \bibinfo{person}{German Ros}, \bibinfo{person}{Felipe Codevilla}, \bibinfo{person}{Antonio Lopez}, {and} \bibinfo{person}{Vladlen Koltun}.} \bibinfo{year}{2017}\natexlab{}.
\newblock \showarticletitle{CARLA: An open urban driving simulator}. In \bibinfo{booktitle}{\emph{Conference on robot learning}}. PMLR, \bibinfo{pages}{1--16}.
\newblock


\bibitem[Etherington(2019)]%
        {etherington2019waymo}
\bibfield{author}{\bibinfo{person}{Darrell Etherington}.} \bibinfo{year}{2019}\natexlab{}.
\newblock \showarticletitle{Waymo has now driven 10 billion autonomous miles in simulation}.
\newblock In \bibinfo{booktitle}{\emph{Techcrunch Sessions: Mobility}}. \bibinfo{publisher}{TechCrunch}.
\newblock


\bibitem[Favaro(2021)]%
        {favaro2021exploring}
\bibfield{author}{\bibinfo{person}{Francesca Favaro}.} \bibinfo{year}{2021}\natexlab{}.
\newblock \showarticletitle{Exploring the Relationship Between" Positive Risk Balance" and" Absence of Unreasonable Risk"}.
\newblock \bibinfo{journal}{\emph{arXiv preprint arXiv:2110.10566}} (\bibinfo{year}{2021}).
\newblock


\bibitem[Favaro et~al\mbox{.}(2023)]%
        {favaro2023building}
\bibfield{author}{\bibinfo{person}{Francesca Favaro}, \bibinfo{person}{Laura Fraade-Blanar}, \bibinfo{person}{Scott Schnelle}, \bibinfo{person}{Trent Victor}, \bibinfo{person}{Mauricio Pe{\~n}a}, \bibinfo{person}{Johan Engstrom}, \bibinfo{person}{John Scanlon}, \bibinfo{person}{Kris Kusano}, {and} \bibinfo{person}{Dan Smith}.} \bibinfo{year}{2023}\natexlab{}.
\newblock \showarticletitle{Building a Credible Case for Safety: Waymo's Approach for the Determination of Absence of Unreasonable Risk}.
\newblock \bibinfo{journal}{\emph{arXiv preprint arXiv:2306.01917}} (\bibinfo{year}{2023}).
\newblock


\bibitem[Foundation(2022)]%
        {autoware2024}
\bibfield{author}{\bibinfo{person}{Autoware Foundation}.} \bibinfo{year}{2022}\natexlab{}.
\newblock \bibinfo{booktitle}{\emph{Autoware - the World’s Leading Open-Source Software project for autonomous driving}}.
\newblock
\urldef\tempurl%
\url{https://github.com/autowarefoundation/autoware}
\showURL{%
\tempurl}


\bibitem[Garcia et~al\mbox{.}(2020)]%
        {garcia2020comprehensive}
\bibfield{author}{\bibinfo{person}{Joshua Garcia}, \bibinfo{person}{Yang Feng}, \bibinfo{person}{Junjie Shen}, \bibinfo{person}{Sumaya Almanee}, \bibinfo{person}{Yuan Xia}, {and} \bibinfo{person}{Qi~Alfred Chen}.} \bibinfo{year}{2020}\natexlab{}.
\newblock \showarticletitle{A comprehensive study of autonomous vehicle bugs}. In \bibinfo{booktitle}{\emph{Proceedings of the ACM/IEEE 42nd international conference on software engineering}}. \bibinfo{pages}{385--396}.
\newblock


\bibitem[Gouraud et~al\mbox{.}(2001)]%
        {gouraud2001new}
\bibfield{author}{\bibinfo{person}{S-D Gouraud}, \bibinfo{person}{Alain Denise}, \bibinfo{person}{M-C Gaudel}, {and} \bibinfo{person}{B Marre}.} \bibinfo{year}{2001}\natexlab{}.
\newblock \showarticletitle{A new way of automating statistical testing methods}. In \bibinfo{booktitle}{\emph{Proceedings 16th Annual International Conference on Automated Software Engineering (ASE 2001)}}. IEEE, \bibinfo{pages}{5--12}.
\newblock


\bibitem[Hasuo(2022)]%
        {hasuo2022responsibility}
\bibfield{author}{\bibinfo{person}{Ichiro Hasuo}.} \bibinfo{year}{2022}\natexlab{}.
\newblock \showarticletitle{Responsibility-sensitive safety: an introduction with an eye to logical foundations and formalization}.
\newblock \bibinfo{journal}{\emph{arXiv preprint arXiv:2206.03418}} (\bibinfo{year}{2022}).
\newblock


\bibitem[Huang et~al\mbox{.}(2016)]%
        {huang2016autonomous}
\bibfield{author}{\bibinfo{person}{WuLing Huang}, \bibinfo{person}{Kunfeng Wang}, \bibinfo{person}{Yisheng Lv}, {and} \bibinfo{person}{FengHua Zhu}.} \bibinfo{year}{2016}\natexlab{}.
\newblock \showarticletitle{Autonomous vehicles testing methods review}. In \bibinfo{booktitle}{\emph{2016 IEEE 19th International Conference on Intelligent Transportation Systems (ITSC)}}. IEEE, \bibinfo{pages}{163--168}.
\newblock


\bibitem[Jard and J{\'e}ron(2005)]%
        {jard2005tgv}
\bibfield{author}{\bibinfo{person}{Claude Jard} {and} \bibinfo{person}{Thierry J{\'e}ron}.} \bibinfo{year}{2005}\natexlab{}.
\newblock \showarticletitle{TGV: theory, principles and algorithms: A tool for the automatic synthesis of conformance test cases for non-deterministic reactive systems}.
\newblock \bibinfo{journal}{\emph{International Journal on Software Tools for Technology Transfer}}  \bibinfo{volume}{7} (\bibinfo{year}{2005}), \bibinfo{pages}{297--315}.
\newblock


\bibitem[Kaur et~al\mbox{.}(2021)]%
        {kaur2021survey}
\bibfield{author}{\bibinfo{person}{Prabhjot Kaur}, \bibinfo{person}{Samira Taghavi}, \bibinfo{person}{Zhaofeng Tian}, {and} \bibinfo{person}{Weisong Shi}.} \bibinfo{year}{2021}\natexlab{}.
\newblock \showarticletitle{A survey on simulators for testing self-driving cars}. In \bibinfo{booktitle}{\emph{2021 Fourth International Conference on Connected and Autonomous Driving (MetroCAD)}}. IEEE, \bibinfo{pages}{62--70}.
\newblock


\bibitem[Koopman and Wagner(2016)]%
        {koopman2016challenges}
\bibfield{author}{\bibinfo{person}{Philip Koopman} {and} \bibinfo{person}{Michael Wagner}.} \bibinfo{year}{2016}\natexlab{}.
\newblock \showarticletitle{Challenges in autonomous vehicle testing and validation}.
\newblock \bibinfo{journal}{\emph{SAE International Journal of Transportation Safety}} \bibinfo{volume}{4}, \bibinfo{number}{1} (\bibinfo{year}{2016}), \bibinfo{pages}{15--24}.
\newblock


\bibitem[Kurakin et~al\mbox{.}(2018)]%
        {kurakin2018adversarial}
\bibfield{author}{\bibinfo{person}{Alexey Kurakin}, \bibinfo{person}{Ian~J Goodfellow}, {and} \bibinfo{person}{Samy Bengio}.} \bibinfo{year}{2018}\natexlab{}.
\newblock \showarticletitle{Adversarial examples in the physical world}.
\newblock In \bibinfo{booktitle}{\emph{Artificial intelligence safety and security}}. \bibinfo{publisher}{Chapman and Hall/CRC}, \bibinfo{pages}{99--112}.
\newblock


\bibitem[Kusano et~al\mbox{.}(2022)]%
        {kusano2022collision}
\bibfield{author}{\bibinfo{person}{Kristofer~D Kusano}, \bibinfo{person}{Kurt Beatty}, \bibinfo{person}{Scott Schnelle}, \bibinfo{person}{Francesca Favaro}, \bibinfo{person}{Cam Crary}, {and} \bibinfo{person}{Trent Victor}.} \bibinfo{year}{2022}\natexlab{}.
\newblock \showarticletitle{Collision avoidance testing of the waymo automated driving system}.
\newblock \bibinfo{journal}{\emph{arXiv preprint arXiv:2212.08148}} (\bibinfo{year}{2022}).
\newblock


\bibitem[Li et~al\mbox{.}(2023)]%
        {li2023simulation}
\bibfield{author}{\bibinfo{person}{Changwen Li}, \bibinfo{person}{Joseph Sifakis}, \bibinfo{person}{Qiang Wang}, \bibinfo{person}{Rongjie Yan}, {and} \bibinfo{person}{Jian Zhang}.} \bibinfo{year}{2023}\natexlab{}.
\newblock \showarticletitle{Simulation-Based Validation for Autonomous Driving Systems}. In \bibinfo{booktitle}{\emph{Proceedings of the 32nd ACM SIGSOFT International Symposium on Software Testing and Analysis}}. \bibinfo{pages}{842--853}.
\newblock


\bibitem[Li et~al\mbox{.}(2021)]%
        {li2021estimation}
\bibfield{author}{\bibinfo{person}{Wenfei Li}, \bibinfo{person}{Huiyun Li}, \bibinfo{person}{Kun Xu}, \bibinfo{person}{Zhejun Huang}, \bibinfo{person}{Ke Li}, {and} \bibinfo{person}{Haiping Du}.} \bibinfo{year}{2021}\natexlab{}.
\newblock \showarticletitle{Estimation of vehicle dynamic parameters based on the two-stage estimation method}.
\newblock \bibinfo{journal}{\emph{Sensors}} \bibinfo{volume}{21}, \bibinfo{number}{11} (\bibinfo{year}{2021}), \bibinfo{pages}{3711}.
\newblock


\bibitem[Lou et~al\mbox{.}(2022)]%
        {lou2022testing}
\bibfield{author}{\bibinfo{person}{Guannan Lou}, \bibinfo{person}{Yao Deng}, \bibinfo{person}{Xi Zheng}, \bibinfo{person}{Mengshi Zhang}, {and} \bibinfo{person}{Tianyi Zhang}.} \bibinfo{year}{2022}\natexlab{}.
\newblock \showarticletitle{Testing of autonomous driving systems: where are we and where should we go?}. In \bibinfo{booktitle}{\emph{Proceedings of the 30th ACM Joint European Software Engineering Conference and Symposium on the Foundations of Software Engineering}}. \bibinfo{pages}{31--43}.
\newblock


\bibitem[Mezali et~al\mbox{.}(2022)]%
        {mezali2022design}
\bibfield{author}{\bibinfo{person}{Yacine Mezali}, \bibinfo{person}{Mohamed~Idriss Khaledi}, \bibinfo{person}{Lo{\"\i}c Coquelin}, \bibinfo{person}{R{\'e}mi R{\'e}gnier}, {and} \bibinfo{person}{Jordan Martin}.} \bibinfo{year}{2022}\natexlab{}.
\newblock \showarticletitle{Design of a new measurable approach for the qualification of the behaviour of an autonomous vehicle}. In \bibinfo{booktitle}{\emph{2022 European Control Conference (ECC)}}. IEEE, \bibinfo{pages}{867--874}.
\newblock


\bibitem[Najm et~al\mbox{.}(2007)]%
        {najm2007pre}
\bibfield{author}{\bibinfo{person}{Wassim~G Najm}, \bibinfo{person}{John~D Smith}, \bibinfo{person}{Mikio Yanagisawa}, {et~al\mbox{.}}} \bibinfo{year}{2007}\natexlab{}.
\newblock \bibinfo{booktitle}{\emph{Pre-crash scenario typology for crash avoidance research}}.
\newblock \bibinfo{type}{{T}echnical {R}eport}. \bibinfo{institution}{United States. Department of Transportation. National Highway Traffic Safety~…}.
\newblock


\bibitem[Newell(1980)]%
        {newell1980physical}
\bibfield{author}{\bibinfo{person}{Allen Newell}.} \bibinfo{year}{1980}\natexlab{}.
\newblock \showarticletitle{Physical symbol systems}.
\newblock \bibinfo{journal}{\emph{Cognitive science}} \bibinfo{volume}{4}, \bibinfo{number}{2} (\bibinfo{year}{1980}), \bibinfo{pages}{135--183}.
\newblock


\bibitem[Rong et~al\mbox{.}(2020)]%
        {rong2020lgsvl}
\bibfield{author}{\bibinfo{person}{Guodong Rong}, \bibinfo{person}{Byung~Hyun Shin}, \bibinfo{person}{Hadi Tabatabaee}, \bibinfo{person}{Qiang Lu}, \bibinfo{person}{Steve Lemke}, \bibinfo{person}{M{\=a}rti{\c{n}}{\v{s}} Mo{\v{z}}eiko}, \bibinfo{person}{Eric Boise}, \bibinfo{person}{Geehoon Uhm}, \bibinfo{person}{Mark Gerow}, \bibinfo{person}{Shalin Mehta}, {et~al\mbox{.}}} \bibinfo{year}{2020}\natexlab{}.
\newblock \showarticletitle{Lgsvl simulator: A high fidelity simulator for autonomous driving}. In \bibinfo{booktitle}{\emph{2020 IEEE 23rd International conference on intelligent transportation systems (ITSC)}}. IEEE, \bibinfo{pages}{1--6}.
\newblock


\bibitem[Shalev-Shwartz et~al\mbox{.}(2017)]%
        {shalev2017formal}
\bibfield{author}{\bibinfo{person}{Shai Shalev-Shwartz}, \bibinfo{person}{Shaked Shammah}, {and} \bibinfo{person}{Amnon Shashua}.} \bibinfo{year}{2017}\natexlab{}.
\newblock \showarticletitle{On a formal model of safe and scalable self-driving cars}.
\newblock \bibinfo{journal}{\emph{arXiv preprint arXiv:1708.06374}} (\bibinfo{year}{2017}).
\newblock


\bibitem[Sifakis(2023)]%
        {sifakis2023testing}
\bibfield{author}{\bibinfo{person}{Joseph Sifakis}.} \bibinfo{year}{2023}\natexlab{}.
\newblock \showarticletitle{Testing System Intelligence}.
\newblock \bibinfo{journal}{\emph{arXiv preprint arXiv:2305.11472}} (\bibinfo{year}{2023}).
\newblock


\bibitem[Tang et~al\mbox{.}(2021)]%
        {tang2021issue}
\bibfield{author}{\bibinfo{person}{Shuncheng Tang}, \bibinfo{person}{Zhenya Zhang}, \bibinfo{person}{Jia Tang}, \bibinfo{person}{Lei Ma}, {and} \bibinfo{person}{Yinxing Xue}.} \bibinfo{year}{2021}\natexlab{}.
\newblock \showarticletitle{Issue categorization and analysis of an open-source driving assistant system}. In \bibinfo{booktitle}{\emph{2021 IEEE International Symposium on Software Reliability Engineering Workshops (ISSREW)}}. IEEE, \bibinfo{pages}{148--153}.
\newblock


\bibitem[Tang et~al\mbox{.}(2023)]%
        {tang2023survey}
\bibfield{author}{\bibinfo{person}{Shuncheng Tang}, \bibinfo{person}{Zhenya Zhang}, \bibinfo{person}{Yi Zhang}, \bibinfo{person}{Jixiang Zhou}, \bibinfo{person}{Yan Guo}, \bibinfo{person}{Shuang Liu}, \bibinfo{person}{Shengjian Guo}, \bibinfo{person}{Yan-Fu Li}, \bibinfo{person}{Lei Ma}, \bibinfo{person}{Yinxing Xue}, {and} \bibinfo{person}{Yang Liu}.} \bibinfo{year}{2023}\natexlab{}.
\newblock \showarticletitle{A survey on automated driving system testing: Landscapes and trends}.
\newblock \bibinfo{journal}{\emph{ACM Transactions on Software Engineering and Methodology}} \bibinfo{volume}{32}, \bibinfo{number}{5} (\bibinfo{year}{2023}), \bibinfo{pages}{1--62}.
\newblock


\bibitem[Team(2017)]%
        {team2017apollo}
\bibfield{author}{\bibinfo{person}{Baidu~Apollo Team}.} \bibinfo{year}{2017}\natexlab{}.
\newblock \bibinfo{title}{Apollo: Open source autonomous driving}.
\newblock
\newblock


\bibitem[Verschure and Althaus(2003)]%
        {verschure2003real}
\bibfield{author}{\bibinfo{person}{Paul~FMJ Verschure} {and} \bibinfo{person}{Philipp Althaus}.} \bibinfo{year}{2003}\natexlab{}.
\newblock \showarticletitle{A real-world rational agent: unifying old and new AI}.
\newblock \bibinfo{journal}{\emph{Cognitive science}} \bibinfo{volume}{27}, \bibinfo{number}{4} (\bibinfo{year}{2003}), \bibinfo{pages}{561--590}.
\newblock


\bibitem[Wegener et~al\mbox{.}(2001)]%
        {wegener2001evolutionary}
\bibfield{author}{\bibinfo{person}{Joachim Wegener}, \bibinfo{person}{Andr{\'e} Baresel}, {and} \bibinfo{person}{Harmen Sthamer}.} \bibinfo{year}{2001}\natexlab{}.
\newblock \showarticletitle{Evolutionary test environment for automatic structural testing}.
\newblock \bibinfo{journal}{\emph{Information and software technology}} \bibinfo{volume}{43}, \bibinfo{number}{14} (\bibinfo{year}{2001}), \bibinfo{pages}{841--854}.
\newblock


\bibitem[Zhong et~al\mbox{.}(2021)]%
        {zhong2021survey}
\bibfield{author}{\bibinfo{person}{Ziyuan Zhong}, \bibinfo{person}{Yun Tang}, \bibinfo{person}{Yuan Zhou}, \bibinfo{person}{Vania de~Oliveira Neves}, \bibinfo{person}{Yang Liu}, {and} \bibinfo{person}{Baishakhi Ray}.} \bibinfo{year}{2021}\natexlab{}.
\newblock \showarticletitle{A survey on scenario-based testing for automated driving systems in high-fidelity simulation}.
\newblock \bibinfo{journal}{\emph{arXiv preprint arXiv:2112.00964}} (\bibinfo{year}{2021}).
\newblock


\end{thebibliography}

\end{document}